\documentclass[pdflatex,sn-mathphys]{sn-jnl}

\usepackage[nameinlink]{cleveref}
\usepackage{ragged2e}
\usepackage{lineno}
\usepackage{tabularx}
\usepackage{ltablex}

\newcommand\python{\texttt{Python}}

\usepackage{comment}
\newcolumntype{L}{>{\raggedright\arraybackslash}X}

\newcommand\micro{\textmu}
\newcommand\micron{\micro m}
\newcommand\microns \micron

\newcommand\msun{M$_{\odot}$}
\usepackage[version=4]{mhchem} 

\jyear{2022}%

\usepackage{astjnlabbrev-jh}

\begin{document}

\title[Zieba --- TRAPPIST-1\,c with JWST]{No thick carbon dioxide atmosphere on the rocky exoplanet TRAPPIST-1\,c}


\author*[1,2]{Sebastian Zieba}\email{zieba@mpia.de}
\author[1]{Laura Kreidberg}
\author[3,4]{Elsa Ducrot}
\author[5]{Michaël Gillon}
\author[6]{Caroline Morley}
\author[7]{Laura Schaefer}
\author[8]{Patrick Tamburo}
\author[9]{Daniel D. B. Koll}
\author[9]{Xintong Lyu}
\author[1,10]{Lorena Acuña}
\author[11,12]{Eric Agol}
\author[13,14]{Aishwarya R. Iyer}
\author[15,16]{Renyu Hu}
\author[11,12]{Andrew P. Lincowski}
\author[11,12]{Victoria S. Meadows}
\author[17]{Franck Selsis}
\author[18,19]{Emeline Bolmont}
\author[20,21]{Avi M. Mandell}
\author[11,12]{Gabrielle Suissa}

\affil[1]{Max-Planck-Institut f\"ur Astronomie, K\"onigstuhl 17, D-69117 Heidelberg, Germany}
\affil[2]{Leiden Observatory, Leiden University, Niels Bohrweg 2, 2333CA Leiden, The Netherlands}
\affil[3]{Université Paris-Saclay, Université Paris-Cité, CEA, CNRS, AIM}
\affil[4]{Paris Region Fellow, Marie Sklodowska-Curie Action}
\affil[5]{Astrobiology Research Unit, University of Liège, Allée du 6 août 19, 4000 Liège, Belgium}
\affil[6]{Department of Astronomy, University of Texas at Austin, 2515 Speedway, Austin TX 78712, USA}
\affil[7]{Department of Earth and Planetary Sciences, Stanford University, Stanford, CA, USA}
\affil[8]{Department of Astronomy \& The Institute for Astrophysical Research, Boston University, 725 Commonwealth Ave., Boston, MA 02215, USA}
\affil[9]{Department of Atmospheric and Oceanic Sciences, Peking University, Beijing, People's Republic of China}
\affil[10]{Aix-Marseille Université, CNRS, CNES, Institut Origines, LAM, Marseille, France}
\affil[11]{Department of Astronomy and Astrobiology Program, University of Washington, Box 351580, Seattle, Washington 98195, USA}
\affil[12]{NASA Nexus for Exoplanet System Science, Virtual Planetary Laboratory Team, Box 351580, University of Washington, Seattle, Washington 98195, USA}
\affil[13]{School of Earth and Space Exploration, Arizona State University, 525 E. University Dr., Tempe AZ 85281}
\affil[14]{NASA FINESST Fellow}
\affil[15]{Jet Propulsion Laboratory, California Institute of Technology, Pasadena, CA, USA}
\affil[16]{Division of Geological and Planetary Sciences, California Institute of Technology, Pasadena, CA, USA}
\affil[17]{Laboratoire d'astrophysique de Bordeaux, Univ. Bordeaux, CNRS, B18N, all{\'e}e Geoffroy Saint-Hilaire, 33615 Pessac, France}
\affil[18]{Observatoire astronomique de l'Universit\'e de Gen\`eve, chemin Pegasi 51, CH-1290 Versoix, Switzerland}
\affil[19]{Centre Vie dans l’Univers, Universit\'e de Gen\`eve, Geneva, Switzerland}
\affil[20]{NASA Goddard Space Flight Center, 8800 Greenbelt Rd, Greenbelt, MD, USA}
\affil[21]{Sellers Exoplanet Environments Collaboration, NASA Goddard}

\maketitle

\textbf{
Seven rocky planets orbit the nearby dwarf star TRAPPIST-1, providing a unique opportunity to search for atmospheres on small planets outside the Solar System \citep{Gillon2017}. Thanks to the recent launch of the James Webb Space Telescope (JWST), possible atmospheric constituents such as carbon dioxide (CO$_2$) are now detectable \citep{Morley2017, Lincowski2018}. Recent JWST observations of the innermost planet TRAPPIST-1\,b showed that it is most probably a bare rock without any CO$_2$ in its atmosphere \citep{Greene2023}. Here we report the detection of thermal emission from the dayside of TRAPPIST-1\,c with the Mid-Infrared Instrument (MIRI) on JWST at 15~\micron. We measure a planet-to-star flux ratio of $f_p/f_* = 421 \pm 94$ parts per million (ppm) which corresponds to an inferred dayside brightness temperature of $380\pm31$ K. This high dayside temperature disfavours a thick, CO$_2$-rich atmosphere on the planet. The data rule out cloud-free O$_2$/CO$_2$ mixtures with surface pressures ranging from 10 bar (with 10 ppm CO$_2$) to 0.1 bar (pure CO$_2$). A Venus-analogue atmosphere with sulfuric acid clouds is also disfavoured at 2.6$\sigma$ confidence. Thinner atmospheres or bare-rock surfaces are consistent with our measured planet-to-star flux ratio. The absence of a thick, CO$_2$-rich atmosphere on TRAPPIST-1\,c suggests a relatively volatile-poor formation history, with less than  $9.5^{+7.5}_{-2.3}$ Earth oceans of water. If all planets in the system formed in the same way, this would indicate a limited reservoir of volatiles for the potentially habitable planets in the system.\\
}

Little is known about the compositions of terrestrial exoplanet atmospheres, or even whether atmospheres are present at all. The atmospheric composition depends on many unknown factors, including the initial inventory of volatiles, outgassing resulting from volcanism, and possible atmospheric escape and collapse \citep[see e.g.,][]{Wordsworth2022}. Atmospheric escape may also depend on the spectral type of the host star: planets around M dwarfs may be particularly vulnerable to atmospheric loss during the long pre-main sequence phase \cite{Luger2015}. The only way to robustly determine whether a terrestrial exoplanet has an atmosphere is to study it directly, through its thermal emission, reflected light, or transmission spectrum. The tightest constraints on atmospheric properties so far have come from observations of the thermal emission of LHS 3844\,b, GJ 1252\,b, and TRAPPIST-1\,b. The measurements revealed dayside temperatures consistent with no redistribution of heat on the planet and no atmospheric absorption from carbon dioxide \cite{Kreidberg2019, Crossfield2022, Greene2023}. These results motivate observations of cooler planets, which may be more likely to retain atmospheres.\\

We observed four eclipses of TRAPPIST-1\,c with MIRI on JWST in imaging mode. The observations took place on 27 October, 30 October, 6 November, and 30 November 2022 as part of General Observer programme 2304. Each visit had a duration of approximately 192 minutes, covering the 42-minute eclipse duration of TRAPPIST-1\,c as well as out-of-eclipse baseline to correct for instrumental systematic noise. The observations used the MIRI F1500W filter, a 3~\micron-wide bandpass centred at 15 \micron, which covers a strong absorption feature from CO$_2$. Across the four visits, we collected 1,190 integrations in total using the FULL subarray. See Methods for further details on the design of the observations.\\

\begin{figure}
\centering
\includegraphics[width=0.85\textwidth]{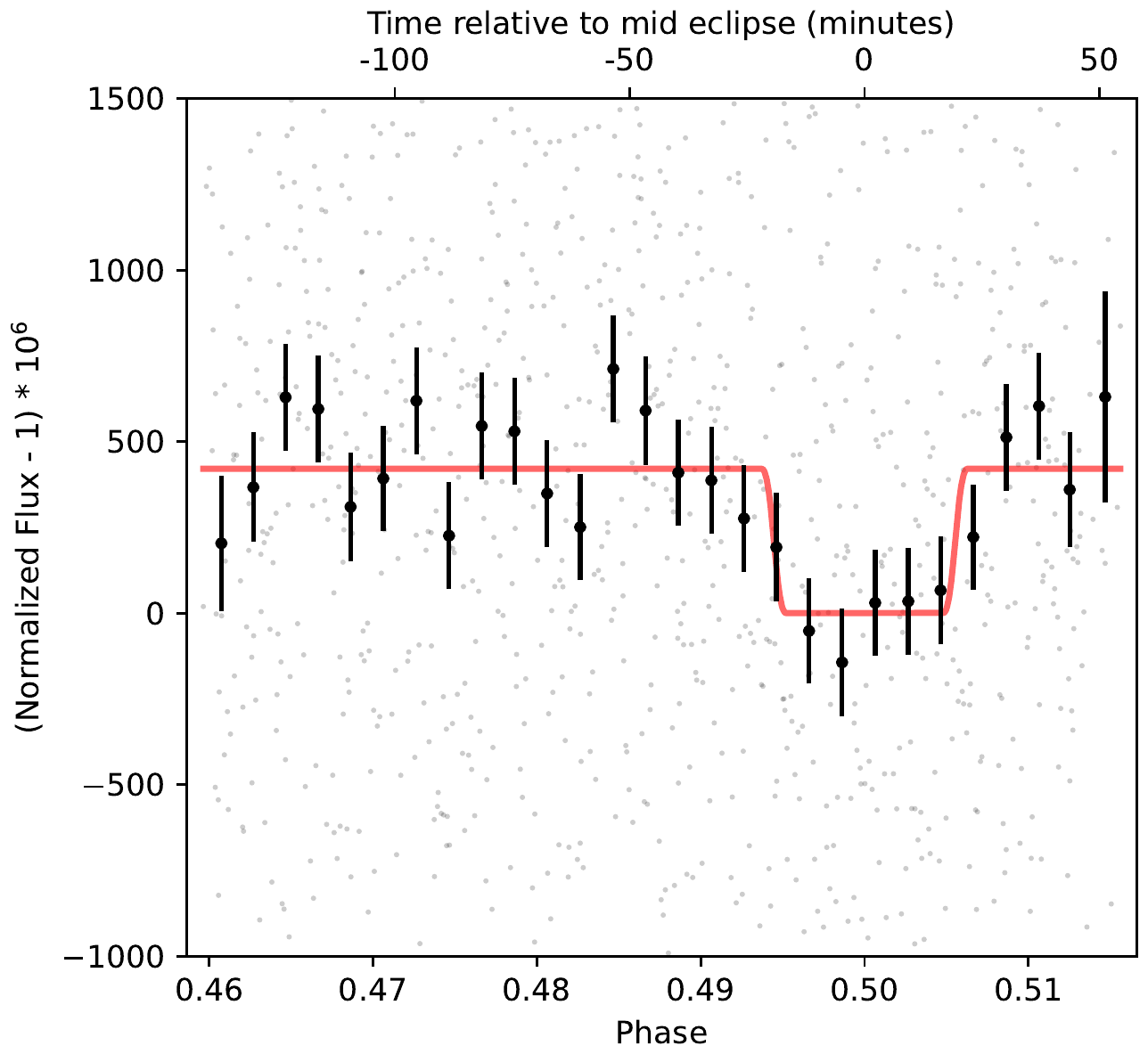}
\caption{\textbf{Eclipse light curve of TRAPPIST-1\,c taken with MIRI F1500W.} The phase-folded secondary eclipse light curve of TRAPPIST-1\,c, measured with the \text{JWST}/MIRI imager at 15 \micron. The eclipse is centred at orbital phase 0.5 and has a measured depth of $f_p/f_* = 421 \pm 94$ ppm. The light curve includes four visits (that is, four eclipses), each spanning approximately 3.2 hours. To make the eclipse more easily visible, we binned the individual integrations (grey points) into 28 orbital phase bins (black points with 1$\sigma$ error bars). The light curve was normalized and divided by the best-fit instrument systematic model. The best-fit eclipse model is shown with the solid red line. The data and fit presented in this figure are based on the \texttt{SZ} reduction, one of the four independent reductions we performed in this work.}
\label{fig:eclipse}
\end{figure}

We performed four independent reductions of the data using the publicly available \texttt{Eureka!} code \cite{Bell2022} as well as several custom software pipelines. Each reduction extracted the light curve of TRAPPIST-1 using aperture photometry (see Methods and Table \ref{tab:reductions}). We then fitted the light curves with an eclipse model and a range of different parameterizations for the instrumental systematics, including a polynomial in time, exponential ramps, and decorrelation against the position and width of the point spread function (PSF). For the different analyses, the scatter of the residuals in the fitted light curves had a root mean square (rms) variability ranging from 938 - 1,079 ppm, within 1.06 - 1.22 times the predicted photon noise limit when using a corrected gain value \cite{Bell2023}. We estimated the eclipse depths using Markov chain Monte Carlo (MCMC) fits to the data, which marginalized over all the free parameters in the analysis. The resulting eclipse depths from the four data analyses are consistent and agree to well within 1$\sigma$ (see Table \ref{tab:analysis}). The phase-folded light curve from one of the reductions can be seen in Figure \ref{fig:eclipse}. To determine the final eclipse depth, we took the mean value and uncertainty from the different reductions. To account for systematic error owing to differences in data reduction and modelling choices, we also added an additional 6 ppm to the uncertainty in quadrature, which corresponds to the standard deviation in the eclipse depth between the four analyses. The resulting eclipse depth is $f_p/f_* = 421 \pm 94$ ppm.\\

\begin{figure}[!htbp]
\centering
\includegraphics[width=0.49\textwidth]{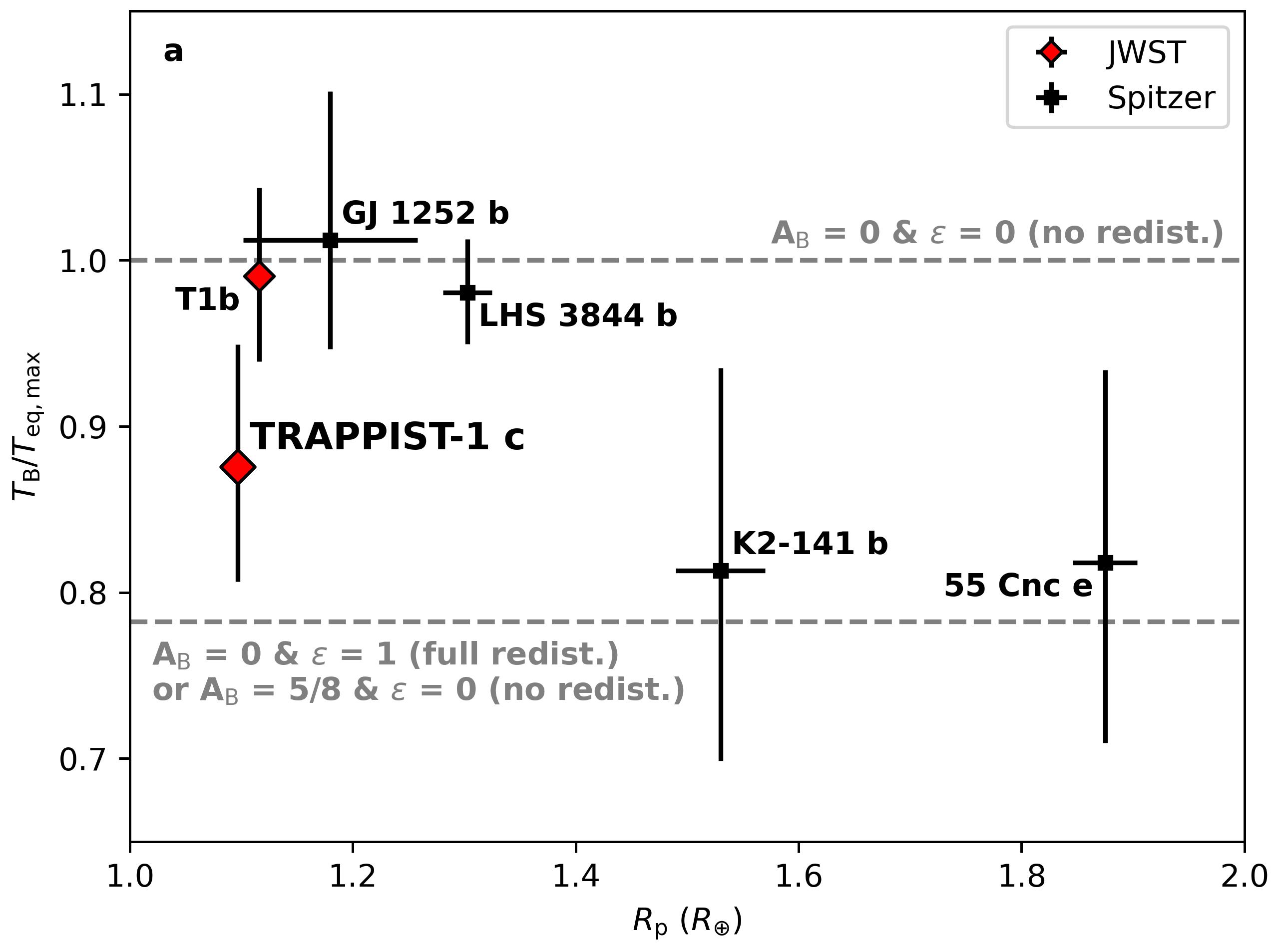}
\includegraphics[width=0.49\textwidth]{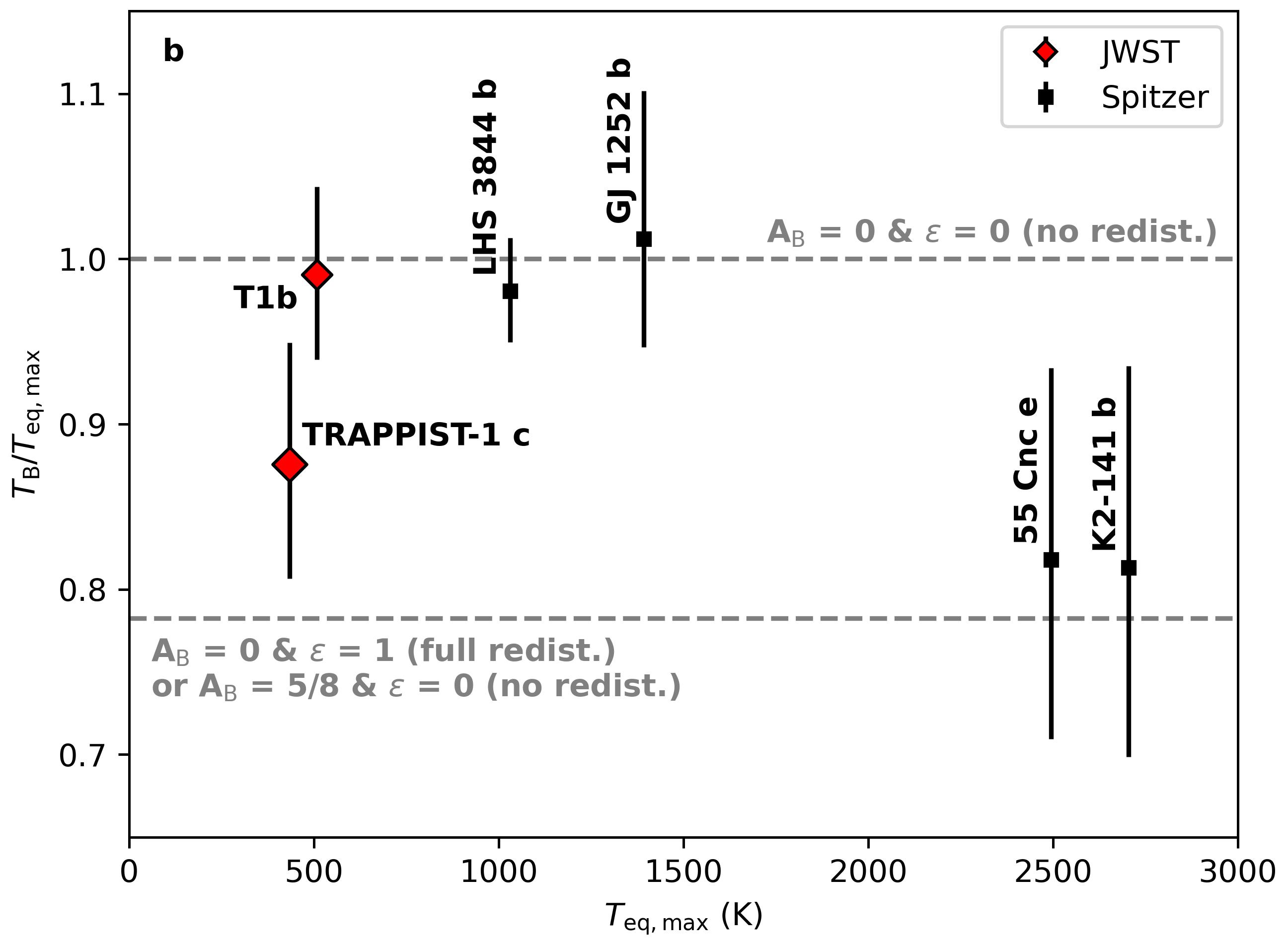}
\caption{\textbf{Comparison of small exoplanets with measured infrared emission.} Following \cite{Crossfield2022}, we show the normalized dayside brightness temperature for super-Earths ($R_p \lt 2R_\oplus$) with measured thermal emission, as a function of planet size (a) and maximum equilibrium temperature (b). The brightness temperatures are normalized relative to predictions for a bare rock with zero albedo and zero heat redistribution, $T_{\textrm{eq,max}}$. The thermal emission of TRAPPIST-1\,c has been detected in this work at 15 \micron. The other planets are TRAPPIST-1\,b (T1b in plot; also at 15 \micron) and planets that have been observed with Spitzer's IRAC Channel 2 at 4.5 \micron. The uncertainties on the radius for the planets in the TRAPPIST-1 system are smaller than the marker symbol. Error bars show 1$\sigma$ uncertainties.}
\label{fig:TB_Teq}
\end{figure}

From the measured eclipse depth, we derive a brightness temperature of $380 \pm 31$ K for TRAPPIST-1\,c. The innermost planet in the system, TRAPPIST-1 b, was found to have a brightness temperature of $503^{+26}_{-27}$ K \citep{Greene2023}. Compared with previous detections of thermal emission from small (R$_p$ $\lt$ 2 $R_\oplus$) rocky planets (see Fig. \ref{fig:TB_Teq}) these temperatures are more than 500 K cooler (the previous lowest measured brightness temperature was $1,040 \pm 40$ K for LHS 3844\,b \citep{Kreidberg2019}). TRAPPIST-1\,c is the first exoplanet with measured thermal emission that is comparable with the inner planets of the Solar System; Mercury and Venus have equilibrium temperatures of 440 K and 227 K, respectively, assuming uniform heat redistribution and taking the measured Bond albedo values ($A_{B,\textrm{Mecury}}$ = 0.068, $A_{B,\textrm{Venus}}$ = 0.76) from \cite{Moroz1985, Mallama2002}. Our measured temperature for TRAPPIST-1\,c is intermediate between the two limiting cases for the atmospheric circulation for a zero-albedo planet: zero heat redistribution (430 K; expected for a fully absorptive bare rock), versus global heat redistribution (340 K; expected for a thick atmosphere). This intermediate value hints at either a moderate amount of heat redistribution by an atmosphere ($\varepsilon = 0.66^{+0.26}_{-0.33}$) or a non-zero Bond albedo for a rocky surface ($A_B = 0.57^{+0.12}_{-0.15}$) (following the parameterization described in \cite{Cowan2011}).\\

\begin{figure}
\centering
\includegraphics[width=0.85\textwidth]{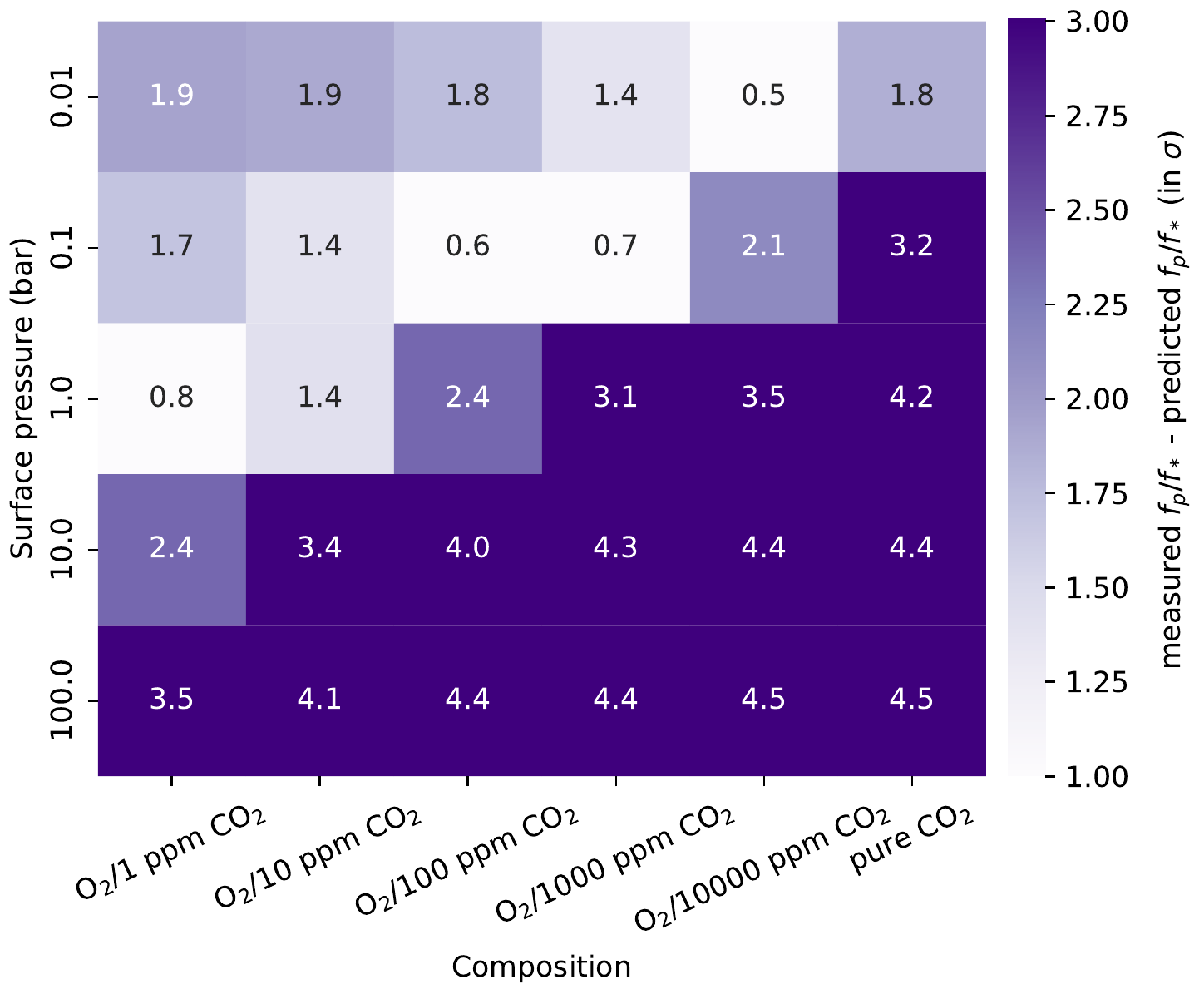}
\caption{\textbf{Grid plot comparing a suite of atmospheric models to the measured eclipse depth.} Comparison between the measured eclipse depth and a suite of different O$_2$/CO$_2$, cloud-free atmospheres for TRAPPIST-1\,c with varying surface pressures and compositions. Darker grid cells indicate that we more significantly rule out this specific atmospheric scenario. The number in each cell is the absolute difference between each model and the observations in units of sigma. The lower the modelled atmosphere is in the grid, the higher its surface pressure. The rightmost column shows pure CO$_2$ atmospheres. The other columns are O$_2$-dominated atmospheres with different amounts of CO$_2$ ranging from 1 ppm (= 0.0001\%) to 10,000 ppm (= 1\%).}
\label{fig:gridplot}
\end{figure}

To further explore which possible atmospheres are consistent with the data, we compared the dayside flux with a grid of cloud-free, O$_2$-dominated models with a range of surface pressures (0.01 bar - 100.0 bar) and CO$_2$ contents (1 ppm - 10,000 ppm). Also, we generated cloud-free, pure CO$_2$ atmospheres using the same surface pressures. The models account for both atmospheric heat redistribution and absorption by constituent gasses \citep{Koll2019, Morley2017, Kreidberg2019} and assume a Bond albedo of 0.1 (see Methods). O$_2$/CO$_2$ mixtures are expected for hot rocky planets orbiting late M-type stars as the planet's H$_2$O photodissociates and escapes over time, leaving a desiccated atmosphere dominated by O$_2$ \citep{Luger2015,Schaefer2016, Bolmont2017}. Substantial CO$_2$ (up to about 100 bar) is expected to accumulate from outgassing and does not escape as easily as H$_2$O \citep{Dorn2018, Kane2020}. For these mixtures, the predicted eclipse depth decreases with increasing surface pressure and with increasing CO$_2$ abundance, owing to the strong CO$_2$ absorption feature centred at 15 \micron. Strong inversions for a planet in this parameter space are not expected \cite{Malik2019}. With our measured eclipse depth, we rule out all thick atmospheres with surface pressures $P_{\textrm{surf}} \geq$ 100 bar (see Fig. \ref{fig:gridplot}). For the conservative assumption that the CO$_2$ content is at least 10 ppm, we rule out $P_{\textrm{surf}} \geq$ 10 bar. For cloud-free, pure CO$_2$ atmospheres we can rule out surface pressures $P_{\textrm{surf}} \geq$ 0.1 bar. As the TRAPPIST-1 planets have precisely measured densities, interior-structure models can give constraints on the atmospheric surface pressures, that is, higher surface pressures would decrease the observed bulk density of the planet. Our findings here agree with these models, which put an upper limit of 160 bar (80 bar) on the surface pressure at a 3$\sigma$ (1$\sigma$) level \citep{Acuna21}.\\

\begin{figure}
\centering
\includegraphics[width=0.99\textwidth]{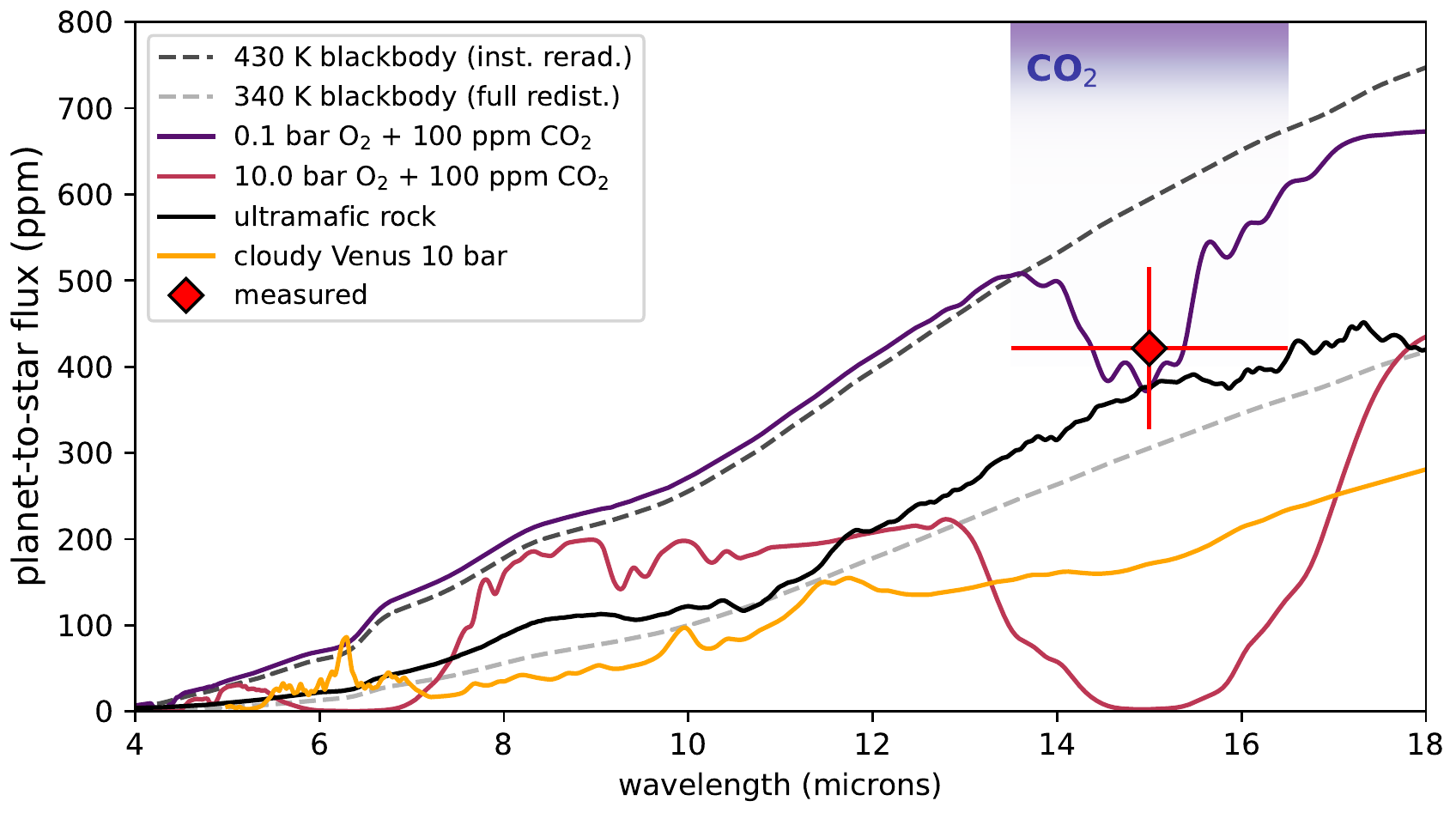}
\caption{\textbf{Observed flux of TRAPPIST-1 c and various emission models.} Simulated emission spectra compared with the measured eclipse depth of TRAPPIST-1\,c (red diamond, with the vertical error bar representing the 1$\sigma$ uncertainty on the measured eclipse depth). The CO$_2$ feature overlaps directly with the MIRI F1500W filter used for these observations. The two limiting cases for the atmospheric circulation for a zero-albedo planet (zero heat redistribution, that is, instant reradiation of incoming flux and global heat redistribution) are marked with dashed lines. Two cloud-free, O$_2$/CO$_2$ mixture atmospheres are shown with purple and red solid lines. They show decreased emission at 15 \micron\ owing to CO$_2$ absorption. A bare-rock model assuming an unweathered ultramafic surface of the planet with a Bond albedo of 0.5 is shown by the solid black line (see text for more information on weathering, including a full comparison of our measurement to a suite of surfaces in Fig. \ref{fig:bare_rocks}). The cloudy Venus forward model with a surface pressure of 10 bars is shown with a solid yellow line.}
\label{fig:data_vs_models}
\end{figure}

We also compared the measured dayside brightness with several physically motivated forward models inspired by Venus. The insolation of TRAPPIST-1\,c is just 8\% greater than that of Venus \cite{Delrez2018}, so it is possible that the two planets could have similar atmospheric chemistry.  We used a coupled climate-photochemistry model to simulate an exact Venus-analogue composition (96.5\%, CO$_2$ 3.5\% N$_2$, and Venus lower atmospheric trace gases), both with and without H$_2$SO$_4$ aerosols \citep{Lincowski2018} (see Methods). The assumed surface pressure was 10 bar, which would produce similar results to a true 93 bar Venus-analogue, because for both cases, the emitting layer and cloud deck lie at similar pressures. We find that these cloudy and cloud-free Venus-like atmospheres are disfavoured at 2.6$\sigma$ and 3.0$\sigma$, respectively (see Fig. \ref{fig:data_vs_models} for the 10 bar cloudy Venus spectrum). The cloudy case is marginally more consistent with the data because the SO$_2$ aerosols locally warm the atmosphere, providing a warmer emission temperature within the core of the 15~\micron\ band, and therefore a larger secondary eclipse depth.\\

Finally, we compared the measured flux with bare-rock models with a variety of surface compositions, including basaltic, feldspathic, Fe-oxidized (50$\%$ nanophase haematite, 50$\%$ basalt), granitoid, metal-rich (FeS$_2$), and ultramafic compositions \citep{hu2012}. We also considered space weathering for these models, as TRAPPIST-1\,c should have been substantially weathered owing to its proximity to the host star. On the Moon and Mercury, space weathering darkens the surface by means of the formation of iron nanoparticles \citep{hapke2001}. On TRAPPIST-1\,c, this process would similarly darken the surface and therefore increase the eclipse depth. We find that all bare-rock surfaces are consistent with the data (see Fig. \ref{fig:data_vs_models} for an unweathered ultramafic surface and Fig. \ref{fig:bare_rocks} for all surfaces that we considered). Overall, fresh low-albedo surfaces (for example basalt) or weathered surfaces are all compatible with the data, comparable with the likely bare-rock exoplanet LHS 3844\,b \citep{Kreidberg2019}.  The highest albedo models, unweathered feldspathic and granitoid surfaces, are a marginally worse fit (consistent at the 2$\sigma$ level).\\

\begin{figure}[ht!]
\centering
\includegraphics[width=0.85\textwidth, trim = 0cm 4.0cm 0cm 5.1cm, clip]{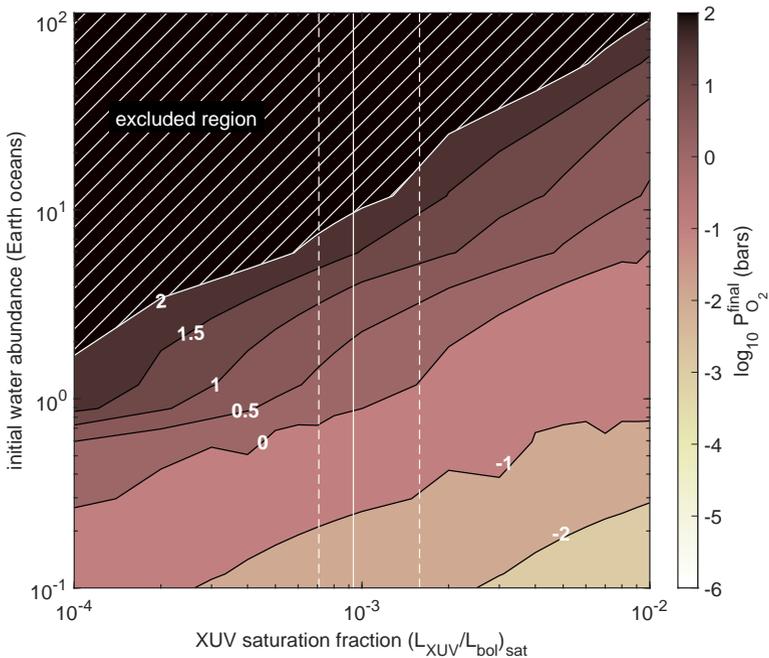}
\caption{\textbf{Final oxygen atmospheric pressure for TRAPPIST-1 c after 7.5 Gyr of energy-limited escape.} We explore different initial planetary water abundances and the amount of XUV the planet receives during the star's saturated activity period \citep{Luger2015}, described as a fraction of its total bolometric luminosity. The vertical lines represent the nominal XUV saturation fraction of $\log_{10}(L_{XUV}/L_{bol}) = {-3.03}^{+0.23}_{-0.12}$ as estimated by \cite{Fleming2020}. We assume an escape efficiency of 0.1. The white numbers are the contour values for the logarithm of the atmospheric pressure in bars. Our upper limit on surface pressure of 10 - 100 bars implies an initial water abundance of approximately 4 - 10 Earth oceans.}
\label{fig:escape}
\end{figure}

To put our results into context with the formation history of the planet, we ran a grid of atmospheric evolution models over a range of initial water inventories (0.1 - 100 Earth oceans) and extreme ultraviolet (XUV) saturation fractions for the host star ($10^{-4}$-${10^{-2}}$) (see Fig. \ref{fig:escape}). The model incorporates outgassing, escape of water vapour and oxygen, and reaction of oxygen with the magma ocean \cite{Schaefer2016}.  For an XUV saturation fraction of 10$^{-3}$ being a typical value for a low-mass star \cite{Chadney2015}, we find that the final surface pressure of oxygen could range over several orders of magnitude (0.1 - 100 bar), depending on the initial water inventory (see Fig. \ref{fig:escape}). Our measured eclipse depth disfavours surface pressures at the high end of this range (greater than 100 bar) for conservative CO$_2$ abundances, implying that TRAPPIST-1\,c most likely formed with a relatively low initial water abundance of less than $9.5^{+7.5}_{-2.3}$ Earth oceans. For higher CO$_2$ abundances ($> 10$ ppm), we rule out surface pressures greater than 10 bars, implying that the planet formed with less than $4 ^{+1.3}_{-0.8}$ Earth oceans. Our result suggests that rocky planets around M-dwarf stars may form with a smaller volatile inventory or experience more atmospheric loss than their counterparts around Sun-like stars. This finding motivates further study of the other planets in the TRAPPIST-1 system to assess whether a low volatile abundance is a typical outcome, particularly for the planets in the habitable zone.\\  

\clearpage
\newpage
\backmatter

\section*{Methods}

\subsection*{JWST MIRI Observations}

\begin{figure}[!htbp]
\centering
\includegraphics[width=0.95\textwidth]{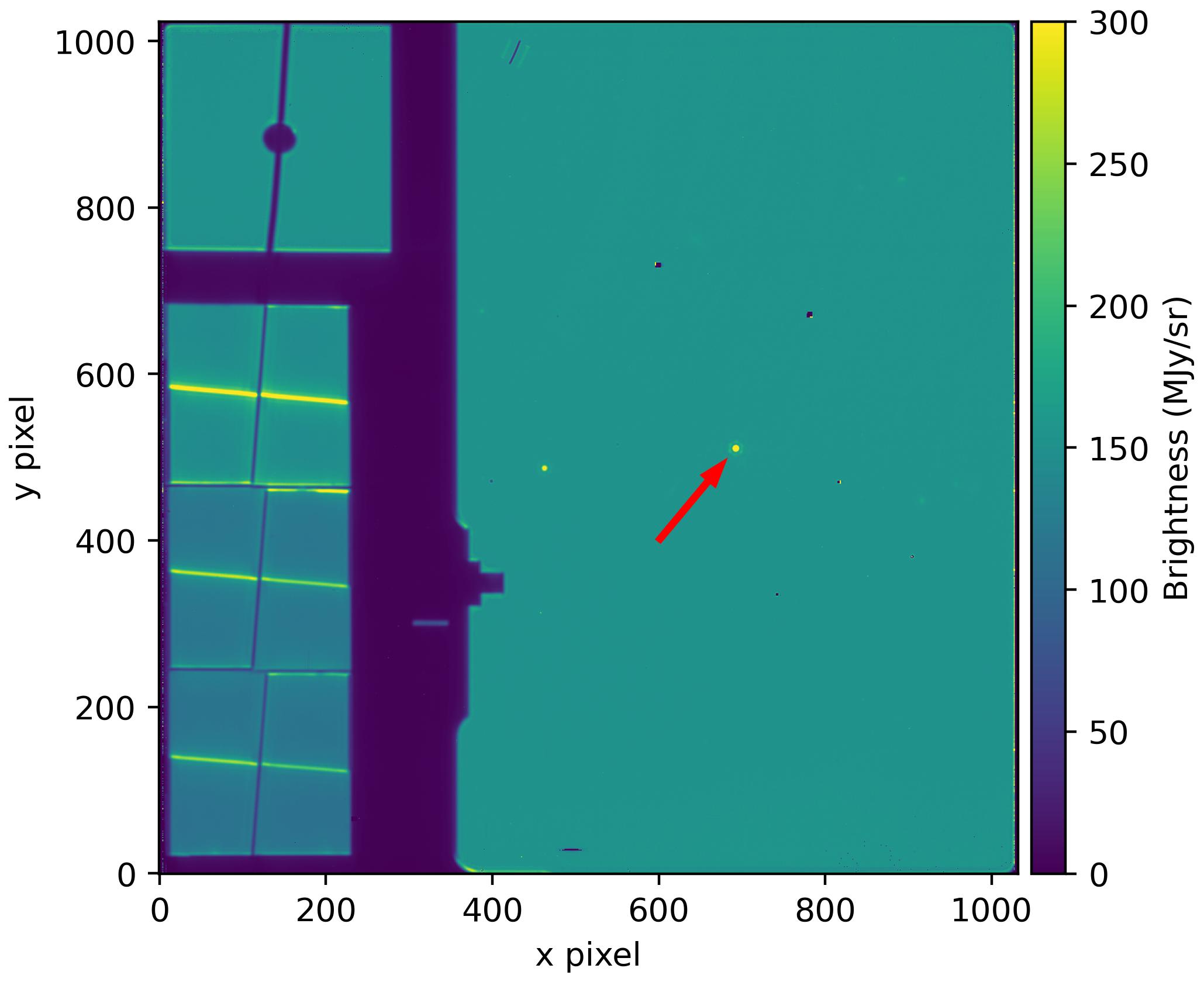}
\caption{\textbf{Example of a MIRI integration using the FULL array.} An integration taken during our observations showing the MIRI imager focal plane. The majority of the FULL array is taken up by the imager field of view on the right side. TRAPPIST-1 is centred on the imager highlighted by the red arrow. The left side of the imager was not used in our analysis and consists out of the Lyot coronagraph (top left) and the three 4-quadrant phase masks coronagraphs (lower left).}
\label{fig:fullarray}
\end{figure}

As part of JWST General Observer (GO) program 2304 (principal investigator (PI): L. Kreidberg) \cite{Kreidberg2021}, we observed four eclipses of the planet TRAPPIST-1\,c (see Table \ref{tab:observations}). They were taken on 25 October, 27 October, 30 October, and 6 November 2022 with JWST's MIRI instrument using the F1500W filter. The observations used the FULL subarray with FASTR1 readout and 13 groups per integration. Each visit had a duration of approximately 3.2 hours. We did not perform target acquisition for any of the visits because it was not enabled for MIRI imaging observations during cycle 1. However, the blind pointing precision of JWST was perfectly sufficient to place the target well centered on the field of view of the full array (74'' x 113''). Figure \ref{fig:fullarray} shows one of the integrations with the FULL array. 

\begin{table}[!h]
\centering
\caption{\textbf{Summary of the observations in JWST program GO 2304.}}
\label{tab:observations}
\begin{tabular}{l|cccc}\hline
                & visit 1       & visit 2       & visit 3      & visit 4       \\\hline
date            & 27. Oct. 2022 & 30. Oct. 2022 & 6. Nov. 2022 & 30. Nov. 2022 \\
start time      & 14:08:35              &  00:09:07             & 06:32:33             &  11:49:52             \\
end time        & 17:21:29              &  03:21:23             & 09:44:49             &  15:02:47             \\
duration (hours)& 3.21          & 3.19          & 3.19        & 3.21         \\
Nint            & 298           & 297           & 297          & 298           \\
Ngroups/int     & 13            & 13            & 13           & 13            \\
stability rms x (pixel) &  0.0032  &  0.0040 &  0.0034 & 0.0031              \\
stability rms y (pixel) &  0.0059  & 0.0074  & 0.0062  & 0.0051             \\\hline
\end{tabular}
\end{table}

\clearpage

\subsection*{Data Reduction}

We performed four different reductions of the data collected for JWST program GO 2304. The assumptions made by the reductions are listed in Table \ref{tab:reductions}. In the following, we describe the individual reductions. 

\begin{table}[h!]
\footnotesize
\centering
\caption{\textbf{Details of the four different data reductions.}}
\label{tab:reductions}
\begin{tabularx}{\textwidth}{L|LLLL}\hline
Step/Parameter &
SZ reduction &
ED reduction &
MG reduction &
PT reduction 
\\\hline\rule{0pt}{12pt}Stage 1 Run? &
Yes &
Yes &
Yes &
- 
\\\rule{0pt}{12pt}Jump correction &
Jump rejection threshold of 7.0, 6.0, 7.0, 5.0 sigma &
No jump correction &
No jump correction &
-
\\\rule{0pt}{12pt}ramp weighting &
default &
uniform &
uniform &
- 
\\\rule{0pt}{12pt}Stage 2 Run? &
Yes &
Yes &
Yes &
-
\\\rule{0pt}{12pt}photom step &
skipped &
skipped &
skipped &
-
\\\rule{0pt}{12pt}Stage 3 notes&
- &
- &
- &
Used Calibration Level 2 data directly from MAST
\\\rule{0pt}{12pt}centroid position determination method &
2D Gaussian fit to target &
2D Gaussian fit to target &
2D Gaussian fit to target &
2D Gaussian fit to target 
\\\rule{0pt}{12pt}target aperture shape &
circle &
circle &
circle &
circle 
\\\rule{0pt}{12pt}aperture radius &
4.0, 4.0, 4.0, 4.0 pixels around the centroid &
3.7, 4.0, 3.6, 3.8 pixels around the centroid &
3.6, 3.6, 3.5, 3.4 pixels around the centroid &
4.4, 4.1, 3.9, 3.5 pixels around the centroid
\\\rule{0pt}{12pt}partial pixels treatment &
pixels were supersampled using a bilinear interpolation &
pixels were supersampled using a bilinear interpolation &
used daophot/phot routine in IRAF
\\\rule{0pt}{12pt}background region shape &
annulus around the centroid &
annulus around the centroid &
annulus around the centroid &
annulus around the centroid 
\\\rule{0pt}{12pt}background aperture size &
25-41 for each visit &
20-35 for each visit &
20-35 for each visit &
30-45 for each visit
\\\rule{0pt}{12pt}background subtraction method &
subtracted the median calculated within the annulus from the whole frame &
subtracted the median calculated within the annulus  from the whole frame &
Computation of the mode of the sky pixel distribution using the mean and median, after 3-sigma clipping of outliers. &
mean of sigma-clipped pixel values within the annulus was subtracted from the whole frame (4-sigma clipping threshold)
\\\rule{0pt}{12pt}Details of outlier rejection/time series clipping &
No outliers removed with sigma clipping. first 10 integrations removed &
sigma clipping set to 4, no exposure removed &
5-Sigma clipping with 20-min moving median. 5, 14, 6, and 4 integrations removed &
No outliers removed with sigma clipping. First exposure removed from each visit
\\\hline
\end{tabularx}
\end{table}
\footnotetext[1]{see \url{https://iraf.net/irafdocs/apspec.pdf}, page 15}

\subsubsection*{Data Reduction SZ}

\begin{figure}
\centering
\includegraphics[width=0.95\textwidth]{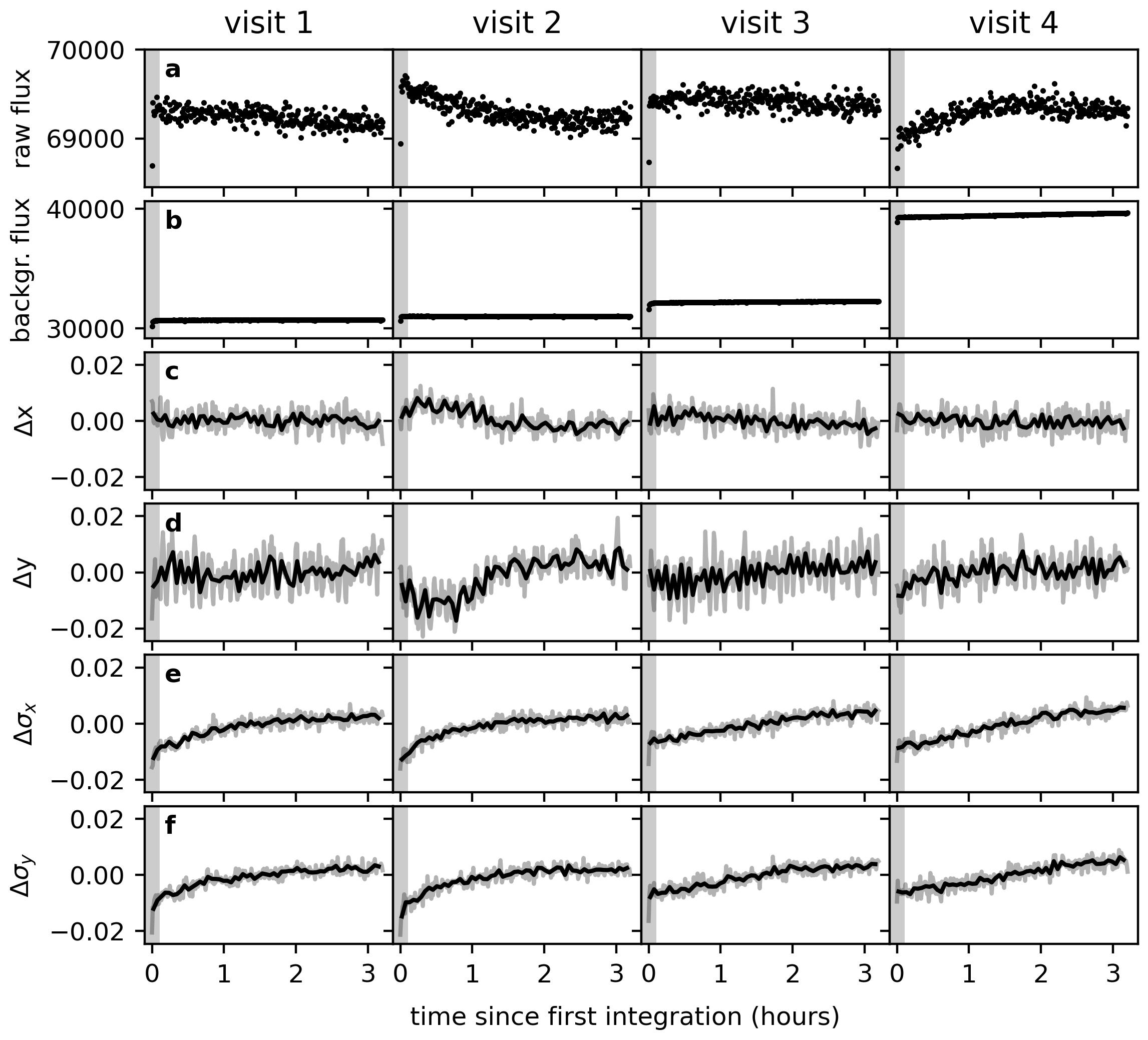}
\caption{\textbf{Diagnostic plot of all four visits taken during JWST program GO 2304 based on the \texttt{SZ} reduction.} Every column corresponds to a visit. \textbf{a and b.} The top and second rows show the raw and background flux in units of electrons per integration per pixel, respectively. The raw flux is referring to the flux level within the target aperture before the subtraction of the background flux. \textbf{c -- f.} The following rows are depicting the properties of the centroid over time. We fitted a 2D Gaussian distribution to the target at every integration to determine its x and y positions on the detector. $\Delta\sigma_x$ and $\Delta\sigma_y$ describe change in the width of the 2D Gaussian with time. The integrations were taken approximately every 40 seconds. The lower four rows were additionally binned to 5 minutes (= 8 integrations) shown with the solid black lines. Due to stronger systematics, we excluded the first 10 integrations in the \texttt{SZ} reduction shown by the grey region at the beginning of each visit.}
\label{fig:systematics}
\end{figure}

For our primary data reduction and data analysis we used the open-source \python{} package \texttt{Eureka!} \citep{Bell2022} which is an end-to-end pipeline for time series observations performed with JWST or the Hubble space telescope (HST). We started our reduction with the raw uncalibrated (``uncal'') FITS files which we downloaded from the Mikulski Archive for Space Telescopes (MAST) and followed the multi-stage approach of \texttt{Eureka!} to generate a light curve for TRAPPIST-1\,c. \texttt{Eureka!} has been previously successfully used to reduce and analyse the first JWST observations of exoplanets \cite{ERS2022, Ahrer2022, Alderson2022, Rustamkulov2022, LustigYaeger2023}.\\
Stages 1 and 2 of \texttt{Eureka!} serve as a wrapper of the JWST Calibration Pipeline \cite{Bushouse2022} (version 1.8.2.). Stage 1 converts groups to slopes and applies basic detector-level corrections. We used the default settings for all steps in this stage but determined a custom ramp-jump detection threshold for each visit by minimizing the median absolute deviation (MAD) of the final light curves. This step detects jumps in the up-the-ramp signal for each pixel by looking for outliers in each integration that might be caused by events such as cosmic rays. We determined a best jump detection threshold of 7$\sigma$, 6$\sigma$, 7$\sigma$ and 5$\sigma$ for visits 1, 2, 3, and 4, respectively, compared with the default value of 4$\sigma$ set in the JWST pipeline. 
In stage 2, we only skipped the \texttt{photom} step to leave the data in units of DN/s and not convert into absolute fluxes. In stage 3 of \texttt{Eureka!}, we first masked pixels in each visit that were flagged with an ``DO NOT USE'' data quality entry, indicating bad pixels identified by the JWST pipeline. Next, we determined the centroid position of the star by fitting a 2D Gaussian to the source. JWST remained very stable during our observations of TRAPPIST-1\,c and our target stayed well within a 0.01-pixel area (see Table \ref{tab:observations} and Fig. \ref{fig:systematics}). We recorded the centroid position in x and y and the width of the 2D Gaussian in x and y over time to be used in the fitting stage. Next, we determined the best target and background apertures by minimizing the rms of the final light curve. We therefore determined a target aperture of 4 pixels and a background annulus from 25 to 41 pixels from the centroid for each visit. The light curves show a ramp-like trend at the beginning of the observations, which has already been observed in previous JWST MIRI observations and is most likely caused by charge trapping (see, for example, \cite{Bell2023}). We decided to remove the first 10 integrations from each visit, corresponding to approximately 6 minutes or 3\% of the data per visit, so that we do not have to also model this initial ramp. Finally, we checked for significant outliers in the final light curves by performing an iterative 5$\sigma$ outlier clipping procedure. However, no integrations were removed during this process, leaving us with 288, 287, 287, and 288 integrations for the four visits, respectively.

\subsubsection*{Data Reduction ED}

For the second data reduction, we also used the \texttt{Eureka!} pipeline \cite{Bell2022} for stages 1 to 5.  We also started from the uncal.fits files and used the default \texttt{jwst pipeline} settings with the exception of the ramp-fitting weighting parameters in stage 1 that we set to uniform instead of default, as it slightly improved the rms of our residuals. This improvement can be explained by the fact that the default ramp-fitting algorithm uses a weighting of the ramp that gives additional weight to the first and last groups of the ramp, which can be problematic when the number of groups is small, such as for TRAPPIST-1\,c (only 13 groups). Indeed, the first and last groups can be affected by effects such as the reset switch charge decay or saturation. Thus, to ensure that we fit the ramp  correctly, we used an unweighted algorithm that applies the same weight to all groups.
Furthermore, in stage 2 we turned off the \texttt{photom} step. Then, in stage 3, we defined a subarray region ([632, 752],[450, 570]), masked the pixels flagged in the DQ array, interpolated bad pixels and performed aperture photometry on the star with an aperture size that minimized the rms of the residuals for each visits. For each integration, we recorded the centre and width of the PSF in the x and y directions after fitting a 2D Gaussian. We computed the background on an annulus of 20 to 35 pixel (centred on the target) and subtracted it. We note that the choice of the background annulus has little impact on the light curve. We did not remove any integrations a priori but, in stage 4 we sigma clipped 4$\sigma$ outliers compared with the median flux calculated using a 10-integrations-width boxcar filter. Then, for each visit for aperture photometry, we chose the aperture radius that led to the smaller rms. These radii were 3.7, 4.0, 3.6, and 3.8 pixels, respectively (see Table \ref{tab:reductions}).

\subsubsection*{Data Reduction MG}
\label{sec:dr_MG}

We reduced the data using the following methodology. Starting from the uncal.fits files, we calibrated them using the two first stages of the \texttt{Eureka!} pipeline \cite{Bell2022}. We performed a systematic exploration of all the combinations of all \texttt{Eureka!} stage 1 options, and we selected the combination resulting in the most precise light curves. Our selected combination corresponds to the default \texttt{jwst pipeline} settings, except for (1) the ramp-fitting weighting parameter set to uniform, and (2) the deactivation of the jump correction. The rest of the reduction was done using a pipeline coded in \texttt{IRAF} and \texttt{Fortran 2003}. It included  for each calibrated image (1) a change of unit from MJy/sr to recorded electrons, (2) the fit of a 2D Gaussian function on the profile of the star to measure the subpixel position of its centroid and its full width at half maximum (FWHM) in both directions, and  (3) the measurement of the stellar and background fluxes  using circular and annular apertures, respectively, with \texttt{IRAF/DAOPHOT} \cite{Stetson1987}. Finally, the resulting light curves were normalized and outliers were  discarded from them using a 5$\sigma$ clipping with 20-min moving median algorithm. For each visit, the radius of the circular aperture used to measure the stellar flux was optimized by minimizing the standard deviation of the residuals. For each stellar flux measurement, the corresponding error was computed taking into account the star and background photon noise, the readout noise, and the dark noise, and assuming a value of 3.1 el/ADU for the gain (E. Ducrot, private communication). See Table \ref{tab:reductions} for more details.

\subsubsection*{Data Reduction PT}
We performed an additional analysis using the level 2 (flux-calibrated) ``calints" science products as processed by the Space Telescope Science Institute and hosted on the MAST archive. We determined centroid positions and average seeing FWHM values in the $x$ and $y$ dimensions with a 2D Gaussian fit to the star. We performed fixed-aperture photometry with circular apertures centred on the source centroids, with radii ranging from 3.2\textendash 5.0 pixels in 0.1-pixel increments. We also performed variable-aperture photometry using circular apertures with radii set to $c$ times a smoothed time series of the measured FWHM values, where $c$ ranged from 0.75\textendash 1.25 in increments of 0.05. We smoothed the FWHM values using a 1D Gaussian kernel with a standard deviation of 2. For both fixed-aperture and variable-aperture photometry, we measured the background using a circular annulus with an inner radius of 30 pixels and an outer radius of 45 pixels. We subtracted the sigma-clipped mean of the pixel values within this annulus from the source counts in each frame, using a clipping level of 4$\sigma$. Finally, we recorded the values of a grid of background-subtracted pixels interior to the average photometric aperture size surrounding the source centroid in each frame. We used normalized time series of these pixel values to test whether pixel-level decorrelation (PLD) methods developed for minimizing intrapixel effects in Spitzer Space Telescope data \citep{Deming2015} are warranted in the analysis of JWST/MIRI time-series data.

We excluded the first integration of each visit from our analysis as the measured source flux in this exposure was found to be significantly lower than the remainder of the time series for each of the four visits. We checked for outliers in each visit by performing sigma clipping with a threshold of 4$\sigma$, but no exposures were flagged with this step. We then selected the aperture size and method (fixed or variable) that minimized the out-of-eclipse scatter for each visit for use in our analysis. We found that fixed-aperture photometry provided the best performance in each case, with optimal radii of 4.4, 4.1, 3.9, and 3.5 pixels for the four visits, respectively.

\subsection*{Data Analysis}

We fitted each of the reductions to extract an eclipse depth value. The different assumptions for the four global fits are listed in Table \ref{tab:analysis}.

\begin{table}[ht!]
\footnotesize
\centering
\caption{\textbf{Details of the four data analyses.} See Methods for more details on the individual fits. The uncertainties on the eclipse depth $f_p/f_*$ are 1$\sigma$.}
\label{tab:analysis}
\begin{tabularx}{\textwidth}{L|LLLL}\hline
Step/Parameter &
  SZ reduction &
  ED reduction &
  MG reduction &
  PT reduction
  \\\hline\rule{0pt}{12pt}Fitting method &
  \texttt{emcee} (MCMC) &
  \texttt{trafit} (MCMC-MH) &
  \texttt{trafit} (MCMC-MH) &
  \texttt{emcee} (MCMC)
  \\\rule{0pt}{12pt}Details for fitting method &
  150,000 steps, 128 walkers, 30,000 as burn-in. Ran sampler for 80 times the autocorrelation length &
  1 chain of 50,000 steps for error correction followed by 2 chains of 100,000 steps &
  2 chains of 100,000 steps, with first 20\% of chains as burn-in. Convergence checked with Gelman-Rubin statistical test. &
  50,000 steps, 64 walkers, 5,000 as burn-in
  \\\rule{0pt}{12pt}total number of free parameters in the joint fit &
  32 &
  35 &
  33 &
  18
  \\\rule{0pt}{12pt}number of free systematic parameters  &
  14 (in time) + 8 (decorr.) + 4 (uncertainty multiplier) &
  14 (in time) + 11 (decorr.) &
  12 (in time) + 5 (decorr.) &
  11 (in time)
  \\\rule{0pt}{12pt}number of free astrophysical parameters &
  6 (4 $f_p/f_*$ + $e$ + $\omega$) &
  10 ($f_p/f_*$, $b$, 4 TTVs, $M_*$, $R_*$, $T_{\textrm{eff}}$, [Fe/H])) &
  16 ($f_p/f_*$, 7 TTVs, $\log{\rho_*}$, $\log{M_*}$, $T_{\textrm{eff}}$, [Fe/H], $\cos{i}$, $(R_p/R_*)^2$, $\sqrt{e} \cos{\omega}$, $\sqrt{e} \sin{\omega}$) &
  7 ($f_p/f_*$ + $P_{orb}$ + $i$ + $a/R_*$ + $e$ + $\omega$ + $t_{sec}$)
  \\\rule{0pt}{12pt}rms of joint fit residuals &
  1020 ppm &
  961 ppm &
  938 ppm &
  1079 ppm
  \\\rule{0pt}{12pt}$f_p/f_*$ &
  $431^{+97}_{-96}$ ppm &
  $423^{+97}_{-95}$ ppm &
  $414 \pm 91$ ppm &
  $418^{+90}_{-91}$ ppm
  \\\hline
\end{tabularx}
\end{table}

\subsubsection*{Data Analysis SZ}

\begin{figure}
\centering
\includegraphics[width=0.99\textwidth]{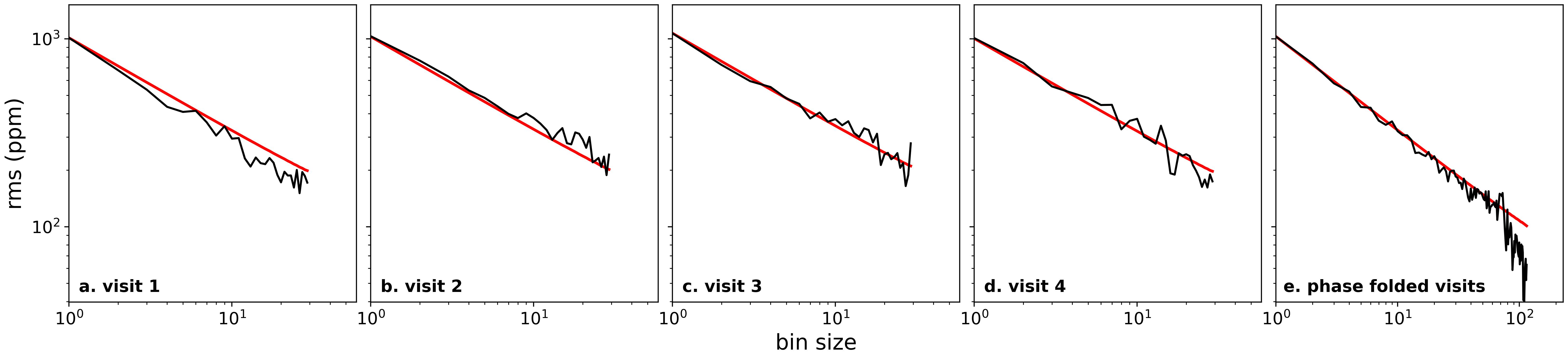}
\caption{\textbf{Allan deviation plots.} \textbf{a -- d.} Allan deviation plots of the individual visits: root-mean-square (rms) of the best-fit residuals from data reduction \texttt{SZ} as a function of the number of data points per bin shown in black. \textbf{e.} The same, but for the combined dataset. A bin size value of one corresponds to no binning. The red line shows the expected behaviour if the residuals are dominated by Gaussian noise. The absolute slope of this line is 1/$\sqrt{\textrm{bin~size}}$, following the inverse square root. The rms of our residuals closely follow this line, showing that our residuals are consistent with uncorrelated photon shot noise.}
\label{fig:allan_devs}
\end{figure}

We fitted the eclipse light curve using the open-source python MCMC sampling routine \texttt{emcee} \cite{ForemanMackey2013}. Our full fitting model, $F(t)$, was the product of a \texttt{batman} \cite{Kreidberg2015} eclipse model, $F_{\textrm{eclipse}}(t)$ and a systematic model, $F_{\textrm{sys}}(t)$. We fit the systematics of JWST with a model of the following form:
\begin{equation} \label{eqn:Fsys_SZ}
    F_{\textrm{sys}}(t) = F_{\textrm{polynom}}(t)\,F_{x}\,(t)F_{y}(t)\,F_{\sigma_x}(t)\,F_{\sigma_y}(t),
\end{equation}
where $F_{\textrm{polynom}}$ is a polynomial in time and $F_{x}(t)$, $F_{y}(t)$, $F_{\sigma_x}(t)$, and $F_{\sigma_y}(t)$ detrend the light curve against a time series of the centroid in x and y and the width of the PSF in x and y, respectively. Before fitting the full light curve consistent out of the four visits, we first determined the best systematic model for each visit by minimizing the Bayesian Information Criterion (BIC) \cite{Schwarz1978, Kass1995, Liddle2007}. We tried a range of polynomials ranging from zeroth order to third order and detrended for the shift in x- and y-pixel positions or for the change in the width of the PSF in time. The best final combination of polynomials and detrending parameters for each visit are listed in Table \ref{tab:analysis}. Our eclipse model used the predicted transit times from \cite{Agol2021} which accounts for the transit-timing variations (TTVs) in the system and we allowed for a non-zero eccentricity. We also accounted for the light travel time, which is approximately 16 seconds for TRAPPIST-1\,c, that is, its semi-major-axis is about 8 light-seconds. We fixed the other parameters of the planet and system, such as the ratio of the semi-major axis to stellar radius $a/R_*$, the ratio of the planetary radius to stellar radius $R_p/R_*$, and the inclination $i$, to the values reported in Ref\cite{Agol2021}. We decided to also supersample the light curve by a factor of 5 in our fitting routine because the sampling of the data ($\approx$ every 40 seconds) is comparable with the ingress/egress duration of 200 seconds \cite{Agol2021}. Our global fit consisted of 32 free parameters: 6 physical (the eccentricity, the argument of periastron, and an eclipse depth for each visit), 22 parameters to fit for the systematics, and 4 free parameters that inflated the uncertainties in the flux for each visit. These four free parameters are necessary because the current gain value on the Calibration References Data System (CRDS) has been empirically shown to be wrong for MIRI data \cite{Bell2023}. For our global MCMC, we used 128 walkers (= 4 times the number of free parameters), 150,000 steps, and discarded the first 20\% of steps (= 30,000 steps) as burn-in. This corresponds to approximately 80 times the autocorrelation length. After calculating a weighted average of the four eclipse depths, we get an eclipse depth of $f_p/f_* = 431^{+97}_{-96}$ ppm for this reduction. Figure \ref{fig:allan_devs} shows the Allan deviation plots of the residuals for each of the visits and the global fit. The rms of the residuals as a function of bin size follows the inverse square root law, which is expected for Gaussian noise.

\subsubsection*{Data Analysis ED}

Once we obtained the light curve for each visit from stage 4 of the \texttt{Eureka!} pipeline we used the \texttt{Fortran} code \texttt{trafit} which is an updated version of the adaptive MCMC code described in \cite{Gillon2010,Gillon2012,Gillon2014}. It uses the eclipse model of \cite{Mandel2002} as a photometric time series, multiplied by a baseline model to represent the other astrophysical and instrumental systematics that could produce photometric variations. 
First, we fit all visits individually. We tested a large range of baseline models to account for different types of external sources of flux variations/modulations (instrumental and stellar effects). This includes polynomials of variable orders in time, background, PSF position on the detector (x, y) and PSF width (in x and y). Once the baseline was chosen, we ran a preliminary analysis with one Markov chain of 50,000 steps to evaluate the need for rescaling the photometric errors through the consideration of a potential underestimation or overestimation of the white noise of each measurement and the presence of time-correlated (red) noise in the light curve. After rescaling the photometric errors, we ran two Markov chains of 100,000 steps each to sample the probability density functions of the parameters of the model and the physical parameters of the system, and assessed the convergence of the MCMC analysis with the Gelman \& Rubin statistical test \citep{GelmanRubin1992}. For each individual analysis, we used the following jump parameters with normal distributions: $M_{\star}$, $R_{\star}$, $T_{\textrm{eff},\star}$, [Fe/H], $t_0$, $b$ all priors were taken from \cite{Ducrot2020} except for the transit timings, which were derived from the dynamical model predictions by \cite{Agol2021}. We fixed $P$, $i$ and $e$ to the literature values given in \cite{Ducrot2020,Agol2021}. The eclipse depth that we computed for each visit individually were $445 \pm 193$ ppm, $418 \pm 173$ ppm, $474 \pm 158$ ppm, and $459 \pm 185$ ppm, respectively. \\
We then performed a global analysis with all four visits, using the baseline models derived from our individual fits for each light curve. Again, we performed a preliminary run of one chain of 50,000 steps to estimate the correction factors that we then apply to the photometric error bars and then a second run with two chains of 100,000 steps. The jump parameters were the same as for the individual fits except for the fact that we fixed $t_0$ and allowed for transit timing variations (TTV) to happen for each visit (each transit TTV has an unconstrained uniform prior). We used the Gelman \& Rubin statistic to assess the convergence of the fit. We measure an eclipse depth of $423_{-95}^{+97}$ ppm from this joint fit.

\subsubsection*{Data Analysis MG}

Our data-analysis methodology was the same as that used by ED, that is, we used the \texttt{Fortran 2003} code \texttt{trafit} to perform a global analysis of the four light curves, adopting the Metropolis-Hasting MCMC algorithm to sample posterior probability distributions of the system's parameters. Here too, we tested for each light curve a large range of baseline models, and we adopted the ones minimizing the BIC. They were (1) a linear polynomial of time for the first visit, (2) a cubic polynomial of time and a linear polynomial of the $y$ position for the second visit, (3) a linear polynomial of time and of the $x$ position for the third visit, and (4) a cubic polynomial of time and of the $y$ position for the fourth visit. We also performed a preliminary analysis (composed of one Markov Chain of 10,000 steps) to assess the need to rescale the photometric errors for white and red noise. We then performed two chains of 500,000 steps each (with the first 20\% as burn-in). The convergence of the analysis was checked using the Gelman \& Rubin statistical test \cite{GelmanRubin1992}. The jump parameters of the analysis, that is, the parameters perturbed at each step of the MCMC chains, were (1) for the star, the logarithm of the mass, the logarithm of the density, the effective temperature, and the metallicity, and (2) for the planet, the planet-to-star radius ratio, the occultation depth, the cosinus of the orbital inclination, the orbital parameters $\sqrt{e} \cos{\omega}$ and $\sqrt{e} \sin{\omega}$ (with $e$ the orbital eccentricity and $\omega$ the argument of pericentre), and the timings of the transits adjacent to each visit.  We assumed normal prior distributions for the following parameters based on the results from reference \cite{Agol2021}: $M_\ast = 0.0898 \pm 0.023$, $R_\ast = 0.1192 \pm 0.0013$, $T_{\textrm{eff}} = 2566 \pm 26$ K, and [Fe/H]  = $0.05 \pm 0.09$ for the star; $(R_p/R_\ast)^2 = 7123 \pm 65$ ppm, $b = 0.11 \pm 0.06$, $e = 0 + 0.003$ (semi-gaussian distribution) for the planet. We also tested the assumption of a circular orbit and obtained similar results. For each visit, we considered for the timings of the two adjacent transits normal prior distributions based on the predictions of  the dynamical model of reference \cite{Agol2021}. At each step of the MCMC, the orbital position of the planet could then be computed for each time of observation from the timings of the two adjacent transits and from $e$ and $\omega$, and taking into account the approximately 16s of light-travel time between occultation and transit. 
This analysis led to the value of $414 \pm 91$ ppm for the occultation, and to an orbital eccentricity of $0.0016_{-0.0008}^{+0.0015}$ consistent with a circular orbit. Under the assumption of a circular orbit, our analysis led to an occultation depth of $397 \pm 92$ ppm, in excellent agreement with the result of the analysis assuming an eccentric orbit. 

We also performed a similar global analysis, but allowing for different occultation depths for each visit. The resulting depths were $400 \pm$ 163 ppm, 374 $\pm$ 184 ppm, $421 \pm 187$ ppm, and $403 \pm 202$ ppm, i.e. they were consistent with a stable thermal emission of the planet's dayside (at this level of precision). Similar to data reduction \texttt{SZ}, we also did create Allan deviation plots for this particular data reduction. The best-fit residuals as a function of bin size from each visit do generally follow the inverse square root law (see Fig. \ref{fig:allan_devs} for the Allan deviation plots of data reduction \texttt{SZ}).

Finally, we computed the brightness temperature of the planet at 15 $\mu$m from our measured occultation depth using the following methodology. We measured the absolute flux density of the star  in all the calibrated images, using an aperture of 25 pixels large enough to encompass the wings of its PSF. We converted these flux densities from MJy/sr to mJy, and computed the mean value of $2.559$ mJy and the standard deviation of 0.016 mJy. We added quadratically to this error of around 0.6\% a systematic error of 3\%, which corresponds to the estimated absolute photometric precision of MIRI (P.-O. Lagage, private communication). It resulted in a total error of $0.079$ mJy. Multiplying the measured flux density by our measured occultation depth led to a planetary flux density of $1.06 \pm 0.23$ $\mu$Jy. Multiplying again this result by the square of the ratio of the distance of the system and the planet' radius, and dividing by $\pi$, led to the mean surface brightness of the planet's dayside. Applying Planck's law, we then computed the brightness temperature of the planet, while its error was obtained from a classical error propagation. Our result, for this specific reduction, was $379 \pm 30$ K, to be compared with an equilibrium temperature of 433 K computed for a null-albedo planet with no heat distribution to the nightside.  

It is also worth mentioning that applying the same computation on the star itself led to a brightness temperature of $1867 \pm 55$ K, which is significantly lower than its effective temperature.

\subsubsection*{Data Analysis PT}
We began our analysis by determining which time-series regressors (if any) should be included for fitting systematics in the photometry on the basis of the BIC. Our total model is the product of a \texttt{batman} eclipse model ($F_{\textrm{eclipse}}$) and a systematics model ($F_{\textrm{syst}}$) to the data, which has a general form of 

\begin{equation}
    F_{\textrm{syst}}(t) = F_{\textrm{polynom}}(t)\,F_x(t)\,F_y(t)\,F_{\textrm{FWHM}}(t)\,F_{\textrm{ramp}}(t)\,F_{\textrm{PLD}}(n,t).
\end{equation}

\noindent Here, $F_{\textrm{polynom}}$ is a polynomial in time, $F_x(t)$ and $F_y(t)$ are time series of the target centroids in $x$ and $y$, $F_{\textrm{FWHM}}(t)$ is the time series of average FWHM values for the source determined with a 2D Gaussian fit, and $F_{\textrm{ramp}}$ is an exponential function that accounts for ramp-up effects. $F_{\textrm{PLD}}(n,t)$ is the linear combination of $n$ basis pixel time series, and it has a form of 

\begin{equation}
    F_{\textrm{PLD}}(n,t) = \sum_{i=1}^{n}{C_i \hat{P}_i(t)}
\end{equation}

\noindent Here, $\hat{P_i}(t)$ is the normalized intensity (from 0\textendash 1) of pixel $i$ at time $t$ and $C_i$ is the coefficient of pixel $i$ determined in the fit. PLD was developed to mitigate systematic intrapixel effects in Spitzer/Infrared Array Camera (IRAC) data \citep{Deming2015}, in which the combination of source PSF motion and intrapixel gain variations introduced percent-level correlated noise in time-series data \citep[e.g.,][]{Ingalls2012}.

In our analysis, we tested forms of $F_{\textrm{polynom}}$ ranging from degree 0\textendash 3 and different sets of PLD basis pixels including the brightest 1, 4, 9, 16, 25, and 36 pixels. For each visit, we explored grids of every possible combination of the components of $F_{\textrm{syst}}(t)$. For each combination, we first initialized the coefficients of each component using linear regression. We then used \texttt{emcee} to perform an MCMC fit of the total eclipse and systematic model to the visit data. We ran 2$\nu$+1 walkers for 10,000 steps in each fit, in which $\nu$ represents the number of free parameters in the total model. The first 1,000 steps of these chains were discarded as burn-in. We fit for seven physical parameters in our calculation of $F_{\textrm{eclipse}}$, these being the orbital period, $a/R_*$, orbital inclination, eccentricity, longitude of periastron, eclipse depth, and time of secondary eclipse. Gaussian priors were assigned to these parameters with means and standard deviations set by their measurements reported in \cite{Agol2021}. We also placed Gaussian priors on the coefficients of the components of $F_{\textrm{syst}}$, with means set by the linear regression fit and standard deviations set to the absolute value of the square root of those values. 

We calculated the BIC of the best-fitting model that resulted from the MCMC analysis, and then selected the form of $F_{\textrm{syst}}$ that minimized the BIC. The form of $F_{\textrm{syst}}$ that we determined for each visit with this approach consisted of only an $F_{\textrm{polynom}}$ component. The first visit was best fit by a linear polynomial, whereas the remaining three were best fit by a quadratic polynomial. 

With the form of $F_{syst}(t)$ determined for each visit, we then performed a joint fit of all four eclipses. This fit included 18 total free parameters: 7 physical and 11 for fitting systematics (see Table~\ref{tab:analysis}). We ran this fit with 64 chains for 50,000 steps, discarding the first 5,000 steps for burn-in. We measured a resulting eclipse depth of 418$^{+90}_{-91}$~ppm from this fit.

\subsection*{Brightness Temperature Calculation}

The following analysis was based on stage 0 (.uncal) data products pre-processed by the JWST data processing software version number 2022\_3b, and calibrated with \texttt{Eureka!} as described above in the ``Data Reduction MG" section. We computed the brightness temperature of the planet at 15 $\mu$m from our measured occultation depth using the following methodology. We measured the absolute flux density of the star in all the calibrated images, using an aperture of 25 pixels large enough to encompass the wings of its PSF. We converted these flux densities from MJy/sr to mJy, and computed the mean value of $2.559$ mJy and the standard deviation of 0.016 mJy. We added quadratically to this error of about 0.6\% a systematic error of 3\%, which corresponds to the estimated absolute photometric precision of MIRI (P.-O. Lagage, private communication). It resulted in a total error of $0.079$ mJy. Multiplying the measured flux density by our measured occultation depth led to a planetary flux density of $1.06 \pm 0.23$ $\mu$Jy. Multiplying again this result by the square of the ratio of the distance of the system and the planet' radius, and dividing by $\pi$, led to the mean surface brightness of the planet's dayside. Applying Planck's law, we then computed the brightness temperature of the planet, whereas its error was obtained from a classical error propagation. Our result, for the \texttt{MG} reduction, was $379 \pm 30$ K, to be compared with an equilibrium temperature of 433 K computed for a null-albedo planet with no heat distribution to the night side. It is also worth mentioning that applying the same computation on the star itself led to a brightness temperature of $1867 \pm 55$ K, which is significantly lower than its effective temperature. 

\subsection*{Emission modelling for TRAPPIST-1\,c}

We generated various emission spectra for TRAPPIST-1\,c to compare them to our measured eclipse depth at 15 \micron. These models include (1) bare-rock spectra, (2) O$_2$/CO$_2$ mixture atmospheres and pure CO$_2$ atmospheres, and (3) coupled climate-photochemical forward models motivated by the composition of Venus. In the following, we describe each of these models. 

\subsubsection*{Bare Rock}

\begin{figure}
\centering
\includegraphics[width=0.49\textwidth]{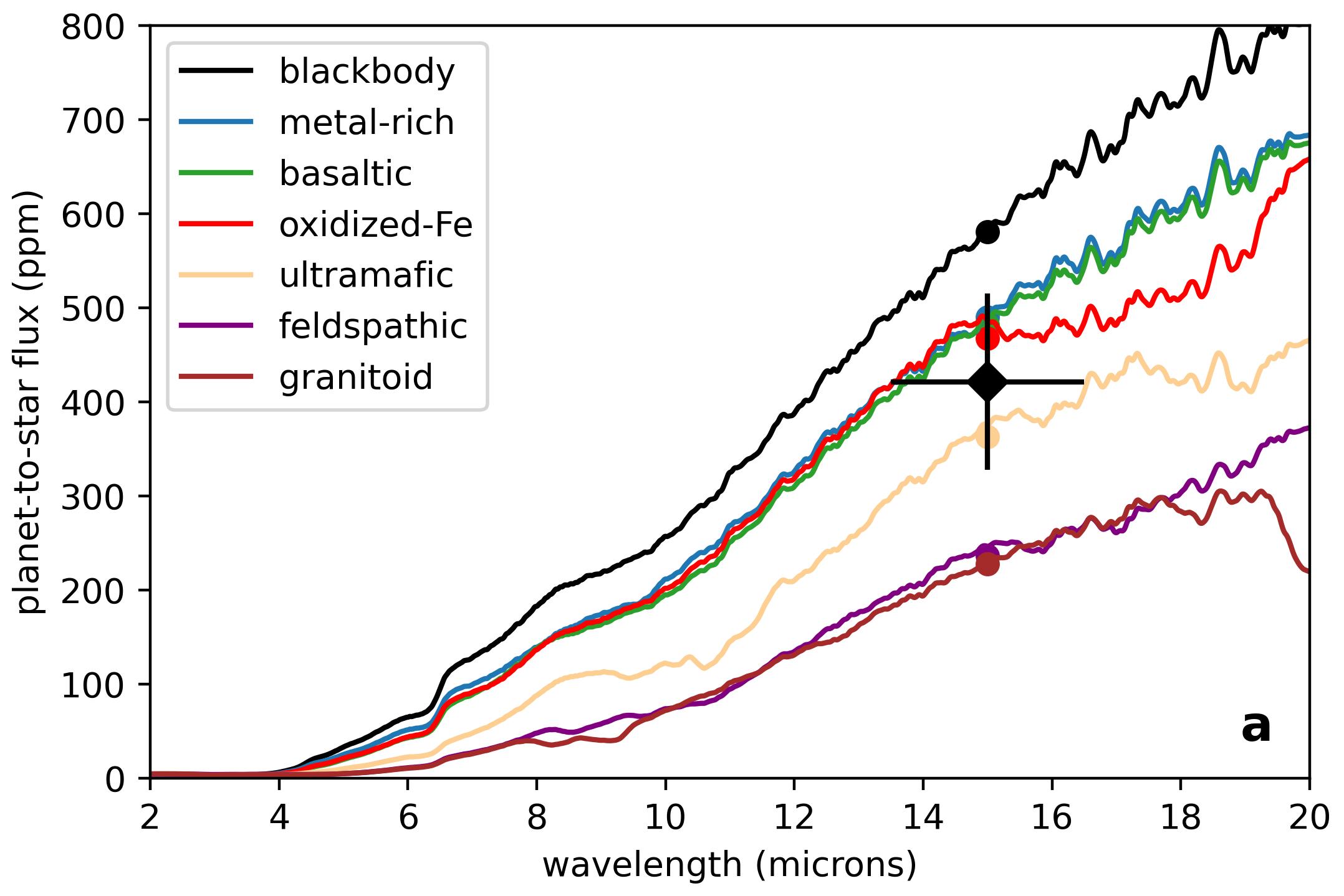}
\includegraphics[width=0.49\textwidth]{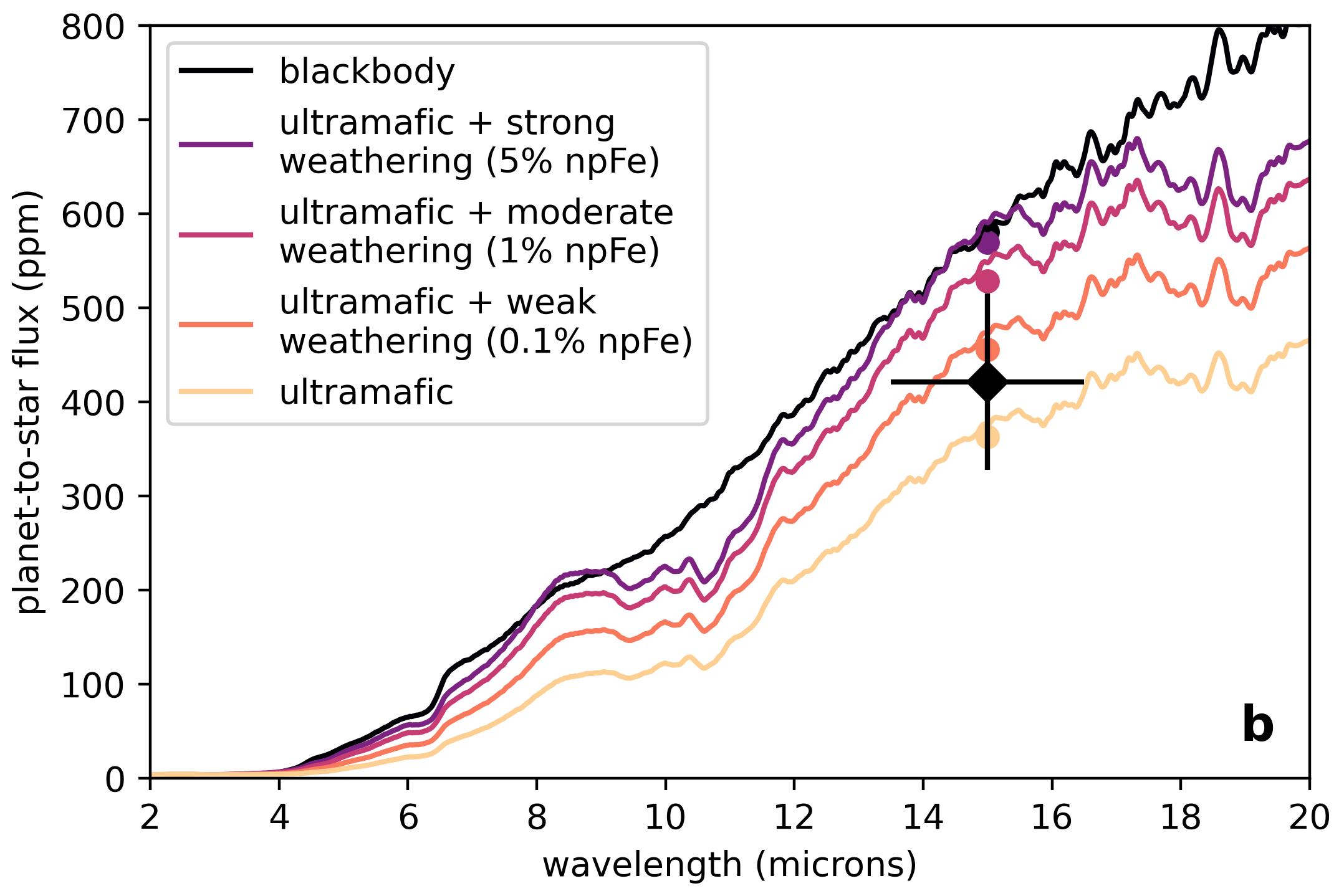}
\caption{\textbf{Measured eclipse depth compared to a suite of simulated bare-rock emission spectra.} \textbf{a.} Secondary eclipse spectra for various fresh surface compositions, assuming that TRAPPIST-1\,c is a bare rock. High-albedo feldspathic and granitoid surfaces are cool and fit the data moderately poorly (2$\sigma$), as does a low-albedo and hot blackbody surface (1.7$\sigma$).
\textbf{b.} Space weathering via formation of iron nanoparticles (npFe) lowers the albedo at short wavelengths, thereby increasing the surface's temperature and its secondary eclipse depth. An intermediate-albedo fresh ultramafic surface would fit the data well, but the fit becomes marginal after taking into account the influence of strong space weathering (1.6$\sigma$, or about $90\%$ confidence). The vertical error bar on our 15 \micron\ measurement represents the 1$\sigma$ uncertainty on the observed eclipse depth.
}
\label{fig:bare_rocks}
\end{figure}

Our bare-rock model is a spatially resolved radiative transfer model and computes scattering and thermal emission for a variety of surface compositions. For each composition, the surface's radiative equilibrium temperature is computed on a 45x90 latitude-longitude grid, assuming TRAPPIST-1\,c is tidally locked. Surface reflectance and emissivity data are from \cite{hu2012}, which were derived from reflectance spectra of rock powders or minerals measured in the laboratory combined with an analytical radiative-transfer model \citep{hapke2002}. These data have previously been used to model surface albedos and emission spectra of bare-rock exoplanets \citep{hu2012,mansfield2019,Kreidberg2019}. Here we consider six compositions as well as a blackbody: basaltic, feldspathic, Fe-oxidized (50$\%$ nanophase haematite, 50$\%$ basalt), granitoid, metal-rich (FeS$_2$), and ultramafic (see Fig. \ref{fig:bare_rocks}).
Given the uncertainty in the measured eclipse depth, we assume a Lambertian surface with isotropic scattering and emission, and neglect the angular dependency of the surface reflectance and emissivity, which would depend on the surface roughness and regolith particle size \cite{hu2012}. Sensitivity tests show that these surface-model assumptions are indistinguishable within the current precision of the TRAPPIST-1\,c measurements (not shown).

Furthermore, albedos and spectra of bare rocks in the Solar System are modified by space weathering, so we also consider the impact of space weathering on TRAPPIST-1\,c. The timescale for lunar space weathering through exposure to the solar wind has been estimated to range from $\sim 10^5$ years to $\sim 10^7$ years \citep{hapke1977,keller2017}. We extrapolate from the lunar value to TRAPPIST-1\,c using scaling relations from a stellar-wind model \cite{johnstone2015}. We find that the space-weathering timescale for TRAPPIST-1\,c is significantly shorter than the lunar value, about $10^2-10^3$ years, largely because of the planet's small semi-major axis. An exposed surface on TRAPPIST-1\,c should therefore have been substantially weathered. To simulate the impact of space weathering on unweathered surfaces, we incorporate the same approach as that in \cite{hapke2001,hapke2012}. The surface composition is modelled as a mixture of a fresh host material (described above) and nanophase metallic iron using Maxwell-Garnett effective medium theory. The refractive index of metallic iron is taken from \cite{rii}.

\subsubsection*{Simple 1+D O$_2$/CO$_2$ Mixtures}

We construct a grid of O$_2$-dominated model atmospheres with a range of surface pressures and mixing ratios of CO$_2$. These are broadly representative of a plausible outcome of planetary atmosphere evolution, in which water in the atmospheres of terrestrial planets orbiting late-type M dwarfs is photolysed and the H is lost, leaving a large O$_2$ reservoir \cite{Wordsworth2015, Schaefer2016}. The atmosphere models we construct are 1D models following the approach presented in \cite{Morley2017}, with adiabatic pressure-temperature profiles in the deep atmosphere and isothermal pressure-temperature profiles above 0.1 bar (for thicker atmospheres, $P>$0.1 bar) or the skin temperature (for thinner atmospheres). This approach uses \texttt{DISORT} \cite{Stamnes1988, Stamnes2000} to calculate radiative transfer in 1D through the atmosphere to generate emission spectra. 

We do consider how an atmosphere can transport heat to the nightside. To include heat transport to the nightside, we implement the analytic approach of \cite{Koll2019}; we use the redistribution factor $f$ calculated in equation (3) of that work for each of the models in the grid. We assume that both the surface Bond albedo and the top-of-the-atmosphere Bond albedo are 0.1. We construct a grid of O$_2$-dominated model atmospheres with surface pressures from 0.01 to 100 bar (in 1-dex steps) and CO$_2$ mixing ratios from 1 ppm to 10,000 ppm (in 1-dex steps). We also generate pure CO$_2$ atmospheres with the same surface pressures. For the thicker atmospheres ($P_{\textrm{surf}} \geq$ 1 bar) we set the thermopause (in which the atmosphere transitions from adiabatic to isothermal) to 0.01 bar.

\subsubsection*{Coupled Climate-Photochemical Venus-like Atmospheres}

We use a 1.5D coupled climate-photochemical forward model \citep[VPL Climate]{Robinson2018,Lincowski2018,LincowskiPhD} that explicitly models day and night hemispheres with layer-by-layer, day-night advective heat transport driven by simplified versions of the 3D primitive equations for atmospheric transport, to simulate plausible atmospheric states for TRAPPIST-1\,c for cloudy Venus-like scenarios. VPL Climate uses SMART \citep{Meadows1996} with DISORT \citep{Stamnes1988,Stamnes2000} for spectrum-resolving radiative transfer for accuracy and versatility for both the climate modelling and the generation of the resulting planetary spectra. The model has been validated for Earth \citep{Robinson2011} and Venus \citep{Arney2014}, but is capable of modelling a range of atmospheric states.

Due to the early luminosity evolution of the star, TRAPPIST-1\,c would have been subjected to very high levels of radiation \citep{Baraffe2015}, and so we would anticipate evolved atmospheres that had undergone atmospheric and possibly ocean loss \citep{Luger2015}. We start with the self-consistently coupled climate-photochemical Venus-like atmospheres generated for an evolved TRAPPIST-1\,c from \citep{Lincowski2018}, with 96.5\% CO$_2$ and 3.5\% N$_2$ and assume Venus lower atmospheric trace gases and self-consistent generated sulfuric acid aerosols. We use these atmospheres as a starting point for 1.5D clear-sky Venus-like atmospheres (0.1, 1, 10 bars) and 1.5D cloudy Venus-like atmospheres (10 bars) with sulfuric acid haze. Note that 10 bar Venus-like atmospheres will produce similar results to a 93 bar Venus-like atmosphere due to the emitting layer being above or at the cloud deck, which is at a similar pressure for the 10 bar and 93 bar cases. All the modelled clear-sky Venus atmospheres produce 15 \micron\ CO$_2$ features with depths spanning 134--143~ppm, with the cloudy 10 bar Venus centered at 181~ppm. Because H$_2$SO$_4$ aerosols are likely to condense in the atmosphere of a Venus-like planet at TRAPPIST-1\,c's orbital distance \citep{Lincowski2018}, we show the dayside spectrum for the 10 bar cloudy Venus for comparison with the data in Figure \ref{fig:data_vs_models}.  The emitting layer (cumulative optical depth 1) for the cloud aerosols occurs at 7 mbars in this atmosphere, although the 15 \micron\ CO$_2$ absorption is sufficiently strong that it emits from a comparable pressure level in the core of the band.  The observations rule out a self-consistent Venus-like atmosphere for TRAPPIST-1\,c to 2.6~sigma. 

\subsection*{Atmospheric Escape models}

We use energy-limited atmospheric escape models \citep{Schaefer2016,Luger2015} from a steam atmosphere to explore the amount of atmospheric escape that TRAPPIST-1\,c may have experienced over its lifetime. The model assumes that escape occurs in the stoichiometric ratios of H/O in water vapour, allows for escape of oxygen, and reaction of oxygen with the magma ocean. The model transitions from magma ocean to passive stagnant-lid outgassing when surface temperatures drop below the silicate melting point. Escape continues throughout all tectonic stages. In Figure \ref{fig:escape}, we show the final amount of O$_{2}$ gas left in the atmosphere after 7.5 Gyr of evolution for a range of planetary water abundances and XUV saturation fractions. For typical saturation fractions of $10^{-3}$ \citep{Wright2018, Fleming2020}, our observations suggest that the planet likely had a relatively low starting volatile abundance. We note that these models are likely upper limits on thermal escape and more detailed models of escape, especially incorporating other gases such as CO$_{2}$ and N$_{2}$, are needed in the future to confirm these results. We also estimate total ion-driven escape fluxes due to stellar wind interactions of a minimum of 1-3 bars over the planet's lifetime, assuming constant stellar wind over time \citep{Dong2018}. We also considered the extended pre-main sequence for a star like TRAPPIST-1. We used the stellar evolution models of \citep{Baraffe2015} for a 0.09 \msun\ star to approximate the pre-main sequence evolution of the star.

\subsection*{Interior structure model}

We use an interior-structure model to perform an MCMC retrieval on the planetary mass and radius of TRAPPIST-1\,c, and the possible stellar Fe/Si of TRAPPIST-1. The estimated Fe/Si mole ratio of TRAPPIST-1 is 0.76$\pm$0.12 \citep{Unterborn18}, which is lower than the Solar value, Fe/Si = 0.97 \citep{Sotin07}. Our interior-structure model solves a set of differential equations to compute the density, pressure, temperature, and gravity as a function of radius in a one-dimensional grid \citep{Brugger16,Brugger17}. The interior model presents two distinct layers: a silicate-rich mantle and an Fe-rich core. On top of the mantle, we couple the interior model with an atmospheric model to compute the emission and the Bond albedo. These two quantities enable us to solve for radiative-convective equilibrium, find the corresponding surface temperature at the bottom of the atmosphere, and find the total atmospheric thickness from the surface up to a transit pressure of 20 mbar \citep{Mousis20,Acuna21}. We consider a H$_{2}$O-dominated atmosphere, with 99\% H$_{2}$O and 1\% CO$_{2}$. Our 1D, k-correlated atmospheric model prescribes a pressure-temperature profile comprised of a near-surface convective layer and an isothermal region on top. In the regions of the atmosphere where the temperature is low enough for water to condense and form clouds, we compute the contribution of these to the optical depth and their reflection properties as described in \cite{Marcq12,Marcq17}.

The posterior distribution function of the surface pressure retrieved by our MCMC indicates a 1$\sigma$ confidence interval of 40$\pm$40 bar for TRAPPIST-1\,c. Surface pressures between 0 and 120 bar would be compatible with our probability density function within 2$\sigma$ \citep{acuna_sub}. Oxygen is more dense than H$_{2}$O. Consequently, for a similar surface pressure, an O$_{2}$-rich atmosphere would be less extended than the H$_{2}$O-dominated envelope we consider in our coupled interior-atmosphere model. This means that the density of TRAPPIST-1\,c could be reproduced with an oxygen-rich atmosphere with a lower surface pressure as low as our H$_{2}$O upper limit, 80 bar.

\subsection*{Stellar Properties}
The stellar properties of TRAPPIST-1 have been constrained with observations of the total luminosity of the star, $L_* = 4\pi D_*^2 F_{bol}$ (based on broadband photometry to obtain the bolometric flux of the star, $F_{bol}$, and a distance measured with GAIA, $D_*$), a mass-luminosity relation \citep{Mann2019} to obtain the stellar mass, $M_*(L_*)$, with uncertainty, as well as a precise stellar density, $\rho_*$, thanks to modelling of the seven transiting planets \citep{vanGrootel2018, Ducrot2020,Agol2021}.  These combine to give the stellar radius and effective temperature, $R_*$ and $T_{\textrm{eff},*}$,
\begin{eqnarray}
R_* &=& \left(\frac{3M_*}{4\pi\rho_*}\right)^{1/3} \propto M_*^{1/3},\\
T_{\textrm{eff}} &=& \left(\frac{L_*}{4\pi R_*^2\sigma}\right)^{1/4} \propto M_*^{-1/6}.
\end{eqnarray}
The planets' properties have also been measured precisely in relation to the star using the depths of transit, yielding $R_p/R_*$, and transit-timing variations, yielding $M_p/M_*$ \citep{Agol2021}. To convert the secondary eclipse depth, $\delta = F_p/F_*$, into a brightness temperature of the planet requires an estimate of the brightness temperature of the star:
\begin{equation}
\delta = \frac{I_{b,p}}{I_{b,*}}\frac{R_p^2}{R_*^2},
\end{equation}
or
\begin{equation} \label{eqn:intensity}
    I_{b,p} = I_{b,*} \delta \frac{R_*^2}{R_p^2},
\end{equation}
in which $I_{b,*},I_{b,p}$ are the mean surface brightness of the star and planet in the MIRI band at full phase (that is, secondary eclipse), respectively.
The ratio $R_p/R_*$ is well constrained from the transit depth, whereas the brightness temperature of the star can be measured with an absolute calibration of the stellar flux in the MIRI band, $F_*$ (e.g. \citep{Ducrot2020}).  The stellar intensity may then be computed as:
\begin{equation}
    I_{b,*} = \frac{F_*}{\Omega_*} = \frac{F_* D_*^2}{\pi R_*^2},
\end{equation}
in which $\Omega_*$ is the solid angle of the star. Because our estimate of $R_*$ is proportional to $M_*^{1/3}$, this means that $I_{b,*} \propto M_*^{-2/3}$.  For a given value of $R_*$, this surface brightness can be translated into a brightness temperature, $T_{b,*}$, and with the equation \ref{eqn:intensity} above, we can compute the intensity and therefore the surface brightness of the planet, $T_{b,p}$ to be 380 $\pm$ 31 K using the eclipse depth and the stellar flux density.
We also estimate the stellar brightness temperature in the MIRI band with an atmospheric model for the star relating T$_{b,*}$ in the MIRI band to the T$_{\textrm{eff}}$, as $\alpha = T_{b,*}/T_{\textrm{eff},*}$. We have accomplished this with the state-of-the-art SPHINX model for low-temperature stars \citep{Iyer2023} and assumed T$_{\textrm{eff}}$ = 2566 K \citep{Agol2021}, yielding $\alpha$ = $0.72$ at $14.87$ \micron. We also compute the $\alpha$ from JWST spectrophotometric observations with a flux of $2.599 \pm 0.079$ mJy at $14.87$ \micron, yielding $\alpha$ = 0.71 $\pm$ 0.02. The MIRI images are flux-calibrated (with an internal error of 3$\%$). We measure the stellar flux in all images within an aperture large enough to encompass the whole PSF, and then compute the mean and the standard deviation. We compute the total error on the measurement to be the 3$\%$ larger than this standard deviation. As the unit of flux in MIRI images is given in Jy/str, we multiply the measured fluxes by the angular area covered by a pixel in str to yield units of Jy.

The stellar brightness temperature scales linearly with effective temperature and metallicity in the MIRI wavelength range, and scales inversely with surface gravity of the star. The effective temperature, however, scales as $T_{\textrm{eff}} \propto M_*^{-1/6}$ (or $R_*^{-1/2}$), with stellar mass (or radius) relative to the estimate based on the measured flux.  The estimate of $\alpha$, therefore, may have a significant imprecision given the possible heterogeneity of the stellar atmosphere, as well as the inherent uncertainties involved in modelling late-type stellar atmospheres. Both the synthetically derived $\alpha$ and those from observations match within 2$\sigma$ uncertainty, lending credence to empirical mass-luminosity relations and synthetic atmosphere-model-derived stellar brightness temperatures. Note, however, that the mass-luminosity relation is only calibrated with a handful of low-mass stars in binaries \citep{Mann2019}, and hence its applicability to TRAPPIST-1 may be tenuous;  this may thus be the weakest link in determining the stellar parameters. Assumption-driven deviations between synthetic models for late-type stars and empirically calibrated methods both still remain a significant challenge in truly understanding these hosts.

\subsection*{Eclipse Timing Variations}

Dynamical modelling of the TRAPPIST-1 system \citep{Agol2021} gives a precise forecast
of the times of transit and eclipse for all seven planets.  These
have been used in the planning of the observations, and can also be compared with
the measured times.

The times of eclipse can be offset from the mid-point between the times of
transit due to four different effects:  1) the light-travel time across
the system \citep{Fabrycky2010}, 2) non-zero eccentricity \citep{Winn2010},
 3) non-uniform emission from an exoplanet \citep[][]{Agol2010}\footnote{This 
 does not change the mid-point of the eclipse, but it does change the shape of 
 ingress/egress, and can lead to an artificial time offset if not accounted for 
 in the modelling.}, and 4) eclipse-timing variations due to perturbations by 
 other planets in the system.  Of these three effects, the second effect is
 typically the largest, which can be used to constrain one component of
 the eccentricity vector of the transiting planet \citep{Winn2010}.

\begin{figure}
\centering
\includegraphics[width=0.95\textwidth]{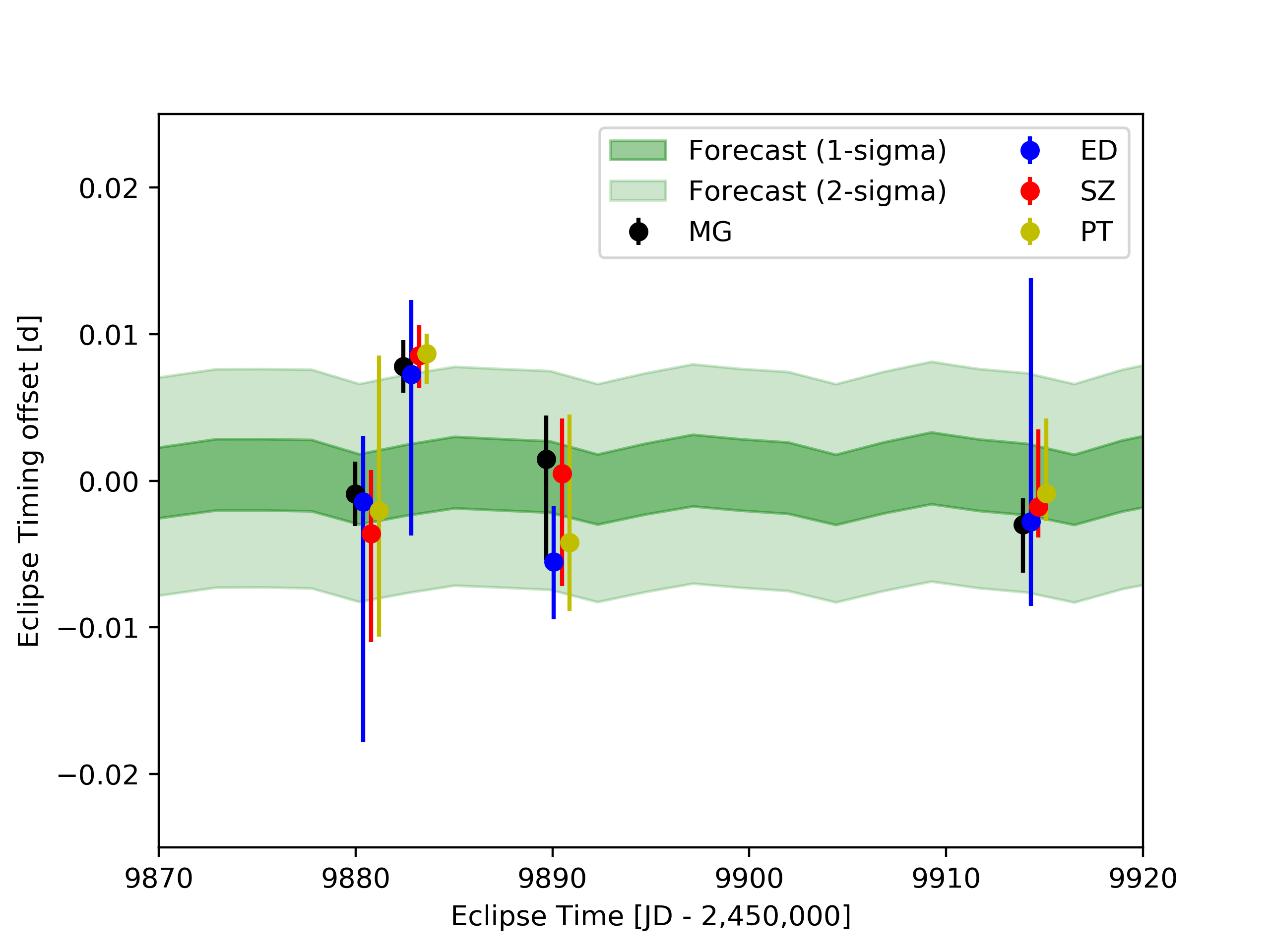}
\caption{\textbf{Measured eclipse times compared with the predicted eclipse times.} The points show the measured eclipse timing offsets (defined as the time of eclipse minus the mean of the two adjacent transit times of planet c) from four different analyses. The error bars correspond to the 16th and 84th percentiles of the eclipse time posterior. The dark (light) green shaded region shows the 1-(2-)sigma confidence intervals forecast from the transit-timing analysis
from \citep{Agol2021}.
}
\label{fig:forecast}
\end{figure}

In Table \ref{tab:eclipse_times} we list the measured eclipse times from the four different reductions and in Figure \ref{fig:forecast} we compare them 
with the forecast from \cite{Agol2021}.
To make the forecast, we used the posterior probability of
the timing model to compute the times of transit and eclipse, and then
we calculate the time of eclipse minus the mean of the two adjacent transits
of planet c to derive an ``Eclipse timing offset".  This offset should be zero for a
circular, unperturbed orbit with negligible light-travel time (which is
about 16 seconds, or 1.8$\times 10^{-4}$ days for TRAPPIST-1\,c).  The dynamical
modelling is constrained by the times of transit, which place some constraint
on the eccentricity of the orbit of planet c (in particular, the mean or
free eccentricity could be non-zero). The uncertainty on the eccentricity
leads to uncertainty on the times of secondary eclipse.  Our forecast models for
the eclipse timing offset have a 1$\sigma$ uncertainty of
$\sim$3.5 minutes at the measured times of eclipse (approximately 0.0024 days).

The measured times were taken from four analyses (by SZ, PT, ED, and MG) in 
which a broad prior was placed on the times of transit, whereas the duration 
and depth were constrained to the measured values of the four eclipses.
The times of each eclipse were then free to vary, and the
posterior times of transit were inferred using MCMC (ED/MG/PT) or nested
sampling (SZ).  The four analyses give good agreement on the values, but
have significant differences between the uncertainties.

Overall, the forecast eclipse timing offsets agree well with the
measured times, within 1-2$\sigma$ offsets.  The uncertainties on the
measured times are comparable with the forecast uncertainties, and so,
in future work, we hope
to use these measured eclipse times to further constrain the eccentricity
vector of the orbit of planet c.  This may help to constrain tidal 
damping models of planet c, but it may also constrain tidal damping
of {\it all} of the planets as the free eccentricity vector of 
planet c is tightly correlated with those of the other planets due 
to the ``eccentricity-eccentricity" degeneracy present in transiting planet systems \cite{Lithwick2012}.

\begin{table}
\centering
\caption{\textbf{Measured eclipse times.} Eclipse times (in BJD$_\textrm{TDB}$) determined by the four different reductions by fitting an eclipse model to each visit. An offset of 2459880.0 days was subtracted from each of the values in the table.}
\label{tab:eclipse_times}
\begin{tabular}{l|llll}\hline
   & visit 1 & visit 2 & visit 3 & visit 4 \\\hline
SZ & $0.1872_{-0.0074}^{+0.0043}$ & $2.6209_{-0.0022}^{+0.0021}$ & $9.8782_{-0.0077}^{+0.0038}$ & $34.0940_{-0.0021}^{+0.0053}$ \\
ED & $0.1894_{-0.0164}^{+0.0452}$ & $2.6197_{-0.0110}^{+0.0051}$ & $9.8722_{-0.0040}^{+0.0038}$ & $34.0930_{-0.0057}^{+0.0166}$ \\
MG & $0.1899 \pm 0.0022$          & $2.6202 \pm 0.0018$          & $9.8792_{-0.0069}^{+0.0033}$ & $34.0928_{-0.0030}^{+0.0018}$ \\
PT & $0.1887^{+0.0106}_{-0.0086}$ & $2.6211^{+0.0014}_{-0.0021}$ & $9.8735^{+0.0087}_{-0.0047}$ & $34.0949^{+0.0051}_{-0.0019}$\\\hline
\end{tabular}
\end{table}

\bmhead{Acknowledgments}
This work is based in part on observations made with the NASA/ESA/CSA James Webb Space Telescope. The data were obtained from the Mikulski Archive for Space Telescopes at the Space Telescope Science Institute, which is operated by the Association of Universities for Research in Astronomy, Inc., under NASA contract NAS 5-03127 for JWST. These observations are associated with program \#2304. MG is F.R.S.-FNRS Research Director, and acknowledges support from the Belgian Federal Science Policy Office BELSPO BRAIN 2.0 (Belgian Research Action through Interdisciplinary Networks) for the  project PORTAL n° B2/212/P1/PORTAL (PhOtotrophy on Rocky habiTAble pLanets). VM and AL are part of the Virtual Planetary Laboratory Team, which is a member of the NASA Nexus for Exoplanet System Science, and funded via NASA Astrobiology Program Grant 80NSSC18K0829. AI acknowledges support from the NASA FINESST Grant 80NSSC21K1846.

\subsection*{Author Contributions}
SZ, LK, MG, PT, ED, AL, VM, DK, CM, LS, EA, LA, and GS contributed significantly to the writing of this manuscript.
SZ, ED, PT, and MG provided a data reduction and data analysis of the four visits for this work and contributed an eclipse depth value. 
CM, DK, XL, RH, AL, and VM ran theoretical models for the planet’s atmosphere and surface.
LA ran models on the planet's interior structure.
AI and EA modeled the stellar spectrum.
LS modeled the atmospheric escape for the planet.
LK, MG, VM, DK, RH, CM, LS, EA, FS EB, and AM contributed to the observing proposal.

\subsection*{Data availability}
The data used in this work were collected by JWST as part of GO program 2304 and will be publicly accessible after the default proprietary period of one year. The most recently taken visit will therefore be publicly accessible on the Mikulski Archive for Space Telescopes (MAST) on Dec, 1st. 2023.

\subsection*{Code availability}
We used the following codes, resources, and \texttt{python} packages to reduce, analyze and interpret our JWST observations of TRAPPIST-1\,c: \texttt{numpy}\citep{numpy2020}, \texttt{matplotlib}\citep{matplotlib2007}, \texttt{astropy}\citep{astropy2022}, \texttt{batman}\citep{Kreidberg2015}, \texttt{Eureka!}\citep{Bell2022}, \texttt{jwst}\citep{Bushouse2022}, \texttt{emcee}\cite{ForemanMackey2013},  \texttt{trafit} \cite{Gillon2010,Gillon2012,Gillon2014}, \texttt{dynesty}\cite{Speagle2020, Koposov2023}, \texttt{SMART}\citep{Meadows1996}, \texttt{VPL Climate}\citep{Robinson2018,Lincowski2018,LincowskiPhD}, \texttt{DISORT} \cite{Stamnes1988, Stamnes2000}, \texttt{IRAF/DAOPHOT} \cite{Stetson1987}. We can share the code used in the data reduction or data analysis upon request.

\subsection*{Competing Interests}
The authors declare no competing interests.



\bibliography{bibliography}


\begin{thebibliography}{85}
\ifx \bisbn   \undefined \def \bisbn  #1{ISBN #1}\fi
\ifx \binits  \undefined \def \binits#1{#1}\fi
\ifx \bauthor  \undefined \def \bauthor#1{#1}\fi
\ifx \batitle  \undefined \def \batitle#1{#1}\fi
\ifx \bjtitle  \undefined \def \bjtitle#1{#1}\fi
\ifx \bvolume  \undefined \def \bvolume#1{\textbf{#1}}\fi
\ifx \byear  \undefined \def \byear#1{#1}\fi
\ifx \bissue  \undefined \def \bissue#1{#1}\fi
\ifx \bfpage  \undefined \def \bfpage#1{#1}\fi
\ifx \blpage  \undefined \def \blpage #1{#1}\fi
\ifx \burl  \undefined \def \burl#1{\textsf{#1}}\fi
\ifx \doiurl  \undefined \def \doiurl#1{\url{https://doi.org/#1}}\fi
\ifx \betal  \undefined \def \betal{\textit{et al.}}\fi
\ifx \binstitute  \undefined \def \binstitute#1{#1}\fi
\ifx \binstitutionaled  \undefined \def \binstitutionaled#1{#1}\fi
\ifx \bctitle  \undefined \def \bctitle#1{#1}\fi
\ifx \beditor  \undefined \def \beditor#1{#1}\fi
\ifx \bpublisher  \undefined \def \bpublisher#1{#1}\fi
\ifx \bbtitle  \undefined \def \bbtitle#1{#1}\fi
\ifx \bedition  \undefined \def \bedition#1{#1}\fi
\ifx \bseriesno  \undefined \def \bseriesno#1{#1}\fi
\ifx \blocation  \undefined \def \blocation#1{#1}\fi
\ifx \bsertitle  \undefined \def \bsertitle#1{#1}\fi
\ifx \bsnm \undefined \def \bsnm#1{#1}\fi
\ifx \bsuffix \undefined \def \bsuffix#1{#1}\fi
\ifx \bparticle \undefined \def \bparticle#1{#1}\fi
\ifx \barticle \undefined \def \barticle#1{#1}\fi
\bibcommenthead
\ifx \bconfdate \undefined \def \bconfdate #1{#1}\fi
\ifx \botherref \undefined \def \botherref #1{#1}\fi
\ifx \url \undefined \def \url#1{\textsf{#1}}\fi
\ifx \bchapter \undefined \def \bchapter#1{#1}\fi
\ifx \bbook \undefined \def \bbook#1{#1}\fi
\ifx \bcomment \undefined \def \bcomment#1{#1}\fi
\ifx \oauthor \undefined \def \oauthor#1{#1}\fi
\ifx \citeauthoryear \undefined \def \citeauthoryear#1{#1}\fi
\ifx \endbibitem  \undefined \def \endbibitem {}\fi
\ifx \bconflocation  \undefined \def \bconflocation#1{#1}\fi
\ifx \arxivurl  \undefined \def \arxivurl#1{\textsf{#1}}\fi
\csname PreBibitemsHook\endcsname

\bibitem{Gillon2017}
\begin{barticle}
\bauthor{\bsnm{{Gillon}}, \binits{M.}},
\bauthor{\bsnm{{Triaud}}, \binits{A.H.M.J.}},
\bauthor{\bsnm{{Demory}}, \binits{B.-O.}},
\bauthor{\bsnm{{Jehin}}, \binits{E.}},
\bauthor{\bsnm{{Agol}}, \binits{E.}},
\bauthor{\bsnm{{Deck}}, \binits{K.M.}},
\bauthor{\bsnm{{Lederer}}, \binits{S.M.}},
\bauthor{\bsnm{{de Wit}}, \binits{J.}},
\bauthor{\bsnm{{Burdanov}}, \binits{A.}},
\bauthor{\bsnm{{Ingalls}}, \binits{J.G.}},
\bauthor{\bsnm{{Bolmont}}, \binits{E.}},
\bauthor{\bsnm{{Leconte}}, \binits{J.}},
\bauthor{\bsnm{{Raymond}}, \binits{S.N.}},
\bauthor{\bsnm{{Selsis}}, \binits{F.}},
\bauthor{\bsnm{{Turbet}}, \binits{M.}},
\bauthor{\bsnm{{Barkaoui}}, \binits{K.}},
\bauthor{\bsnm{{Burgasser}}, \binits{A.}},
\bauthor{\bsnm{{Burleigh}}, \binits{M.R.}},
\bauthor{\bsnm{{Carey}}, \binits{S.J.}},
\bauthor{\bsnm{{Chaushev}}, \binits{A.}},
\bauthor{\bsnm{{Copperwheat}}, \binits{C.M.}},
\bauthor{\bsnm{{Delrez}}, \binits{L.}},
\bauthor{\bsnm{{Fernandes}}, \binits{C.S.}},
\bauthor{\bsnm{{Holdsworth}}, \binits{D.L.}},
\bauthor{\bsnm{{Kotze}}, \binits{E.J.}},
\bauthor{\bsnm{{Van Grootel}}, \binits{V.}},
\bauthor{\bsnm{{Almleaky}}, \binits{Y.}},
\bauthor{\bsnm{{Benkhaldoun}}, \binits{Z.}},
\bauthor{\bsnm{{Magain}}, \binits{P.}},
\bauthor{\bsnm{{Queloz}}, \binits{D.}}:
\batitle{{Seven temperate terrestrial planets around the nearby ultracool dwarf
  star TRAPPIST-1}}.
\bjtitle{\nat}
\bvolume{542}(\bissue{7642}),
\bfpage{456}--\blpage{460}
(\byear{2017})
{\href{https://arxiv.org/abs/1703.01424}{{arXiv:1703.01424}}}
{[astro-ph.EP]}.
\doiurl{10.1038/nature21360}
\end{barticle}
\endbibitem

\bibitem{Morley2017}
\begin{barticle}
\bauthor{\bsnm{{Morley}}, \binits{C.V.}},
\bauthor{\bsnm{{Kreidberg}}, \binits{L.}},
\bauthor{\bsnm{{Rustamkulov}}, \binits{Z.}},
\bauthor{\bsnm{{Robinson}}, \binits{T.}},
\bauthor{\bsnm{{Fortney}}, \binits{J.J.}}:
\batitle{{Observing the Atmospheres of Known Temperate Earth-sized Planets with
  JWST}}.
\bjtitle{\apj}
\bvolume{850}(\bissue{2}),
\bfpage{121}
(\byear{2017})
{\href{https://arxiv.org/abs/1708.04239}{{arXiv:1708.04239}}}
{[astro-ph.EP]}.
\doiurl{10.3847/1538-4357/aa927b}
\end{barticle}
\endbibitem

\bibitem{Lincowski2018}
\begin{barticle}
\bauthor{\bsnm{{Lincowski}}, \binits{A.P.}},
\bauthor{\bsnm{{Meadows}}, \binits{V.S.}},
\bauthor{\bsnm{{Crisp}}, \binits{D.}},
\bauthor{\bsnm{{Robinson}}, \binits{T.D.}},
\bauthor{\bsnm{{Luger}}, \binits{R.}},
\bauthor{\bsnm{{Lustig-Yaeger}}, \binits{J.}},
\bauthor{\bsnm{{Arney}}, \binits{G.N.}}:
\batitle{{Evolved Climates and Observational Discriminants for the TRAPPIST-1
  Planetary System}}.
\bjtitle{\apj}
\bvolume{867}(\bissue{1}),
\bfpage{76}
(\byear{2018})
{\href{https://arxiv.org/abs/1809.07498}{{arXiv:1809.07498}}}
{[astro-ph.EP]}.
\doiurl{10.3847/1538-4357/aae36a}
\end{barticle}
\endbibitem

\bibitem{Greene2023}
\begin{botherref}
\oauthor{\bsnm{{Greene}}, \binits{T.P.}},
\oauthor{\bsnm{{Bell}}, \binits{T.J.}},
\oauthor{\bsnm{{Ducrot}}, \binits{E.}},
\oauthor{\bsnm{{Dyrek}}, \binits{A.}},
\oauthor{\bsnm{{Lagage}}, \binits{P.-O.}},
\oauthor{\bsnm{{Fortney}}, \binits{J.J.}}:
{Thermal emission from the Earth-sized exoplanet TRAPPIST-1 b using JWST}.
arXiv e-prints,
2303--14849
(2023)
{\href{https://arxiv.org/abs/2303.14849}{{arXiv:2303.14849}}}
{[astro-ph.EP]}.
\doiurl{10.48550/arXiv.2303.14849}
\end{botherref}
\endbibitem

\bibitem{Wordsworth2022}
\begin{barticle}
\bauthor{\bsnm{{Wordsworth}}, \binits{R.}},
\bauthor{\bsnm{{Kreidberg}}, \binits{L.}}:
\batitle{{Atmospheres of Rocky Exoplanets}}.
\bjtitle{\araa}
\bvolume{60},
\bfpage{159}--\blpage{201}
(\byear{2022})
{\href{https://arxiv.org/abs/2112.04663}{{arXiv:2112.04663}}}
{[astro-ph.EP]}.
\doiurl{10.1146/annurev-astro-052920-125632}
\end{barticle}
\endbibitem

\bibitem{Luger2015}
\begin{barticle}
\bauthor{\bsnm{{Luger}}, \binits{R.}},
\bauthor{\bsnm{{Barnes}}, \binits{R.}}:
\batitle{{Extreme Water Loss and Abiotic O2Buildup on Planets Throughout the
  Habitable Zones of M Dwarfs}}.
\bjtitle{Astrobiology}
\bvolume{15}(\bissue{2}),
\bfpage{119}--\blpage{143}
(\byear{2015})
{\href{https://arxiv.org/abs/1411.7412}{{arXiv:1411.7412}}}
{[astro-ph.EP]}.
\doiurl{10.1089/ast.2014.1231}
\end{barticle}
\endbibitem

\bibitem{Kreidberg2019}
\begin{barticle}
\bauthor{\bsnm{{Kreidberg}}, \binits{L.}},
\bauthor{\bsnm{{Koll}}, \binits{D.D.B.}},
\bauthor{\bsnm{{Morley}}, \binits{C.}},
\bauthor{\bsnm{{Hu}}, \binits{R.}},
\bauthor{\bsnm{{Schaefer}}, \binits{L.}},
\bauthor{\bsnm{{Deming}}, \binits{D.}},
\bauthor{\bsnm{{Stevenson}}, \binits{K.B.}},
\bauthor{\bsnm{{Dittmann}}, \binits{J.}},
\bauthor{\bsnm{{Vanderburg}}, \binits{A.}},
\bauthor{\bsnm{{Berardo}}, \binits{D.}},
\bauthor{\bsnm{{Guo}}, \binits{X.}},
\bauthor{\bsnm{{Stassun}}, \binits{K.}},
\bauthor{\bsnm{{Crossfield}}, \binits{I.}},
\bauthor{\bsnm{{Charbonneau}}, \binits{D.}},
\bauthor{\bsnm{{Latham}}, \binits{D.W.}},
\bauthor{\bsnm{{Loeb}}, \binits{A.}},
\bauthor{\bsnm{{Ricker}}, \binits{G.}},
\bauthor{\bsnm{{Seager}}, \binits{S.}},
\bauthor{\bsnm{{Vanderspek}}, \binits{R.}}:
\batitle{{Absence of a thick atmosphere on the terrestrial exoplanet LHS
  3844b}}.
\bjtitle{\nat}
\bvolume{573}(\bissue{7772}),
\bfpage{87}--\blpage{90}
(\byear{2019})
{\href{https://arxiv.org/abs/1908.06834}{{arXiv:1908.06834}}}
{[astro-ph.EP]}.
\doiurl{10.1038/s41586-019-1497-4}
\end{barticle}
\endbibitem

\bibitem{Crossfield2022}
\begin{barticle}
\bauthor{\bsnm{{Crossfield}}, \binits{I.J.M.}},
\bauthor{\bsnm{{Malik}}, \binits{M.}},
\bauthor{\bsnm{{Hill}}, \binits{M.L.}},
\bauthor{\bsnm{{Kane}}, \binits{S.R.}},
\bauthor{\bsnm{{Foley}}, \binits{B.}},
\bauthor{\bsnm{{Polanski}}, \binits{A.S.}},
\bauthor{\bsnm{{Coria}}, \binits{D.}},
\bauthor{\bsnm{{Brande}}, \binits{J.}},
\bauthor{\bsnm{{Zhang}}, \binits{Y.}},
\bauthor{\bsnm{{Wienke}}, \binits{K.}},
\bauthor{\bsnm{{Kreidberg}}, \binits{L.}},
\bauthor{\bsnm{{Cowan}}, \binits{N.B.}},
\bauthor{\bsnm{{Dragomir}}, \binits{D.}},
\bauthor{\bsnm{{Gorjian}}, \binits{V.}},
\bauthor{\bsnm{{Mikal-Evans}}, \binits{T.}},
\bauthor{\bsnm{{Benneke}}, \binits{B.}},
\bauthor{\bsnm{{Christiansen}}, \binits{J.L.}},
\bauthor{\bsnm{{Deming}}, \binits{D.}},
\bauthor{\bsnm{{Morales}}, \binits{F.Y.}}:
\batitle{{GJ 1252b: A Hot Terrestrial Super-Earth with No Atmosphere}}.
\bjtitle{\apjl}
\bvolume{937}(\bissue{1}),
\bfpage{17}
(\byear{2022})
{\href{https://arxiv.org/abs/2208.09479}{{arXiv:2208.09479}}}
{[astro-ph.EP]}.
\doiurl{10.3847/2041-8213/ac886b}
\end{barticle}
\endbibitem

\bibitem{Bell2022}
\begin{barticle}
\bauthor{\bsnm{{Bell}}, \binits{T.}},
\bauthor{\bsnm{{Ahrer}}, \binits{E.-M.}},
\bauthor{\bsnm{{Brande}}, \binits{J.}},
\bauthor{\bsnm{{Carter}}, \binits{A.}},
\bauthor{\bsnm{{Feinstein}}, \binits{A.}},
\bauthor{\bsnm{{Caloca}}, \binits{G.}},
\bauthor{\bsnm{{Mansfield}}, \binits{M.}},
\bauthor{\bsnm{{Zieba}}, \binits{S.}},
\bauthor{\bsnm{{Piaulet}}, \binits{C.}},
\bauthor{\bsnm{{Benneke}}, \binits{B.}},
\bauthor{\bsnm{{Filippazzo}}, \binits{J.}},
\bauthor{\bsnm{{May}}, \binits{E.}},
\bauthor{\bsnm{{Roy}}, \binits{P.-A.}},
\bauthor{\bsnm{{Kreidberg}}, \binits{L.}},
\bauthor{\bsnm{{Stevenson}}, \binits{K.}}:
\batitle{{Eureka!: An End-to-End Pipeline for JWST Time-Series Observations}}.
\bjtitle{The Journal of Open Source Software}
\bvolume{7}(\bissue{79}),
\bfpage{4503}
(\byear{2022})
{\href{https://arxiv.org/abs/2207.03585}{{arXiv:2207.03585}}}
{[astro-ph.IM]}.
\doiurl{10.21105/joss.04503}
\end{barticle}
\endbibitem

\bibitem{Bell2023}
\begin{botherref}
\oauthor{\bsnm{{Bell}}, \binits{T.J.}},
\oauthor{\bsnm{{Kreidberg}}, \binits{L.}},
\oauthor{\bsnm{{Kendrew}}, \binits{S.}},
\oauthor{\bsnm{{Bean}}, \binits{J.}},
\oauthor{\bsnm{{Crouzet}}, \binits{N.}},
\oauthor{\bsnm{{Ducrot}}, \binits{E.}},
\oauthor{\bsnm{{Dyrek}}, \binits{A.}},
\oauthor{\bsnm{{Gao}}, \binits{P.}},
\oauthor{\bsnm{{Lagage}}, \binits{P.-O.}},
\oauthor{\bsnm{{Moses}}, \binits{J.I.}}:
{A First Look at the JWST MIRI/LRS Phase Curve of WASP-43b}.
arXiv e-prints,
2301--06350
(2023)
{\href{https://arxiv.org/abs/2301.06350}{{arXiv:2301.06350}}}
{[astro-ph.IM]}.
\doiurl{10.48550/arXiv.2301.06350}
\end{botherref}
\endbibitem

\bibitem{Moroz1985}
\begin{barticle}
\bauthor{\bsnm{{Moroz}}, \binits{V.I.}},
\bauthor{\bsnm{{Ekonomov}}, \binits{A.P.}},
\bauthor{\bsnm{{Moshkin}}, \binits{B.E.}},
\bauthor{\bsnm{{Revercomb}}, \binits{H.E.}},
\bauthor{\bsnm{{Sromovsky}}, \binits{L.A.}},
\bauthor{\bsnm{{Schofield}}, \binits{J.T.}},
\bauthor{\bsnm{{Sp{\"a}nkuch}}, \binits{D.}},
\bauthor{\bsnm{{Taylor}}, \binits{F.W.}},
\bauthor{\bsnm{{Tomasko}}, \binits{M.G.}}:
\batitle{{Solar and thermal radiation in the Venus atmosphere}}.
\bjtitle{Advances in Space Research}
\bvolume{5}(\bissue{11}),
\bfpage{197}--\blpage{232}
(\byear{1985}).
\doiurl{10.1016/0273-1177(85)90202-9}
\end{barticle}
\endbibitem

\bibitem{Mallama2002}
\begin{barticle}
\bauthor{\bsnm{{Mallama}}, \binits{A.}},
\bauthor{\bsnm{{Wang}}, \binits{D.}},
\bauthor{\bsnm{{Howard}}, \binits{R.A.}}:
\batitle{{Photometry of Mercury from SOHO/LASCO and Earth. The Phase Function
  from 2 to 170 deg.}}
\bjtitle{\icarus}
\bvolume{155}(\bissue{2}),
\bfpage{253}--\blpage{264}
(\byear{2002}).
\doiurl{10.1006/icar.2001.6723}
\end{barticle}
\endbibitem

\bibitem{Cowan2011}
\begin{barticle}
\bauthor{\bsnm{{Cowan}}, \binits{N.B.}},
\bauthor{\bsnm{{Agol}}, \binits{E.}}:
\batitle{{The Statistics of Albedo and Heat Recirculation on Hot Exoplanets}}.
\bjtitle{\apj}
\bvolume{729}(\bissue{1}),
\bfpage{54}
(\byear{2011})
{\href{https://arxiv.org/abs/1001.0012}{{arXiv:1001.0012}}}
{[astro-ph.EP]}.
\doiurl{10.1088/0004-637X/729/1/54}
\end{barticle}
\endbibitem

\bibitem{Koll2019}
\begin{barticle}
\bauthor{\bsnm{{Koll}}, \binits{D.D.B.}},
\bauthor{\bsnm{{Malik}}, \binits{M.}},
\bauthor{\bsnm{{Mansfield}}, \binits{M.}},
\bauthor{\bsnm{{Kempton}}, \binits{E.M.-R.}},
\bauthor{\bsnm{{Kite}}, \binits{E.}},
\bauthor{\bsnm{{Abbot}}, \binits{D.}},
\bauthor{\bsnm{{Bean}}, \binits{J.L.}}:
\batitle{{Identifying Candidate Atmospheres on Rocky M Dwarf Planets via
  Eclipse Photometry}}.
\bjtitle{\apj}
\bvolume{886}(\bissue{2}),
\bfpage{140}
(\byear{2019})
{\href{https://arxiv.org/abs/1907.13138}{{arXiv:1907.13138}}}
{[astro-ph.EP]}.
\doiurl{10.3847/1538-4357/ab4c91}
\end{barticle}
\endbibitem

\bibitem{Schaefer2016}
\begin{barticle}
\bauthor{\bsnm{{Schaefer}}, \binits{L.}},
\bauthor{\bsnm{{Wordsworth}}, \binits{R.D.}},
\bauthor{\bsnm{{Berta-Thompson}}, \binits{Z.}},
\bauthor{\bsnm{{Sasselov}}, \binits{D.}}:
\batitle{{Predictions of the Atmospheric Composition of GJ 1132b}}.
\bjtitle{\apj}
\bvolume{829}(\bissue{2}),
\bfpage{63}
(\byear{2016})
{\href{https://arxiv.org/abs/1607.03906}{{arXiv:1607.03906}}}
{[astro-ph.EP]}.
\doiurl{10.3847/0004-637X/829/2/63}
\end{barticle}
\endbibitem

\bibitem{Bolmont2017}
\begin{barticle}
\bauthor{\bsnm{{Bolmont}}, \binits{E.}},
\bauthor{\bsnm{{Selsis}}, \binits{F.}},
\bauthor{\bsnm{{Owen}}, \binits{J.E.}},
\bauthor{\bsnm{{Ribas}}, \binits{I.}},
\bauthor{\bsnm{{Raymond}}, \binits{S.N.}},
\bauthor{\bsnm{{Leconte}}, \binits{J.}},
\bauthor{\bsnm{{Gillon}}, \binits{M.}}:
\batitle{{Water loss from terrestrial planets orbiting ultracool dwarfs:
  implications for the planets of TRAPPIST-1}}.
\bjtitle{\mnras}
\bvolume{464}(\bissue{3}),
\bfpage{3728}--\blpage{3741}
(\byear{2017})
{\href{https://arxiv.org/abs/1605.00616}{{arXiv:1605.00616}}}
{[astro-ph.EP]}.
\doiurl{10.1093/mnras/stw2578}
\end{barticle}
\endbibitem

\bibitem{Dorn2018}
\begin{barticle}
\bauthor{\bsnm{{Dorn}}, \binits{C.}},
\bauthor{\bsnm{{Noack}}, \binits{L.}},
\bauthor{\bsnm{{Rozel}}, \binits{A.B.}}:
\batitle{{Outgassing on stagnant-lid super-Earths}}.
\bjtitle{\aap}
\bvolume{614},
\bfpage{18}
(\byear{2018})
{\href{https://arxiv.org/abs/1802.09264}{{arXiv:1802.09264}}}
{[astro-ph.EP]}.
\doiurl{10.1051/0004-6361/201731513}
\end{barticle}
\endbibitem

\bibitem{Kane2020}
\begin{barticle}
\bauthor{\bsnm{{Kane}}, \binits{S.R.}},
\bauthor{\bsnm{{Roettenbacher}}, \binits{R.M.}},
\bauthor{\bsnm{{Unterborn}}, \binits{C.T.}},
\bauthor{\bsnm{{Foley}}, \binits{B.J.}},
\bauthor{\bsnm{{Hill}}, \binits{M.L.}}:
\batitle{{A Volatile-poor Formation of LHS 3844b Based on Its Lack of
  Significant Atmosphere}}.
\bjtitle{\psj}
\bvolume{1}(\bissue{2}),
\bfpage{36}
(\byear{2020})
{\href{https://arxiv.org/abs/2007.14493}{{arXiv:2007.14493}}}
{[astro-ph.EP]}.
\doiurl{10.3847/PSJ/abaab5}
\end{barticle}
\endbibitem

\bibitem{Malik2019}
\begin{barticle}
\bauthor{\bsnm{{Malik}}, \binits{M.}},
\bauthor{\bsnm{{Kempton}}, \binits{E.M.-R.}},
\bauthor{\bsnm{{Koll}}, \binits{D.D.B.}},
\bauthor{\bsnm{{Mansfield}}, \binits{M.}},
\bauthor{\bsnm{{Bean}}, \binits{J.L.}},
\bauthor{\bsnm{{Kite}}, \binits{E.}}:
\batitle{{Analyzing Atmospheric Temperature Profiles and Spectra of M Dwarf
  Rocky Planets}}.
\bjtitle{\apj}
\bvolume{886}(\bissue{2}),
\bfpage{142}
(\byear{2019})
{\href{https://arxiv.org/abs/1907.13135}{{arXiv:1907.13135}}}
{[astro-ph.EP]}.
\doiurl{10.3847/1538-4357/ab4a05}
\end{barticle}
\endbibitem

\bibitem{Acuna21}
\begin{barticle}
\bauthor{\bsnm{{Acu{\~n}a}}, \binits{L.}},
\bauthor{\bsnm{{Deleuil}}, \binits{M.}},
\bauthor{\bsnm{{Mousis}}, \binits{O.}},
\bauthor{\bsnm{{Marcq}}, \binits{E.}},
\bauthor{\bsnm{{Levesque}}, \binits{M.}},
\bauthor{\bsnm{{Aguichine}}, \binits{A.}}:
\batitle{{Characterisation of the hydrospheres of TRAPPIST-1 planets}}.
\bjtitle{\aap}
\bvolume{647},
\bfpage{53}
(\byear{2021})
{\href{https://arxiv.org/abs/2101.08172}{{arXiv:2101.08172}}}
{[astro-ph.EP]}.
\doiurl{10.1051/0004-6361/202039885}
\end{barticle}
\endbibitem

\bibitem{Delrez2018}
\begin{barticle}
\bauthor{\bsnm{{Delrez}}, \binits{L.}},
\bauthor{\bsnm{{Gillon}}, \binits{M.}},
\bauthor{\bsnm{{Triaud}}, \binits{A.H.M.J.}},
\bauthor{\bsnm{{Demory}}, \binits{B.-O.}},
\bauthor{\bsnm{{de Wit}}, \binits{J.}},
\bauthor{\bsnm{{Ingalls}}, \binits{J.G.}},
\bauthor{\bsnm{{Agol}}, \binits{E.}},
\bauthor{\bsnm{{Bolmont}}, \binits{E.}},
\bauthor{\bsnm{{Burdanov}}, \binits{A.}},
\bauthor{\bsnm{{Burgasser}}, \binits{A.J.}},
\bauthor{\bsnm{{Carey}}, \binits{S.J.}},
\bauthor{\bsnm{{Jehin}}, \binits{E.}},
\bauthor{\bsnm{{Leconte}}, \binits{J.}},
\bauthor{\bsnm{{Lederer}}, \binits{S.}},
\bauthor{\bsnm{{Queloz}}, \binits{D.}},
\bauthor{\bsnm{{Selsis}}, \binits{F.}},
\bauthor{\bsnm{{Van Grootel}}, \binits{V.}}:
\batitle{{Early 2017 observations of TRAPPIST-1 with Spitzer}}.
\bjtitle{\mnras}
\bvolume{475}(\bissue{3}),
\bfpage{3577}--\blpage{3597}
(\byear{2018})
{\href{https://arxiv.org/abs/1801.02554}{{arXiv:1801.02554}}}
{[astro-ph.EP]}.
\doiurl{10.1093/mnras/sty051}
\end{barticle}
\endbibitem

\bibitem{hu2012}
\begin{barticle}
\bauthor{\bsnm{{Hu}}, \binits{R.}},
\bauthor{\bsnm{{Ehlmann}}, \binits{B.L.}},
\bauthor{\bsnm{{Seager}}, \binits{S.}}:
\batitle{{Theoretical Spectra of Terrestrial Exoplanet Surfaces}}.
\bjtitle{\apj}
\bvolume{752}(\bissue{1}),
\bfpage{7}
(\byear{2012})
{\href{https://arxiv.org/abs/1204.1544}{{arXiv:1204.1544}}}
{[astro-ph.EP]}.
\doiurl{10.1088/0004-637X/752/1/7}
\end{barticle}
\endbibitem

\bibitem{hapke2001}
\begin{barticle}
\bauthor{\bsnm{{Hapke}}, \binits{B.}}:
\batitle{{Space weathering from Mercury to the asteroid belt}}.
\bjtitle{\jgr}
\bvolume{106}(\bissue{E5}),
\bfpage{10039}--\blpage{10074}
(\byear{2001}).
\doiurl{10.1029/2000JE001338}
\end{barticle}
\endbibitem

\bibitem{Chadney2015}
\begin{barticle}
\bauthor{\bsnm{{Chadney}}, \binits{J.M.}},
\bauthor{\bsnm{{Galand}}, \binits{M.}},
\bauthor{\bsnm{{Unruh}}, \binits{Y.C.}},
\bauthor{\bsnm{{Koskinen}}, \binits{T.T.}},
\bauthor{\bsnm{{Sanz-Forcada}}, \binits{J.}}:
\batitle{{XUV-driven mass loss from extrasolar giant planets orbiting active
  stars}}.
\bjtitle{\icarus}
\bvolume{250},
\bfpage{357}--\blpage{367}
(\byear{2015})
{\href{https://arxiv.org/abs/1412.3380}{{arXiv:1412.3380}}}
{[astro-ph.SR]}.
\doiurl{10.1016/j.icarus.2014.12.012}
\end{barticle}
\endbibitem

\bibitem{Fleming2020}
\begin{barticle}
\bauthor{\bsnm{{Fleming}}, \binits{D.P.}},
\bauthor{\bsnm{{Barnes}}, \binits{R.}},
\bauthor{\bsnm{{Luger}}, \binits{R.}},
\bauthor{\bsnm{{VanderPlas}}, \binits{J.T.}}:
\batitle{{On the XUV Luminosity Evolution of TRAPPIST-1}}.
\bjtitle{\apj}
\bvolume{891}(\bissue{2}),
\bfpage{155}
(\byear{2020})
{\href{https://arxiv.org/abs/1906.05250}{{arXiv:1906.05250}}}
{[astro-ph.SR]}.
\doiurl{10.3847/1538-4357/ab77ad}
\end{barticle}
\endbibitem

\bibitem{Kreidberg2021}
\begin{botherref}
\oauthor{\bsnm{{Kreidberg}}, \binits{L.}},
\oauthor{\bsnm{{Agol}}, \binits{E.}},
\oauthor{\bsnm{{Bolmont}}, \binits{E.}},
\oauthor{\bsnm{{Gillon}}, \binits{M.}},
\oauthor{\bsnm{{Hu}}, \binits{R.}},
\oauthor{\bsnm{{Koll}}, \binits{D.}},
\oauthor{\bsnm{{Mandell}}, \binits{A.}},
\oauthor{\bsnm{{Meadows}}, \binits{V.S.}},
\oauthor{\bsnm{{Morley}}, \binits{C.}},
\oauthor{\bsnm{{Schaefer}}, \binits{L.}},
\oauthor{\bsnm{{Selsis}}, \binits{F.}},
\oauthor{\bsnm{{de Wit}}, \binits{J.}}:
{Hot Take on a Cool World: Does Trappist-1c Have an Atmosphere?}
JWST Proposal. Cycle 1, ID. \#2304
(2021)
\end{botherref}
\endbibitem

\bibitem{ERS2022}
\begin{barticle}
\bauthor{\bsnm{{JWST Transiting Exoplanet Community Early Release Science
  Team}}},
\bauthor{\bsnm{{Ahrer}}, \binits{E.-M.}},
\bauthor{\bsnm{{Alderson}}, \binits{L.}},
\bauthor{\bsnm{{Batalha}}, \binits{N.M.}},
\bauthor{\bsnm{{Batalha}}, \binits{N.E.}},
\bauthor{\bsnm{{Bean}}, \binits{J.L.}},
\bauthor{\bsnm{{Beatty}}, \binits{T.G.}},
\bauthor{\bsnm{{Bell}}, \binits{T.J.}},
\bauthor{\bsnm{{Benneke}}, \binits{B.}},
\bauthor{\bsnm{{Berta-Thompson}}, \binits{Z.K.}},
\bauthor{\bsnm{{Carter}}, \binits{A.L.}},
\bauthor{\bsnm{{Crossfield}}, \binits{I.J.M.}},
\bauthor{\bsnm{{Espinoza}}, \binits{N.}},
\bauthor{\bsnm{{Feinstein}}, \binits{A.D.}},
\bauthor{\bsnm{{Fortney}}, \binits{J.J.}},
\bauthor{\bsnm{{Gibson}}, \binits{N.P.}},
\bauthor{\bsnm{{Goyal}}, \binits{J.M.}},
\bauthor{\bsnm{{Kempton}}, \binits{E.M.-R.}},
\bauthor{\bsnm{{Kirk}}, \binits{J.}},
\bauthor{\bsnm{{Kreidberg}}, \binits{L.}},
\bauthor{\bsnm{{L{\'o}pez-Morales}}, \binits{M.}},
\bauthor{\bsnm{{Line}}, \binits{M.R.}},
\bauthor{\bsnm{{Lothringer}}, \binits{J.D.}},
\bauthor{\bsnm{{Moran}}, \binits{S.E.}},
\bauthor{\bsnm{{Mukherjee}}, \binits{S.}},
\bauthor{\bsnm{{Ohno}}, \binits{K.}},
\bauthor{\bsnm{{Parmentier}}, \binits{V.}},
\bauthor{\bsnm{{Piaulet}}, \binits{C.}},
\bauthor{\bsnm{{Rustamkulov}}, \binits{Z.}},
\bauthor{\bsnm{{Schlawin}}, \binits{E.}},
\bauthor{\bsnm{{Sing}}, \binits{D.K.}},
\bauthor{\bsnm{{Stevenson}}, \binits{K.B.}},
\bauthor{\bsnm{{Wakeford}}, \binits{H.R.}},
\bauthor{\bsnm{{Allen}}, \binits{N.H.}},
\bauthor{\bsnm{{Birkmann}}, \binits{S.M.}},
\bauthor{\bsnm{{Brande}}, \binits{J.}},
\bauthor{\bsnm{{Crouzet}}, \binits{N.}},
\bauthor{\bsnm{{Cubillos}}, \binits{P.E.}},
\bauthor{\bsnm{{Damiano}}, \binits{M.}},
\bauthor{\bsnm{{D{\'e}sert}}, \binits{J.-M.}},
\bauthor{\bsnm{{Gao}}, \binits{P.}},
\bauthor{\bsnm{{Harrington}}, \binits{J.}},
\bauthor{\bsnm{{Hu}}, \binits{R.}},
\bauthor{\bsnm{{Kendrew}}, \binits{S.}},
\bauthor{\bsnm{{Knutson}}, \binits{H.A.}},
\bauthor{\bsnm{{Lagage}}, \binits{P.-O.}},
\bauthor{\bsnm{{Leconte}}, \binits{J.}},
\bauthor{\bsnm{{Lendl}}, \binits{M.}},
\bauthor{\bsnm{{MacDonald}}, \binits{R.J.}},
\bauthor{\bsnm{{May}}, \binits{E.M.}},
\bauthor{\bsnm{{Miguel}}, \binits{Y.}},
\bauthor{\bsnm{{Molaverdikhani}}, \binits{K.}},
\bauthor{\bsnm{{Moses}}, \binits{J.I.}},
\bauthor{\bsnm{{Murray}}, \binits{C.A.}},
\bauthor{\bsnm{{Nehring}}, \binits{M.}},
\bauthor{\bsnm{{Nikolov}}, \binits{N.K.}},
\bauthor{\bsnm{{Petit dit de la Roche}}, \binits{D.J.M.}},
\bauthor{\bsnm{{Radica}}, \binits{M.}},
\bauthor{\bsnm{{Roy}}, \binits{P.-A.}},
\bauthor{\bsnm{{Stassun}}, \binits{K.G.}},
\bauthor{\bsnm{{Taylor}}, \binits{J.}},
\bauthor{\bsnm{{Waalkes}}, \binits{W.C.}},
\bauthor{\bsnm{{Wachiraphan}}, \binits{P.}},
\bauthor{\bsnm{{Welbanks}}, \binits{L.}},
\bauthor{\bsnm{{Wheatley}}, \binits{P.J.}},
\bauthor{\bsnm{{Aggarwal}}, \binits{K.}},
\bauthor{\bsnm{{Alam}}, \binits{M.K.}},
\bauthor{\bsnm{{Banerjee}}, \binits{A.}},
\bauthor{\bsnm{{Barstow}}, \binits{J.K.}},
\bauthor{\bsnm{{Blecic}}, \binits{J.}},
\bauthor{\bsnm{{Casewell}}, \binits{S.L.}},
\bauthor{\bsnm{{Changeat}}, \binits{Q.}},
\bauthor{\bsnm{{Chubb}}, \binits{K.L.}},
\bauthor{\bsnm{{Col{\'o}n}}, \binits{K.D.}},
\bauthor{\bsnm{{Coulombe}}, \binits{L.-P.}},
\bauthor{\bsnm{{Daylan}}, \binits{T.}},
\bauthor{\bsnm{{de Val-Borro}}, \binits{M.}},
\bauthor{\bsnm{{Decin}}, \binits{L.}},
\bauthor{\bsnm{{Dos Santos}}, \binits{L.A.}},
\bauthor{\bsnm{{Flagg}}, \binits{L.}},
\bauthor{\bsnm{{France}}, \binits{K.}},
\bauthor{\bsnm{{Fu}}, \binits{G.}},
\bauthor{\bsnm{{Garc{\'\i}a Mu{\~n}oz}}, \binits{A.}},
\bauthor{\bsnm{{Gizis}}, \binits{J.E.}},
\bauthor{\bsnm{{Glidden}}, \binits{A.}},
\bauthor{\bsnm{{Grant}}, \binits{D.}},
\bauthor{\bsnm{{Heng}}, \binits{K.}},
\bauthor{\bsnm{{Henning}}, \binits{T.}},
\bauthor{\bsnm{{Hong}}, \binits{Y.-C.}},
\bauthor{\bsnm{{Inglis}}, \binits{J.}},
\bauthor{\bsnm{{Iro}}, \binits{N.}},
\bauthor{\bsnm{{Kataria}}, \binits{T.}},
\bauthor{\bsnm{{Komacek}}, \binits{T.D.}},
\bauthor{\bsnm{{Krick}}, \binits{J.E.}},
\bauthor{\bsnm{{Lee}}, \binits{E.K.H.}},
\bauthor{\bsnm{{Lewis}}, \binits{N.K.}},
\bauthor{\bsnm{{Lillo-Box}}, \binits{J.}},
\bauthor{\bsnm{{Lustig-Yaeger}}, \binits{J.}},
\bauthor{\bsnm{{Mancini}}, \binits{L.}},
\bauthor{\bsnm{{Mandell}}, \binits{A.M.}},
\bauthor{\bsnm{{Mansfield}}, \binits{M.}},
\bauthor{\bsnm{{Marley}}, \binits{M.S.}},
\bauthor{\bsnm{{Mikal-Evans}}, \binits{T.}},
\bauthor{\bsnm{{Morello}}, \binits{G.}},
\bauthor{\bsnm{{Nixon}}, \binits{M.C.}},
\bauthor{\bsnm{{Ortiz Ceballos}}, \binits{K.}},
\bauthor{\bsnm{{Piette}}, \binits{A.A.A.}},
\bauthor{\bsnm{{Powell}}, \binits{D.}},
\bauthor{\bsnm{{Rackham}}, \binits{B.V.}},
\bauthor{\bsnm{{Ramos-Rosado}}, \binits{L.}},
\bauthor{\bsnm{{Rauscher}}, \binits{E.}},
\bauthor{\bsnm{{Redfield}}, \binits{S.}},
\bauthor{\bsnm{{Rogers}}, \binits{L.K.}},
\bauthor{\bsnm{{Roman}}, \binits{M.T.}},
\bauthor{\bsnm{{Roudier}}, \binits{G.M.}},
\bauthor{\bsnm{{Scarsdale}}, \binits{N.}},
\bauthor{\bsnm{{Shkolnik}}, \binits{E.L.}},
\bauthor{\bsnm{{Southworth}}, \binits{J.}},
\bauthor{\bsnm{{Spake}}, \binits{J.J.}},
\bauthor{\bsnm{{Steinrueck}}, \binits{M.E.}},
\bauthor{\bsnm{{Tan}}, \binits{X.}},
\bauthor{\bsnm{{Teske}}, \binits{J.K.}},
\bauthor{\bsnm{{Tremblin}}, \binits{P.}},
\bauthor{\bsnm{{Tsai}}, \binits{S.-M.}},
\bauthor{\bsnm{{Tucker}}, \binits{G.S.}},
\bauthor{\bsnm{{Turner}}, \binits{J.D.}},
\bauthor{\bsnm{{Valenti}}, \binits{J.A.}},
\bauthor{\bsnm{{Venot}}, \binits{O.}},
\bauthor{\bsnm{{Waldmann}}, \binits{I.P.}},
\bauthor{\bsnm{{Wallack}}, \binits{N.L.}},
\bauthor{\bsnm{{Zhang}}, \binits{X.}},
\bauthor{\bsnm{{Zieba}}, \binits{S.}}:
\batitle{{Identification of carbon dioxide in an exoplanet atmosphere}}.
\bjtitle{\nat}
\bvolume{614}(\bissue{7949}),
\bfpage{649}--\blpage{652}
(\byear{2023})
{\href{https://arxiv.org/abs/2208.11692}{{arXiv:2208.11692}}}
{[astro-ph.EP]}.
\doiurl{10.1038/s41586-022-05269-w}
\end{barticle}
\endbibitem

\bibitem{Ahrer2022}
\begin{barticle}
\bauthor{\bsnm{{Ahrer}}, \binits{E.-M.}},
\bauthor{\bsnm{{Stevenson}}, \binits{K.B.}},
\bauthor{\bsnm{{Mansfield}}, \binits{M.}},
\bauthor{\bsnm{{Moran}}, \binits{S.E.}},
\bauthor{\bsnm{{Brande}}, \binits{J.}},
\bauthor{\bsnm{{Morello}}, \binits{G.}},
\bauthor{\bsnm{{Murray}}, \binits{C.A.}},
\bauthor{\bsnm{{Nikolov}}, \binits{N.K.}},
\bauthor{\bsnm{{Petit dit de la Roche}}, \binits{D.J.M.}},
\bauthor{\bsnm{{Schlawin}}, \binits{E.}},
\bauthor{\bsnm{{Wheatley}}, \binits{P.J.}},
\bauthor{\bsnm{{Zieba}}, \binits{S.}},
\bauthor{\bsnm{{Batalha}}, \binits{N.E.}},
\bauthor{\bsnm{{Damiano}}, \binits{M.}},
\bauthor{\bsnm{{Goyal}}, \binits{J.M.}},
\bauthor{\bsnm{{Lendl}}, \binits{M.}},
\bauthor{\bsnm{{Lothringer}}, \binits{J.D.}},
\bauthor{\bsnm{{Mukherjee}}, \binits{S.}},
\bauthor{\bsnm{{Ohno}}, \binits{K.}},
\bauthor{\bsnm{{Batalha}}, \binits{N.M.}},
\bauthor{\bsnm{{Battley}}, \binits{M.P.}},
\bauthor{\bsnm{{Bean}}, \binits{J.L.}},
\bauthor{\bsnm{{Beatty}}, \binits{T.G.}},
\bauthor{\bsnm{{Benneke}}, \binits{B.}},
\bauthor{\bsnm{{Berta-Thompson}}, \binits{Z.K.}},
\bauthor{\bsnm{{Carter}}, \binits{A.L.}},
\bauthor{\bsnm{{Cubillos}}, \binits{P.E.}},
\bauthor{\bsnm{{Daylan}}, \binits{T.}},
\bauthor{\bsnm{{Espinoza}}, \binits{N.}},
\bauthor{\bsnm{{Gao}}, \binits{P.}},
\bauthor{\bsnm{{Gibson}}, \binits{N.P.}},
\bauthor{\bsnm{{Gill}}, \binits{S.}},
\bauthor{\bsnm{{Harrington}}, \binits{J.}},
\bauthor{\bsnm{{Hu}}, \binits{R.}},
\bauthor{\bsnm{{Kreidberg}}, \binits{L.}},
\bauthor{\bsnm{{Lewis}}, \binits{N.K.}},
\bauthor{\bsnm{{Line}}, \binits{M.R.}},
\bauthor{\bsnm{{L{\'o}pez-Morales}}, \binits{M.}},
\bauthor{\bsnm{{Parmentier}}, \binits{V.}},
\bauthor{\bsnm{{Powell}}, \binits{D.K.}},
\bauthor{\bsnm{{Sing}}, \binits{D.K.}},
\bauthor{\bsnm{{Tsai}}, \binits{S.-M.}},
\bauthor{\bsnm{{Wakeford}}, \binits{H.R.}},
\bauthor{\bsnm{{Welbanks}}, \binits{L.}},
\bauthor{\bsnm{{Alam}}, \binits{M.K.}},
\bauthor{\bsnm{{Alderson}}, \binits{L.}},
\bauthor{\bsnm{{Allen}}, \binits{N.H.}},
\bauthor{\bsnm{{Anderson}}, \binits{D.R.}},
\bauthor{\bsnm{{Barstow}}, \binits{J.K.}},
\bauthor{\bsnm{{Bayliss}}, \binits{D.}},
\bauthor{\bsnm{{Bell}}, \binits{T.J.}},
\bauthor{\bsnm{{Blecic}}, \binits{J.}},
\bauthor{\bsnm{{Bryant}}, \binits{E.M.}},
\bauthor{\bsnm{{Burleigh}}, \binits{M.R.}},
\bauthor{\bsnm{{Carone}}, \binits{L.}},
\bauthor{\bsnm{{Casewell}}, \binits{S.L.}},
\bauthor{\bsnm{{Changeat}}, \binits{Q.}},
\bauthor{\bsnm{{Chubb}}, \binits{K.L.}},
\bauthor{\bsnm{{Crossfield}}, \binits{I.J.M.}},
\bauthor{\bsnm{{Crouzet}}, \binits{N.}},
\bauthor{\bsnm{{Decin}}, \binits{L.}},
\bauthor{\bsnm{{D{\'e}sert}}, \binits{J.-M.}},
\bauthor{\bsnm{{Feinstein}}, \binits{A.D.}},
\bauthor{\bsnm{{Flagg}}, \binits{L.}},
\bauthor{\bsnm{{Fortney}}, \binits{J.J.}},
\bauthor{\bsnm{{Gizis}}, \binits{J.E.}},
\bauthor{\bsnm{{Heng}}, \binits{K.}},
\bauthor{\bsnm{{Iro}}, \binits{N.}},
\bauthor{\bsnm{{Kempton}}, \binits{E.M.-R.}},
\bauthor{\bsnm{{Kendrew}}, \binits{S.}},
\bauthor{\bsnm{{Kirk}}, \binits{J.}},
\bauthor{\bsnm{{Knutson}}, \binits{H.A.}},
\bauthor{\bsnm{{Komacek}}, \binits{T.D.}},
\bauthor{\bsnm{{Lagage}}, \binits{P.-O.}},
\bauthor{\bsnm{{Leconte}}, \binits{J.}},
\bauthor{\bsnm{{Lustig-Yaeger}}, \binits{J.}},
\bauthor{\bsnm{{MacDonald}}, \binits{R.J.}},
\bauthor{\bsnm{{Mancini}}, \binits{L.}},
\bauthor{\bsnm{{May}}, \binits{E.M.}},
\bauthor{\bsnm{{Mayne}}, \binits{N.J.}},
\bauthor{\bsnm{{Miguel}}, \binits{Y.}},
\bauthor{\bsnm{{Mikal-Evans}}, \binits{T.}},
\bauthor{\bsnm{{Molaverdikhani}}, \binits{K.}},
\bauthor{\bsnm{{Palle}}, \binits{E.}},
\bauthor{\bsnm{{Piaulet}}, \binits{C.}},
\bauthor{\bsnm{{Rackham}}, \binits{B.V.}},
\bauthor{\bsnm{{Redfield}}, \binits{S.}},
\bauthor{\bsnm{{Rogers}}, \binits{L.K.}},
\bauthor{\bsnm{{Roy}}, \binits{P.-A.}},
\bauthor{\bsnm{{Rustamkulov}}, \binits{Z.}},
\bauthor{\bsnm{{Shkolnik}}, \binits{E.L.}},
\bauthor{\bsnm{{Sotzen}}, \binits{K.S.}},
\bauthor{\bsnm{{Taylor}}, \binits{J.}},
\bauthor{\bsnm{{Tremblin}}, \binits{P.}},
\bauthor{\bsnm{{Tucker}}, \binits{G.S.}},
\bauthor{\bsnm{{Turner}}, \binits{J.D.}},
\bauthor{\bsnm{{de Val-Borro}}, \binits{M.}},
\bauthor{\bsnm{{Venot}}, \binits{O.}},
\bauthor{\bsnm{{Zhang}}, \binits{X.}}:
\batitle{{Early Release Science of the exoplanet WASP-39b with JWST NIRCam}}.
\bjtitle{\nat}
\bvolume{614}(\bissue{7949}),
\bfpage{653}--\blpage{658}
(\byear{2023})
{\href{https://arxiv.org/abs/2211.10489}{{arXiv:2211.10489}}}
{[astro-ph.EP]}.
\doiurl{10.1038/s41586-022-05590-4}
\end{barticle}
\endbibitem

\bibitem{Alderson2022}
\begin{barticle}
\bauthor{\bsnm{{Alderson}}, \binits{L.}},
\bauthor{\bsnm{{Wakeford}}, \binits{H.R.}},
\bauthor{\bsnm{{Alam}}, \binits{M.K.}},
\bauthor{\bsnm{{Batalha}}, \binits{N.E.}},
\bauthor{\bsnm{{Lothringer}}, \binits{J.D.}},
\bauthor{\bsnm{{Adams Redai}}, \binits{J.}},
\bauthor{\bsnm{{Barat}}, \binits{S.}},
\bauthor{\bsnm{{Brande}}, \binits{J.}},
\bauthor{\bsnm{{Damiano}}, \binits{M.}},
\bauthor{\bsnm{{Daylan}}, \binits{T.}},
\bauthor{\bsnm{{Espinoza}}, \binits{N.}},
\bauthor{\bsnm{{Flagg}}, \binits{L.}},
\bauthor{\bsnm{{Goyal}}, \binits{J.M.}},
\bauthor{\bsnm{{Grant}}, \binits{D.}},
\bauthor{\bsnm{{Hu}}, \binits{R.}},
\bauthor{\bsnm{{Inglis}}, \binits{J.}},
\bauthor{\bsnm{{Lee}}, \binits{E.K.H.}},
\bauthor{\bsnm{{Mikal-Evans}}, \binits{T.}},
\bauthor{\bsnm{{Ramos-Rosado}}, \binits{L.}},
\bauthor{\bsnm{{Roy}}, \binits{P.-A.}},
\bauthor{\bsnm{{Wallack}}, \binits{N.L.}},
\bauthor{\bsnm{{Batalha}}, \binits{N.M.}},
\bauthor{\bsnm{{Bean}}, \binits{J.L.}},
\bauthor{\bsnm{{Benneke}}, \binits{B.}},
\bauthor{\bsnm{{Berta-Thompson}}, \binits{Z.K.}},
\bauthor{\bsnm{{Carter}}, \binits{A.L.}},
\bauthor{\bsnm{{Changeat}}, \binits{Q.}},
\bauthor{\bsnm{{Col{\'o}n}}, \binits{K.D.}},
\bauthor{\bsnm{{Crossfield}}, \binits{I.J.M.}},
\bauthor{\bsnm{{D{\'e}sert}}, \binits{J.-M.}},
\bauthor{\bsnm{{Foreman-Mackey}}, \binits{D.}},
\bauthor{\bsnm{{Gibson}}, \binits{N.P.}},
\bauthor{\bsnm{{Kreidberg}}, \binits{L.}},
\bauthor{\bsnm{{Line}}, \binits{M.R.}},
\bauthor{\bsnm{{L{\'o}pez-Morales}}, \binits{M.}},
\bauthor{\bsnm{{Molaverdikhani}}, \binits{K.}},
\bauthor{\bsnm{{Moran}}, \binits{S.E.}},
\bauthor{\bsnm{{Morello}}, \binits{G.}},
\bauthor{\bsnm{{Moses}}, \binits{J.I.}},
\bauthor{\bsnm{{Mukherjee}}, \binits{S.}},
\bauthor{\bsnm{{Schlawin}}, \binits{E.}},
\bauthor{\bsnm{{Sing}}, \binits{D.K.}},
\bauthor{\bsnm{{Stevenson}}, \binits{K.B.}},
\bauthor{\bsnm{{Taylor}}, \binits{J.}},
\bauthor{\bsnm{{Aggarwal}}, \binits{K.}},
\bauthor{\bsnm{{Ahrer}}, \binits{E.-M.}},
\bauthor{\bsnm{{Allen}}, \binits{N.H.}},
\bauthor{\bsnm{{Barstow}}, \binits{J.K.}},
\bauthor{\bsnm{{Bell}}, \binits{T.J.}},
\bauthor{\bsnm{{Blecic}}, \binits{J.}},
\bauthor{\bsnm{{Casewell}}, \binits{S.L.}},
\bauthor{\bsnm{{Chubb}}, \binits{K.L.}},
\bauthor{\bsnm{{Crouzet}}, \binits{N.}},
\bauthor{\bsnm{{Cubillos}}, \binits{P.E.}},
\bauthor{\bsnm{{Decin}}, \binits{L.}},
\bauthor{\bsnm{{Feinstein}}, \binits{A.D.}},
\bauthor{\bsnm{{Fortney}}, \binits{J.J.}},
\bauthor{\bsnm{{Harrington}}, \binits{J.}},
\bauthor{\bsnm{{Heng}}, \binits{K.}},
\bauthor{\bsnm{{Iro}}, \binits{N.}},
\bauthor{\bsnm{{Kempton}}, \binits{E.M.-R.}},
\bauthor{\bsnm{{Kirk}}, \binits{J.}},
\bauthor{\bsnm{{Knutson}}, \binits{H.A.}},
\bauthor{\bsnm{{Krick}}, \binits{J.}},
\bauthor{\bsnm{{Leconte}}, \binits{J.}},
\bauthor{\bsnm{{Lendl}}, \binits{M.}},
\bauthor{\bsnm{{MacDonald}}, \binits{R.J.}},
\bauthor{\bsnm{{Mancini}}, \binits{L.}},
\bauthor{\bsnm{{Mansfield}}, \binits{M.}},
\bauthor{\bsnm{{May}}, \binits{E.M.}},
\bauthor{\bsnm{{Mayne}}, \binits{N.J.}},
\bauthor{\bsnm{{Miguel}}, \binits{Y.}},
\bauthor{\bsnm{{Nikolov}}, \binits{N.K.}},
\bauthor{\bsnm{{Ohno}}, \binits{K.}},
\bauthor{\bsnm{{Palle}}, \binits{E.}},
\bauthor{\bsnm{{Parmentier}}, \binits{V.}},
\bauthor{\bsnm{{Petit dit de la Roche}}, \binits{D.J.M.}},
\bauthor{\bsnm{{Piaulet}}, \binits{C.}},
\bauthor{\bsnm{{Powell}}, \binits{D.}},
\bauthor{\bsnm{{Rackham}}, \binits{B.V.}},
\bauthor{\bsnm{{Redfield}}, \binits{S.}},
\bauthor{\bsnm{{Rogers}}, \binits{L.K.}},
\bauthor{\bsnm{{Rustamkulov}}, \binits{Z.}},
\bauthor{\bsnm{{Tan}}, \binits{X.}},
\bauthor{\bsnm{{Tremblin}}, \binits{P.}},
\bauthor{\bsnm{{Tsai}}, \binits{S.-M.}},
\bauthor{\bsnm{{Turner}}, \binits{J.D.}},
\bauthor{\bsnm{{de Val-Borro}}, \binits{M.}},
\bauthor{\bsnm{{Venot}}, \binits{O.}},
\bauthor{\bsnm{{Welbanks}}, \binits{L.}},
\bauthor{\bsnm{{Wheatley}}, \binits{P.J.}},
\bauthor{\bsnm{{Zhang}}, \binits{X.}}:
\batitle{{Early Release Science of the exoplanet WASP-39b with JWST NIRSpec
  G395H}}.
\bjtitle{\nat}
\bvolume{614}(\bissue{7949}),
\bfpage{664}--\blpage{669}
(\byear{2023})
{\href{https://arxiv.org/abs/2211.10488}{{arXiv:2211.10488}}}
{[astro-ph.EP]}.
\doiurl{10.1038/s41586-022-05591-3}
\end{barticle}
\endbibitem

\bibitem{Rustamkulov2022}
\begin{barticle}
\bauthor{\bsnm{{Rustamkulov}}, \binits{Z.}},
\bauthor{\bsnm{{Sing}}, \binits{D.K.}},
\bauthor{\bsnm{{Mukherjee}}, \binits{S.}},
\bauthor{\bsnm{{May}}, \binits{E.M.}},
\bauthor{\bsnm{{Kirk}}, \binits{J.}},
\bauthor{\bsnm{{Schlawin}}, \binits{E.}},
\bauthor{\bsnm{{Line}}, \binits{M.R.}},
\bauthor{\bsnm{{Piaulet}}, \binits{C.}},
\bauthor{\bsnm{{Carter}}, \binits{A.L.}},
\bauthor{\bsnm{{Batalha}}, \binits{N.E.}},
\bauthor{\bsnm{{Goyal}}, \binits{J.M.}},
\bauthor{\bsnm{{L{\'o}pez-Morales}}, \binits{M.}},
\bauthor{\bsnm{{Lothringer}}, \binits{J.D.}},
\bauthor{\bsnm{{MacDonald}}, \binits{R.J.}},
\bauthor{\bsnm{{Moran}}, \binits{S.E.}},
\bauthor{\bsnm{{Stevenson}}, \binits{K.B.}},
\bauthor{\bsnm{{Wakeford}}, \binits{H.R.}},
\bauthor{\bsnm{{Espinoza}}, \binits{N.}},
\bauthor{\bsnm{{Bean}}, \binits{J.L.}},
\bauthor{\bsnm{{Batalha}}, \binits{N.M.}},
\bauthor{\bsnm{{Benneke}}, \binits{B.}},
\bauthor{\bsnm{{Berta-Thompson}}, \binits{Z.K.}},
\bauthor{\bsnm{{Crossfield}}, \binits{I.J.M.}},
\bauthor{\bsnm{{Gao}}, \binits{P.}},
\bauthor{\bsnm{{Kreidberg}}, \binits{L.}},
\bauthor{\bsnm{{Powell}}, \binits{D.K.}},
\bauthor{\bsnm{{Cubillos}}, \binits{P.E.}},
\bauthor{\bsnm{{Gibson}}, \binits{N.P.}},
\bauthor{\bsnm{{Leconte}}, \binits{J.}},
\bauthor{\bsnm{{Molaverdikhani}}, \binits{K.}},
\bauthor{\bsnm{{Nikolov}}, \binits{N.K.}},
\bauthor{\bsnm{{Parmentier}}, \binits{V.}},
\bauthor{\bsnm{{Roy}}, \binits{P.}},
\bauthor{\bsnm{{Taylor}}, \binits{J.}},
\bauthor{\bsnm{{Turner}}, \binits{J.D.}},
\bauthor{\bsnm{{Wheatley}}, \binits{P.J.}},
\bauthor{\bsnm{{Aggarwal}}, \binits{K.}},
\bauthor{\bsnm{{Ahrer}}, \binits{E.}},
\bauthor{\bsnm{{Alam}}, \binits{M.K.}},
\bauthor{\bsnm{{Alderson}}, \binits{L.}},
\bauthor{\bsnm{{Allen}}, \binits{N.H.}},
\bauthor{\bsnm{{Banerjee}}, \binits{A.}},
\bauthor{\bsnm{{Barat}}, \binits{S.}},
\bauthor{\bsnm{{Barrado}}, \binits{D.}},
\bauthor{\bsnm{{Barstow}}, \binits{J.K.}},
\bauthor{\bsnm{{Bell}}, \binits{T.J.}},
\bauthor{\bsnm{{Blecic}}, \binits{J.}},
\bauthor{\bsnm{{Brande}}, \binits{J.}},
\bauthor{\bsnm{{Casewell}}, \binits{S.}},
\bauthor{\bsnm{{Changeat}}, \binits{Q.}},
\bauthor{\bsnm{{Chubb}}, \binits{K.L.}},
\bauthor{\bsnm{{Crouzet}}, \binits{N.}},
\bauthor{\bsnm{{Daylan}}, \binits{T.}},
\bauthor{\bsnm{{Decin}}, \binits{L.}},
\bauthor{\bsnm{{D{\'e}sert}}, \binits{J.}},
\bauthor{\bsnm{{Mikal-Evans}}, \binits{T.}},
\bauthor{\bsnm{{Feinstein}}, \binits{A.D.}},
\bauthor{\bsnm{{Flagg}}, \binits{L.}},
\bauthor{\bsnm{{Fortney}}, \binits{J.J.}},
\bauthor{\bsnm{{Harrington}}, \binits{J.}},
\bauthor{\bsnm{{Heng}}, \binits{K.}},
\bauthor{\bsnm{{Hong}}, \binits{Y.}},
\bauthor{\bsnm{{Hu}}, \binits{R.}},
\bauthor{\bsnm{{Iro}}, \binits{N.}},
\bauthor{\bsnm{{Kataria}}, \binits{T.}},
\bauthor{\bsnm{{Kempton}}, \binits{E.M.-R.}},
\bauthor{\bsnm{{Krick}}, \binits{J.}},
\bauthor{\bsnm{{Lendl}}, \binits{M.}},
\bauthor{\bsnm{{Lillo-Box}}, \binits{J.}},
\bauthor{\bsnm{{Louca}}, \binits{A.}},
\bauthor{\bsnm{{Lustig-Yaeger}}, \binits{J.}},
\bauthor{\bsnm{{Mancini}}, \binits{L.}},
\bauthor{\bsnm{{Mansfield}}, \binits{M.}},
\bauthor{\bsnm{{Mayne}}, \binits{N.J.}},
\bauthor{\bsnm{{Miguel}}, \binits{Y.}},
\bauthor{\bsnm{{Morello}}, \binits{G.}},
\bauthor{\bsnm{{Ohno}}, \binits{K.}},
\bauthor{\bsnm{{Palle}}, \binits{E.}},
\bauthor{\bsnm{{Petit dit de la Roche}}, \binits{D.J.M.}},
\bauthor{\bsnm{{Rackham}}, \binits{B.V.}},
\bauthor{\bsnm{{Radica}}, \binits{M.}},
\bauthor{\bsnm{{Ramos-Rosado}}, \binits{L.}},
\bauthor{\bsnm{{Redfield}}, \binits{S.}},
\bauthor{\bsnm{{Rogers}}, \binits{L.K.}},
\bauthor{\bsnm{{Shkolnik}}, \binits{E.L.}},
\bauthor{\bsnm{{Southworth}}, \binits{J.}},
\bauthor{\bsnm{{Teske}}, \binits{J.}},
\bauthor{\bsnm{{Tremblin}}, \binits{P.}},
\bauthor{\bsnm{{Tucker}}, \binits{G.S.}},
\bauthor{\bsnm{{Venot}}, \binits{O.}},
\bauthor{\bsnm{{Waalkes}}, \binits{W.C.}},
\bauthor{\bsnm{{Welbanks}}, \binits{L.}},
\bauthor{\bsnm{{Zhang}}, \binits{X.}},
\bauthor{\bsnm{{Zieba}}, \binits{S.}}:
\batitle{{Early Release Science of the exoplanet WASP-39b with JWST NIRSpec
  PRISM}}.
\bjtitle{\nat}
\bvolume{614}(\bissue{7949}),
\bfpage{659}--\blpage{663}
(\byear{2023})
{\href{https://arxiv.org/abs/2211.10487}{{arXiv:2211.10487}}}
{[astro-ph.EP]}.
\doiurl{10.1038/s41586-022-05677-y}
\end{barticle}
\endbibitem

\bibitem{LustigYaeger2023}
\begin{botherref}
\oauthor{\bsnm{{Lustig-Yaeger}}, \binits{J.}},
\oauthor{\bsnm{{Fu}}, \binits{G.}},
\oauthor{\bsnm{{May}}, \binits{E.M.}},
\oauthor{\bsnm{{Ortiz Ceballos}}, \binits{K.N.}},
\oauthor{\bsnm{{Moran}}, \binits{S.E.}},
\oauthor{\bsnm{{Peacock}}, \binits{S.}},
\oauthor{\bsnm{{Stevenson}}, \binits{K.B.}},
\oauthor{\bsnm{{L{\'o}pez-Morales}}, \binits{M.}},
\oauthor{\bsnm{{MacDonald}}, \binits{R.J.}},
\oauthor{\bsnm{{Mayorga}}, \binits{L.C.}},
\oauthor{\bsnm{{Sing}}, \binits{D.K.}},
\oauthor{\bsnm{{Sotzen}}, \binits{K.S.}},
\oauthor{\bsnm{{Valenti}}, \binits{J.A.}},
\oauthor{\bsnm{{Adams}}, \binits{J.}},
\oauthor{\bsnm{{Alam}}, \binits{M.K.}},
\oauthor{\bsnm{{Batalha}}, \binits{N.E.}},
\oauthor{\bsnm{{Bennett}}, \binits{K.A.}},
\oauthor{\bsnm{{Gonzalez-Quiles}}, \binits{J.}},
\oauthor{\bsnm{{Kirk}}, \binits{J.}},
\oauthor{\bsnm{{Kruse}}, \binits{E.}},
\oauthor{\bsnm{{Lothringer}}, \binits{J.D.}},
\oauthor{\bsnm{{Rustamkulov}}, \binits{Z.}},
\oauthor{\bsnm{{Wakeford}}, \binits{H.R.}}:
{A JWST transmission spectrum of a nearby Earth-sized exoplanet}.
arXiv e-prints,
2301--04191
(2023)
{\href{https://arxiv.org/abs/2301.04191}{{arXiv:2301.04191}}}
{[astro-ph.EP]}.
\doiurl{10.48550/arXiv.2301.04191}
\end{botherref}
\endbibitem

\bibitem{Bushouse2022}
\begin{botherref}
\oauthor{\bsnm{{Bushouse}}, \binits{H.}},
\oauthor{\bsnm{{Eisenhamer}}, \binits{J.}},
\oauthor{\bsnm{{Dencheva}}, \binits{N.}},
\oauthor{\bsnm{{Davies}}, \binits{J.}},
\oauthor{\bsnm{{Greenfield}}, \binits{P.}},
\oauthor{\bsnm{{Morrison}}, \binits{J.}},
\oauthor{\bsnm{{Hodge}}, \binits{P.}},
\oauthor{\bsnm{{Simon}}, \binits{B.}},
\oauthor{\bsnm{{Grumm}}, \binits{D.}},
\oauthor{\bsnm{{Droettboom}}, \binits{M.}},
\oauthor{\bsnm{{Slavich}}, \binits{E.}},
\oauthor{\bsnm{{Sosey}}, \binits{M.}},
\oauthor{\bsnm{{Pauly}}, \binits{T.}},
\oauthor{\bsnm{{Miller}}, \binits{T.}},
\oauthor{\bsnm{{Jedrzejewski}}, \binits{R.}},
\oauthor{\bsnm{{Hack}}, \binits{W.}},
\oauthor{\bsnm{{Davis}}, \binits{D.}},
\oauthor{\bsnm{{Crawford}}, \binits{S.}},
\oauthor{\bsnm{{Law}}, \binits{D.}},
\oauthor{\bsnm{{Gordon}}, \binits{K.}},
\oauthor{\bsnm{{Regan}}, \binits{M.}},
\oauthor{\bsnm{{Cara}}, \binits{M.}},
\oauthor{\bsnm{{MacDonald}}, \binits{K.}},
\oauthor{\bsnm{{Bradley}}, \binits{L.}},
\oauthor{\bsnm{{Shanahan}}, \binits{C.}},
\oauthor{\bsnm{{Jamieson}}, \binits{W.}},
\oauthor{\bsnm{{Teodoro}}, \binits{M.}},
\oauthor{\bsnm{{Williams}}, \binits{T.}}:
{JWST Calibration Pipeline}.
Zenodo
(2022).
\doiurl{10.5281/zenodo.7325378}
\end{botherref}
\endbibitem

\bibitem{Stetson1987}
\begin{barticle}
\bauthor{\bsnm{{Stetson}}, \binits{P.B.}}:
\batitle{{DAOPHOT: A Computer Program for Crowded-Field Stellar Photometry}}.
\bjtitle{\pasp}
\bvolume{99},
\bfpage{191}
(\byear{1987}).
\doiurl{10.1086/131977}
\end{barticle}
\endbibitem

\bibitem{Deming2015}
\begin{barticle}
\bauthor{\bsnm{{Deming}}, \binits{D.}},
\bauthor{\bsnm{{Knutson}}, \binits{H.}},
\bauthor{\bsnm{{Kammer}}, \binits{J.}},
\bauthor{\bsnm{{Fulton}}, \binits{B.J.}},
\bauthor{\bsnm{{Ingalls}}, \binits{J.}},
\bauthor{\bsnm{{Carey}}, \binits{S.}},
\bauthor{\bsnm{{Burrows}}, \binits{A.}},
\bauthor{\bsnm{{Fortney}}, \binits{J.J.}},
\bauthor{\bsnm{{Todorov}}, \binits{K.}},
\bauthor{\bsnm{{Agol}}, \binits{E.}},
\bauthor{\bsnm{{Cowan}}, \binits{N.}},
\bauthor{\bsnm{{Desert}}, \binits{J.-M.}},
\bauthor{\bsnm{{Fraine}}, \binits{J.}},
\bauthor{\bsnm{{Langton}}, \binits{J.}},
\bauthor{\bsnm{{Morley}}, \binits{C.}},
\bauthor{\bsnm{{Showman}}, \binits{A.P.}}:
\batitle{{Spitzer Secondary Eclipses of the Dense, Modestly-irradiated, Giant
  Exoplanet HAT-P-20b Using Pixel-level Decorrelation}}.
\bjtitle{\apj}
\bvolume{805}(\bissue{2}),
\bfpage{132}
(\byear{2015})
{\href{https://arxiv.org/abs/1411.7404}{{arXiv:1411.7404}}}
{[astro-ph.EP]}.
\doiurl{10.1088/0004-637X/805/2/132}
\end{barticle}
\endbibitem

\bibitem{ForemanMackey2013}
\begin{barticle}
\bauthor{\bsnm{{Foreman-Mackey}}, \binits{D.}},
\bauthor{\bsnm{{Hogg}}, \binits{D.W.}},
\bauthor{\bsnm{{Lang}}, \binits{D.}},
\bauthor{\bsnm{{Goodman}}, \binits{J.}}:
\batitle{{emcee: The MCMC Hammer}}.
\bjtitle{\pasp}
\bvolume{125}(\bissue{925}),
\bfpage{306}
(\byear{2013})
{\href{https://arxiv.org/abs/1202.3665}{{arXiv:1202.3665}}}
{[astro-ph.IM]}.
\doiurl{10.1086/670067}
\end{barticle}
\endbibitem

\bibitem{Kreidberg2015}
\begin{barticle}
\bauthor{\bsnm{{Kreidberg}}, \binits{L.}}:
\batitle{{batman: BAsic Transit Model cAlculatioN in Python}}.
\bjtitle{\pasp}
\bvolume{127}(\bissue{957}),
\bfpage{1161}
(\byear{2015})
{\href{https://arxiv.org/abs/1507.08285}{{arXiv:1507.08285}}}
{[astro-ph.EP]}.
\doiurl{10.1086/683602}
\end{barticle}
\endbibitem

\bibitem{Schwarz1978}
\begin{barticle}
\bauthor{\bsnm{{Schwarz}}, \binits{G.}}:
\batitle{{Estimating the Dimension of a Model}}.
\bjtitle{Annals of Statistics}
\bvolume{6}(\bissue{2}),
\bfpage{461}--\blpage{464}
(\byear{1978})
\end{barticle}
\endbibitem

\bibitem{Kass1995}
\begin{barticle}
\bauthor{\bsnm{Kass}, \binits{R.E.}},
\bauthor{\bsnm{Raftery}, \binits{A.E.}}:
\batitle{Bayes factors}.
\bjtitle{Journal of the American Statistical Association}
\bvolume{90},
\bfpage{773}--\blpage{795}
(\byear{1995})
\end{barticle}
\endbibitem

\bibitem{Liddle2007}
\begin{barticle}
\bauthor{\bsnm{{Liddle}}, \binits{A.R.}}:
\batitle{{Information criteria for astrophysical model selection}}.
\bjtitle{\mnras}
\bvolume{377}(\bissue{1}),
\bfpage{74}--\blpage{78}
(\byear{2007})
{\href{https://arxiv.org/abs/astro-ph/0701113}{{arXiv:astro-ph/0701113}}}
{[astro-ph]}.
\doiurl{10.1111/j.1745-3933.2007.00306.x}
\end{barticle}
\endbibitem

\bibitem{Agol2021}
\begin{barticle}
\bauthor{\bsnm{{Agol}}, \binits{E.}},
\bauthor{\bsnm{{Dorn}}, \binits{C.}},
\bauthor{\bsnm{{Grimm}}, \binits{S.L.}},
\bauthor{\bsnm{{Turbet}}, \binits{M.}},
\bauthor{\bsnm{{Ducrot}}, \binits{E.}},
\bauthor{\bsnm{{Delrez}}, \binits{L.}},
\bauthor{\bsnm{{Gillon}}, \binits{M.}},
\bauthor{\bsnm{{Demory}}, \binits{B.-O.}},
\bauthor{\bsnm{{Burdanov}}, \binits{A.}},
\bauthor{\bsnm{{Barkaoui}}, \binits{K.}},
\bauthor{\bsnm{{Benkhaldoun}}, \binits{Z.}},
\bauthor{\bsnm{{Bolmont}}, \binits{E.}},
\bauthor{\bsnm{{Burgasser}}, \binits{A.}},
\bauthor{\bsnm{{Carey}}, \binits{S.}},
\bauthor{\bsnm{{de Wit}}, \binits{J.}},
\bauthor{\bsnm{{Fabrycky}}, \binits{D.}},
\bauthor{\bsnm{{Foreman-Mackey}}, \binits{D.}},
\bauthor{\bsnm{{Haldemann}}, \binits{J.}},
\bauthor{\bsnm{{Hernandez}}, \binits{D.M.}},
\bauthor{\bsnm{{Ingalls}}, \binits{J.}},
\bauthor{\bsnm{{Jehin}}, \binits{E.}},
\bauthor{\bsnm{{Langford}}, \binits{Z.}},
\bauthor{\bsnm{{Leconte}}, \binits{J.}},
\bauthor{\bsnm{{Lederer}}, \binits{S.M.}},
\bauthor{\bsnm{{Luger}}, \binits{R.}},
\bauthor{\bsnm{{Malhotra}}, \binits{R.}},
\bauthor{\bsnm{{Meadows}}, \binits{V.S.}},
\bauthor{\bsnm{{Morris}}, \binits{B.M.}},
\bauthor{\bsnm{{Pozuelos}}, \binits{F.J.}},
\bauthor{\bsnm{{Queloz}}, \binits{D.}},
\bauthor{\bsnm{{Raymond}}, \binits{S.N.}},
\bauthor{\bsnm{{Selsis}}, \binits{F.}},
\bauthor{\bsnm{{Sestovic}}, \binits{M.}},
\bauthor{\bsnm{{Triaud}}, \binits{A.H.M.J.}},
\bauthor{\bsnm{{Van Grootel}}, \binits{V.}}:
\batitle{{Refining the Transit-timing and Photometric Analysis of TRAPPIST-1:
  Masses, Radii, Densities, Dynamics, and Ephemerides}}.
\bjtitle{\psj}
\bvolume{2}(\bissue{1}),
\bfpage{1}
(\byear{2021})
{\href{https://arxiv.org/abs/2010.01074}{{arXiv:2010.01074}}}
{[astro-ph.EP]}.
\doiurl{10.3847/PSJ/abd022}
\end{barticle}
\endbibitem

\bibitem{Gillon2010}
\begin{barticle}
\bauthor{\bsnm{{Gillon}}, \binits{M.}},
\bauthor{\bsnm{{Deming}}, \binits{D.}},
\bauthor{\bsnm{{Demory}}, \binits{B.-O.}},
\bauthor{\bsnm{{Lovis}}, \binits{C.}},
\bauthor{\bsnm{{Seager}}, \binits{S.}},
\bauthor{\bsnm{{Mayor}}, \binits{M.}},
\bauthor{\bsnm{{Pepe}}, \binits{F.}},
\bauthor{\bsnm{{Queloz}}, \binits{D.}},
\bauthor{\bsnm{{Segransan}}, \binits{D.}},
\bauthor{\bsnm{{Udry}}, \binits{S.}},
\bauthor{\bsnm{{Delmelle}}, \binits{S.}},
\bauthor{\bsnm{{Magain}}, \binits{P.}}:
\batitle{{The Spitzer search for the transits of HARPS low-mass planets. I. No
  transit for the super-Earth HD 40307b}}.
\bjtitle{\aap}
\bvolume{518},
\bfpage{25}
(\byear{2010})
{\href{https://arxiv.org/abs/1002.4707}{{arXiv:1002.4707}}}
{[astro-ph.EP]}.
\doiurl{10.1051/0004-6361/201014144}
\end{barticle}
\endbibitem

\bibitem{Gillon2012}
\begin{barticle}
\bauthor{\bsnm{{Gillon}}, \binits{M.}},
\bauthor{\bsnm{{Triaud}}, \binits{A.H.M.J.}},
\bauthor{\bsnm{{Fortney}}, \binits{J.J.}},
\bauthor{\bsnm{{Demory}}, \binits{B.-O.}},
\bauthor{\bsnm{{Jehin}}, \binits{E.}},
\bauthor{\bsnm{{Lendl}}, \binits{M.}},
\bauthor{\bsnm{{Magain}}, \binits{P.}},
\bauthor{\bsnm{{Kabath}}, \binits{P.}},
\bauthor{\bsnm{{Queloz}}, \binits{D.}},
\bauthor{\bsnm{{Alonso}}, \binits{R.}},
\bauthor{\bsnm{{Anderson}}, \binits{D.R.}},
\bauthor{\bsnm{{Collier Cameron}}, \binits{A.}},
\bauthor{\bsnm{{Fumel}}, \binits{A.}},
\bauthor{\bsnm{{Hebb}}, \binits{L.}},
\bauthor{\bsnm{{Hellier}}, \binits{C.}},
\bauthor{\bsnm{{Lanotte}}, \binits{A.}},
\bauthor{\bsnm{{Maxted}}, \binits{P.F.L.}},
\bauthor{\bsnm{{Mowlavi}}, \binits{N.}},
\bauthor{\bsnm{{Smalley}}, \binits{B.}}:
\batitle{{The TRAPPIST survey of southern transiting planets. I. Thirty
  eclipses of the ultra-short period planet WASP-43 b}}.
\bjtitle{\aap}
\bvolume{542},
\bfpage{4}
(\byear{2012})
{\href{https://arxiv.org/abs/1201.2789}{{arXiv:1201.2789}}}
{[astro-ph.EP]}.
\doiurl{10.1051/0004-6361/201218817}
\end{barticle}
\endbibitem

\bibitem{Gillon2014}
\begin{barticle}
\bauthor{\bsnm{{Gillon}}, \binits{M.}},
\bauthor{\bsnm{{Demory}}, \binits{B.-O.}},
\bauthor{\bsnm{{Madhusudhan}}, \binits{N.}},
\bauthor{\bsnm{{Deming}}, \binits{D.}},
\bauthor{\bsnm{{Seager}}, \binits{S.}},
\bauthor{\bsnm{{Zsom}}, \binits{A.}},
\bauthor{\bsnm{{Knutson}}, \binits{H.A.}},
\bauthor{\bsnm{{Lanotte}}, \binits{A.A.}},
\bauthor{\bsnm{{Bonfils}}, \binits{X.}},
\bauthor{\bsnm{{D{\'e}sert}}, \binits{J.-M.}},
\bauthor{\bsnm{{Delrez}}, \binits{L.}},
\bauthor{\bsnm{{Jehin}}, \binits{E.}},
\bauthor{\bsnm{{Fraine}}, \binits{J.D.}},
\bauthor{\bsnm{{Magain}}, \binits{P.}},
\bauthor{\bsnm{{Triaud}}, \binits{A.H.M.J.}}:
\batitle{{Search for a habitable terrestrial planet transiting the nearby red
  dwarf GJ 1214}}.
\bjtitle{\aap}
\bvolume{563},
\bfpage{21}
(\byear{2014})
{\href{https://arxiv.org/abs/1307.6722}{{arXiv:1307.6722}}}
{[astro-ph.EP]}.
\doiurl{10.1051/0004-6361/201322362}
\end{barticle}
\endbibitem

\bibitem{Mandel2002}
\begin{barticle}
\bauthor{\bsnm{{Mandel}}, \binits{K.}},
\bauthor{\bsnm{{Agol}}, \binits{E.}}:
\batitle{{Analytic Light Curves for Planetary Transit Searches}}.
\bjtitle{\apjl}
\bvolume{580}(\bissue{2}),
\bfpage{171}--\blpage{175}
(\byear{2002})
{\href{https://arxiv.org/abs/astro-ph/0210099}{{arXiv:astro-ph/0210099}}}
{[astro-ph]}.
\doiurl{10.1086/345520}
\end{barticle}
\endbibitem

\bibitem{GelmanRubin1992}
\begin{barticle}
\bauthor{\bsnm{{Gelman}}, \binits{A.}},
\bauthor{\bsnm{{Rubin}}, \binits{D.B.}}:
\batitle{{Inference from Iterative Simulation Using Multiple Sequences}}.
\bjtitle{Statistical Science}
\bvolume{7},
\bfpage{457}--\blpage{472}
(\byear{1992}).
\doiurl{10.1214/ss/1177011136}
\end{barticle}
\endbibitem

\bibitem{Ducrot2020}
\begin{barticle}
\bauthor{\bsnm{{Ducrot}}, \binits{E.}},
\bauthor{\bsnm{{Gillon}}, \binits{M.}},
\bauthor{\bsnm{{Delrez}}, \binits{L.}},
\bauthor{\bsnm{{Agol}}, \binits{E.}},
\bauthor{\bsnm{{Rimmer}}, \binits{P.}},
\bauthor{\bsnm{{Turbet}}, \binits{M.}},
\bauthor{\bsnm{{G{\"u}nther}}, \binits{M.N.}},
\bauthor{\bsnm{{Demory}}, \binits{B.-O.}},
\bauthor{\bsnm{{Triaud}}, \binits{A.H.M.J.}},
\bauthor{\bsnm{{Bolmont}}, \binits{E.}},
\bauthor{\bsnm{{Burgasser}}, \binits{A.}},
\bauthor{\bsnm{{Carey}}, \binits{S.J.}},
\bauthor{\bsnm{{Ingalls}}, \binits{J.G.}},
\bauthor{\bsnm{{Jehin}}, \binits{E.}},
\bauthor{\bsnm{{Leconte}}, \binits{J.}},
\bauthor{\bsnm{{Lederer}}, \binits{S.M.}},
\bauthor{\bsnm{{Queloz}}, \binits{D.}},
\bauthor{\bsnm{{Raymond}}, \binits{S.N.}},
\bauthor{\bsnm{{Selsis}}, \binits{F.}},
\bauthor{\bsnm{{Van Grootel}}, \binits{V.}},
\bauthor{\bsnm{{de Wit}}, \binits{J.}}:
\batitle{{TRAPPIST-1: Global results of the Spitzer Exploration Science Program
  Red Worlds}}.
\bjtitle{\aap}
\bvolume{640},
\bfpage{112}
(\byear{2020})
{\href{https://arxiv.org/abs/2006.13826}{{arXiv:2006.13826}}}
{[astro-ph.EP]}.
\doiurl{10.1051/0004-6361/201937392}
\end{barticle}
\endbibitem

\bibitem{Ingalls2012}
\begin{bchapter}
\bauthor{\bsnm{{Ingalls}}, \binits{J.G.}},
\bauthor{\bsnm{{Krick}}, \binits{J.E.}},
\bauthor{\bsnm{{Carey}}, \binits{S.J.}},
\bauthor{\bsnm{{Laine}}, \binits{S.}},
\bauthor{\bsnm{{Surace}}, \binits{J.A.}},
\bauthor{\bsnm{{Glaccum}}, \binits{W.J.}},
\bauthor{\bsnm{{Grillmair}}, \binits{C.C.}},
\bauthor{\bsnm{{Lowrance}}, \binits{P.J.}}:
\bctitle{{Intra-pixel gain variations and high-precision photometry with the
  Infrared Array Camera (IRAC)}}.
In: \beditor{\bsnm{{Clampin}}, \binits{M.C.}},
\beditor{\bsnm{{Fazio}}, \binits{G.G.}},
\beditor{\bsnm{{MacEwen}}, \binits{H.A.}},
\beditor{\bsnm{{Oschmann}}, \binits{J.} \bsuffix{Jacobus~M.}} (eds.)
\bbtitle{Space Telescopes and Instrumentation 2012: Optical, Infrared, and
  Millimeter Wave}.
\bsertitle{Society of Photo-Optical Instrumentation Engineers (SPIE) Conference
  Series},
vol. \bseriesno{8442},
p. \bfpage{84421}
(\byear{2012}).
\doiurl{10.1117/12.926947}
\end{bchapter}
\endbibitem

\bibitem{hapke2002}
\begin{barticle}
\bauthor{\bsnm{{Hapke}}, \binits{B.}}:
\batitle{{Bidirectional Reflectance Spectroscopy. 5. The Coherent Backscatter
  Opposition Effect and Anisotropic Scattering}}.
\bjtitle{\icarus}
\bvolume{157}(\bissue{2}),
\bfpage{523}--\blpage{534}
(\byear{2002}).
\doiurl{10.1006/icar.2002.6853}
\end{barticle}
\endbibitem

\bibitem{mansfield2019}
\begin{barticle}
\bauthor{\bsnm{{Mansfield}}, \binits{M.}},
\bauthor{\bsnm{{Kite}}, \binits{E.S.}},
\bauthor{\bsnm{{Hu}}, \binits{R.}},
\bauthor{\bsnm{{Koll}}, \binits{D.D.B.}},
\bauthor{\bsnm{{Malik}}, \binits{M.}},
\bauthor{\bsnm{{Bean}}, \binits{J.L.}},
\bauthor{\bsnm{{Kempton}}, \binits{E.M.-R.}}:
\batitle{{Identifying Atmospheres on Rocky Exoplanets through Inferred High
  Albedo}}.
\bjtitle{\apj}
\bvolume{886}(\bissue{2}),
\bfpage{141}
(\byear{2019})
{\href{https://arxiv.org/abs/1907.13150}{{arXiv:1907.13150}}}
{[astro-ph.EP]}.
\doiurl{10.3847/1538-4357/ab4c90}
\end{barticle}
\endbibitem

\bibitem{hapke1977}
\begin{barticle}
\bauthor{\bsnm{{Hapke}}, \binits{B.}}:
\batitle{{Interpretations of optical observations of Mercury and the moon}}.
\bjtitle{Physics of the Earth and Planetary Interiors}
\bvolume{15}(\bissue{2-3}),
\bfpage{264}--\blpage{274}
(\byear{1977}).
\doiurl{10.1016/0031-9201(77)90035-8}
\end{barticle}
\endbibitem

\bibitem{keller2017}
\begin{bchapter}
\bauthor{\bsnm{Keller}, \binits{L.P.}},
\bauthor{\bsnm{Berger}, \binits{E.L.}}:
\bctitle{Space {{Weathering Rates}} in {{Lunar}} and {{Itokawa Samples}}}.
In: \bbtitle{Asteroids, {{Comets}}, {{Meteors}} ({{ACM}}) 2017 {{Meeting}}},
\bconflocation{{Montevideo}}
(\byear{2017})
\end{bchapter}
\endbibitem

\bibitem{johnstone2015}
\begin{barticle}
\bauthor{\bsnm{{Johnstone}}, \binits{C.P.}},
\bauthor{\bsnm{{G{\"u}del}}, \binits{M.}},
\bauthor{\bsnm{{Brott}}, \binits{I.}},
\bauthor{\bsnm{{L{\"u}ftinger}}, \binits{T.}}:
\batitle{{Stellar winds on the main-sequence. II. The evolution of rotation and
  winds}}.
\bjtitle{\aap}
\bvolume{577},
\bfpage{28}
(\byear{2015})
{\href{https://arxiv.org/abs/1503.07494}{{arXiv:1503.07494}}}
{[astro-ph.SR]}.
\doiurl{10.1051/0004-6361/201425301}
\end{barticle}
\endbibitem

\bibitem{hapke2012}
\begin{bbook}
\bauthor{\bsnm{Hapke}, \binits{B.}}:
\bbtitle{Theory of Reflectance and Emittance Spectroscopy},
\bedition{2}nd edn.
\bpublisher{{Cambridge University Press}},
\blocation{{Cambridge}}
(\byear{2012}).
\doiurl{10.1017/CBO9781139025683}
\end{bbook}
\endbibitem

\bibitem{rii}
\begin{botherref}
\oauthor{\bsnm{Polyanskiy}, \binits{M.N.}}:
Refractive index database.
\url{https://refractiveindex.info}.
Accessed on 2022-06-23
\end{botherref}
\endbibitem

\bibitem{Wordsworth2015}
\begin{barticle}
\bauthor{\bsnm{{Wordsworth}}, \binits{R.}}:
\batitle{{Atmospheric Heat Redistribution and Collapse on Tidally Locked Rocky
  Planets}}.
\bjtitle{\apj}
\bvolume{806}(\bissue{2}),
\bfpage{180}
(\byear{2015})
{\href{https://arxiv.org/abs/1412.5575}{{arXiv:1412.5575}}}
{[astro-ph.EP]}.
\doiurl{10.1088/0004-637X/806/2/180}
\end{barticle}
\endbibitem

\bibitem{Stamnes1988}
\begin{barticle}
\bauthor{\bsnm{{Stamnes}}, \binits{K.}},
\bauthor{\bsnm{{Tsay}}, \binits{S.-C.}},
\bauthor{\bsnm{{Jayaweera}}, \binits{K.}},
\bauthor{\bsnm{{Wiscombe}}, \binits{W.}}:
\batitle{{Numerically stable algorithm for discrete-ordinate-method radiative
  transfer in multiple scattering and emitting layered media}}.
\bjtitle{\ao}
\bvolume{27},
\bfpage{2502}--\blpage{2509}
(\byear{1988}).
\doiurl{10.1364/AO.27.002502}
\end{barticle}
\endbibitem

\bibitem{Stamnes2000}
\begin{botherref}
\oauthor{\bsnm{{Stamnes}}, \binits{K.}},
\oauthor{\bsnm{{Tsay}}, \binits{S.C.}},
\oauthor{\bsnm{{Wiscombe}}, \binits{W.}},
\oauthor{\bsnm{{Laszlo}}, \binits{I.}}:
{DISORT, a general-purpose Fortran program for discrete-ordinate-method
  radiative transfer in scattering and emitting layered media: documentation of
  methodology.}
\url{ftp://climate.gsfc.nasa.gov/pub/wiscombe/MultipleScatt/}
(2000)
\end{botherref}
\endbibitem

\bibitem{Robinson2018}
\begin{barticle}
\bauthor{\bsnm{{Robinson}}, \binits{T.D.}},
\bauthor{\bsnm{{Crisp}}, \binits{D.}}:
\batitle{{Linearized Flux Evolution (LiFE): A technique for rapidly adapting
  fluxes from full-physics radiative transfer models}}.
\bjtitle{\jqsrt}
\bvolume{211},
\bfpage{78}--\blpage{95}
(\byear{2018})
{\href{https://arxiv.org/abs/1803.02378}{{arXiv:1803.02378}}}
{[astro-ph.EP]}.
\doiurl{10.1016/j.jqsrt.2018.03.002}
\end{barticle}
\endbibitem

\bibitem{LincowskiPhD}
\begin{botherref}
\oauthor{\bsnm{{Lincowski}}, \binits{A.P.}}:
{The nature and characterization of M dwarf terrestrial planetary atmospheres:
  A theoretical case study of the TRAPPIST-1 planetary system}.
PhD thesis,
University of Washington, Seattle
(January 2020)
\end{botherref}
\endbibitem

\bibitem{Meadows1996}
\begin{barticle}
\bauthor{\bsnm{{Meadows}}, \binits{V.S.}},
\bauthor{\bsnm{{Crisp}}, \binits{D.}}:
\batitle{{Ground-based near-infrared observations of the Venus nightside: The
  thermal structure and water abundance near the surface}}.
\bjtitle{\jgr}
\bvolume{101}(\bissue{E2}),
\bfpage{4595}--\blpage{4622}
(\byear{1996}).
\doiurl{10.1029/95JE03567}
\end{barticle}
\endbibitem

\bibitem{Robinson2011}
\begin{barticle}
\bauthor{\bsnm{{Robinson}}, \binits{T.D.}},
\bauthor{\bsnm{{Meadows}}, \binits{V.S.}},
\bauthor{\bsnm{{Crisp}}, \binits{D.}},
\bauthor{\bsnm{{Deming}}, \binits{D.}},
\bauthor{\bsnm{{A'Hearn}}, \binits{M.F.}},
\bauthor{\bsnm{{Charbonneau}}, \binits{D.}},
\bauthor{\bsnm{{Livengood}}, \binits{T.A.}},
\bauthor{\bsnm{{Seager}}, \binits{S.}},
\bauthor{\bsnm{{Barry}}, \binits{R.K.}},
\bauthor{\bsnm{{Hearty}}, \binits{T.}},
\bauthor{\bsnm{{Hewagama}}, \binits{T.}},
\bauthor{\bsnm{{Lisse}}, \binits{C.M.}},
\bauthor{\bsnm{{McFadden}}, \binits{L.A.}},
\bauthor{\bsnm{{Wellnitz}}, \binits{D.D.}}:
\batitle{{Earth as an Extrasolar Planet: Earth Model Validation Using EPOXI
  Earth Observations}}.
\bjtitle{Astrobiology}
\bvolume{11}(\bissue{5}),
\bfpage{393}--\blpage{408}
(\byear{2011}).
\doiurl{10.1089/ast.2011.0642}
\end{barticle}
\endbibitem

\bibitem{Arney2014}
\begin{barticle}
\bauthor{\bsnm{{Arney}}, \binits{G.}},
\bauthor{\bsnm{{Meadows}}, \binits{V.}},
\bauthor{\bsnm{{Crisp}}, \binits{D.}},
\bauthor{\bsnm{{Schmidt}}, \binits{S.J.}},
\bauthor{\bsnm{{Bailey}}, \binits{J.}},
\bauthor{\bsnm{{Robinson}}, \binits{T.}}:
\batitle{{Spatially resolved measurements of H$_{2}$O, HCl, CO, OCS, SO$_{2}$,
  cloud opacity, and acid concentration in the Venus near-infrared spectral
  windows}}.
\bjtitle{Journal of Geophysical Research (Planets)}
\bvolume{119}(\bissue{8}),
\bfpage{1860}--\blpage{1891}
(\byear{2014}).
\doiurl{10.1002/2014JE004662}
\end{barticle}
\endbibitem

\bibitem{Baraffe2015}
\begin{barticle}
\bauthor{\bsnm{{Baraffe}}, \binits{I.}},
\bauthor{\bsnm{{Homeier}}, \binits{D.}},
\bauthor{\bsnm{{Allard}}, \binits{F.}},
\bauthor{\bsnm{{Chabrier}}, \binits{G.}}:
\batitle{{New evolutionary models for pre-main sequence and main sequence
  low-mass stars down to the hydrogen-burning limit}}.
\bjtitle{\aap}
\bvolume{577},
\bfpage{42}
(\byear{2015})
{\href{https://arxiv.org/abs/1503.04107}{{arXiv:1503.04107}}}
{[astro-ph.SR]}.
\doiurl{10.1051/0004-6361/201425481}
\end{barticle}
\endbibitem

\bibitem{Wright2018}
\begin{barticle}
\bauthor{\bsnm{{Wright}}, \binits{N.J.}},
\bauthor{\bsnm{{Newton}}, \binits{E.R.}},
\bauthor{\bsnm{{Williams}}, \binits{P.K.G.}},
\bauthor{\bsnm{{Drake}}, \binits{J.J.}},
\bauthor{\bsnm{{Yadav}}, \binits{R.K.}}:
\batitle{{The stellar rotation-activity relationship in fully convective M
  dwarfs}}.
\bjtitle{\mnras}
\bvolume{479}(\bissue{2}),
\bfpage{2351}--\blpage{2360}
(\byear{2018})
{\href{https://arxiv.org/abs/1807.03304}{{arXiv:1807.03304}}}
{[astro-ph.SR]}.
\doiurl{10.1093/mnras/sty1670}
\end{barticle}
\endbibitem

\bibitem{Dong2018}
\begin{barticle}
\bauthor{\bsnm{{Dong}}, \binits{C.}},
\bauthor{\bsnm{{Jin}}, \binits{M.}},
\bauthor{\bsnm{{Lingam}}, \binits{M.}},
\bauthor{\bsnm{{Airapetian}}, \binits{V.S.}},
\bauthor{\bsnm{{Ma}}, \binits{Y.}},
\bauthor{\bsnm{{van der Holst}}, \binits{B.}}:
\batitle{{Atmospheric escape from the TRAPPIST-1 planets and implications for
  habitability}}.
\bjtitle{Proceedings of the National Academy of Science}
\bvolume{115}(\bissue{2}),
\bfpage{260}--\blpage{265}
(\byear{2018})
{\href{https://arxiv.org/abs/1705.05535}{{arXiv:1705.05535}}}
{[astro-ph.EP]}.
\doiurl{10.1073/pnas.1708010115}
\end{barticle}
\endbibitem

\bibitem{Unterborn18}
\begin{barticle}
\bauthor{\bsnm{{Unterborn}}, \binits{C.T.}},
\bauthor{\bsnm{{Desch}}, \binits{S.J.}},
\bauthor{\bsnm{{Hinkel}}, \binits{N.R.}},
\bauthor{\bsnm{{Lorenzo}}, \binits{A.}}:
\batitle{{Inward migration of the TRAPPIST-1 planets as inferred from their
  water-rich compositions}}.
\bjtitle{Nature Astronomy}
\bvolume{2},
\bfpage{297}--\blpage{302}
(\byear{2018})
{\href{https://arxiv.org/abs/1706.02689}{{arXiv:1706.02689}}}
{[astro-ph.EP]}.
\doiurl{10.1038/s41550-018-0411-6}
\end{barticle}
\endbibitem

\bibitem{Sotin07}
\begin{barticle}
\bauthor{\bsnm{{Sotin}}, \binits{C.}},
\bauthor{\bsnm{{Grasset}}, \binits{O.}},
\bauthor{\bsnm{{Mocquet}}, \binits{A.}}:
\batitle{{Mass radius curve for extrasolar Earth-like planets and ocean
  planets}}.
\bjtitle{\icarus}
\bvolume{191}(\bissue{1}),
\bfpage{337}--\blpage{351}
(\byear{2007}).
\doiurl{10.1016/j.icarus.2007.04.006}
\end{barticle}
\endbibitem

\bibitem{Brugger16}
\begin{barticle}
\bauthor{\bsnm{{Brugger}}, \binits{B.}},
\bauthor{\bsnm{{Mousis}}, \binits{O.}},
\bauthor{\bsnm{{Deleuil}}, \binits{M.}},
\bauthor{\bsnm{{Lunine}}, \binits{J.I.}}:
\batitle{{Possible Internal Structures and Compositions of Proxima Centauri
  b}}.
\bjtitle{\apjl}
\bvolume{831}(\bissue{2}),
\bfpage{16}
(\byear{2016})
{\href{https://arxiv.org/abs/1609.09757}{{arXiv:1609.09757}}}
{[astro-ph.EP]}.
\doiurl{10.3847/2041-8205/831/2/L16}
\end{barticle}
\endbibitem

\bibitem{Brugger17}
\begin{barticle}
\bauthor{\bsnm{{Brugger}}, \binits{B.}},
\bauthor{\bsnm{{Mousis}}, \binits{O.}},
\bauthor{\bsnm{{Deleuil}}, \binits{M.}},
\bauthor{\bsnm{{Deschamps}}, \binits{F.}}:
\batitle{{Constraints on Super-Earth Interiors from Stellar Abundances}}.
\bjtitle{\apj}
\bvolume{850}(\bissue{1}),
\bfpage{93}
(\byear{2017})
{\href{https://arxiv.org/abs/1710.09776}{{arXiv:1710.09776}}}
{[astro-ph.EP]}.
\doiurl{10.3847/1538-4357/aa965a}
\end{barticle}
\endbibitem

\bibitem{Mousis20}
\begin{barticle}
\bauthor{\bsnm{{Mousis}}, \binits{O.}},
\bauthor{\bsnm{{Deleuil}}, \binits{M.}},
\bauthor{\bsnm{{Aguichine}}, \binits{A.}},
\bauthor{\bsnm{{Marcq}}, \binits{E.}},
\bauthor{\bsnm{{Naar}}, \binits{J.}},
\bauthor{\bsnm{{Aguirre}}, \binits{L.A.}},
\bauthor{\bsnm{{Brugger}}, \binits{B.}},
\bauthor{\bsnm{{Gon{\c{c}}alves}}, \binits{T.}}:
\batitle{{Irradiated Ocean Planets Bridge Super-Earth and Sub-Neptune
  Populations}}.
\bjtitle{\apjl}
\bvolume{896}(\bissue{2}),
\bfpage{22}
(\byear{2020})
{\href{https://arxiv.org/abs/2002.05243}{{arXiv:2002.05243}}}
{[astro-ph.EP]}.
\doiurl{10.3847/2041-8213/ab9530}
\end{barticle}
\endbibitem

\bibitem{Marcq12}
\begin{barticle}
\bauthor{\bsnm{{Marcq}}, \binits{E.}}:
\batitle{{A simple 1-D radiative-convective atmospheric model designed for
  integration into coupled models of magma ocean planets}}.
\bjtitle{Journal of Geophysical Research (Planets)}
\bvolume{117}(\bissue{E1}),
\bfpage{01001}
(\byear{2012}).
\doiurl{10.1029/2011JE003912}
\end{barticle}
\endbibitem

\bibitem{Marcq17}
\begin{barticle}
\bauthor{\bsnm{{Marcq}}, \binits{E.}},
\bauthor{\bsnm{{Salvador}}, \binits{A.}},
\bauthor{\bsnm{{Massol}}, \binits{H.}},
\bauthor{\bsnm{{Davaille}}, \binits{A.}}:
\batitle{{Thermal radiation of magma ocean planets using a 1-D
  radiative-convective model of H$_{2}$O-CO$_{2}$ atmospheres}}.
\bjtitle{Journal of Geophysical Research (Planets)}
\bvolume{122}(\bissue{7}),
\bfpage{1539}--\blpage{1553}
(\byear{2017}).
\doiurl{10.1002/2016JE005224}
\end{barticle}
\endbibitem

\bibitem{acuna_sub}
\begin{botherref}
\oauthor{\bsnm{{Acuna}}, \binits{L.}},
\oauthor{\bsnm{{Deleuil}}, \binits{M.}},
\oauthor{\bsnm{{Mousis}}, \binits{O.}}:
{Interior-atmosphere modelling to assess the observability of rocky planets
  with JWST}.
arXiv e-prints,
2305--01250
(2023)
{\href{https://arxiv.org/abs/2305.01250}{{arXiv:2305.01250}}}
{[astro-ph.EP]}.
\doiurl{10.48550/arXiv.2305.01250}
\end{botherref}
\endbibitem

\bibitem{Mann2019}
\begin{barticle}
\bauthor{\bsnm{{Mann}}, \binits{A.W.}},
\bauthor{\bsnm{{Dupuy}}, \binits{T.}},
\bauthor{\bsnm{{Kraus}}, \binits{A.L.}},
\bauthor{\bsnm{{Gaidos}}, \binits{E.}},
\bauthor{\bsnm{{Ansdell}}, \binits{M.}},
\bauthor{\bsnm{{Ireland}}, \binits{M.}},
\bauthor{\bsnm{{Rizzuto}}, \binits{A.C.}},
\bauthor{\bsnm{{Hung}}, \binits{C.-L.}},
\bauthor{\bsnm{{Dittmann}}, \binits{J.}},
\bauthor{\bsnm{{Factor}}, \binits{S.}},
\bauthor{\bsnm{{Feiden}}, \binits{G.}},
\bauthor{\bsnm{{Martinez}}, \binits{R.A.}},
\bauthor{\bsnm{{Ru{\'\i}z-Rodr{\'\i}guez}}, \binits{D.}},
\bauthor{\bsnm{{Thao}}, \binits{P.C.}}:
\batitle{{How to Constrain Your M Dwarf. II. The Mass-Luminosity-Metallicity
  Relation from 0.075 to 0.70 Solar Masses}}.
\bjtitle{\apj}
\bvolume{871}(\bissue{1}),
\bfpage{63}
(\byear{2019})
{\href{https://arxiv.org/abs/1811.06938}{{arXiv:1811.06938}}}
{[astro-ph.SR]}.
\doiurl{10.3847/1538-4357/aaf3bc}
\end{barticle}
\endbibitem

\bibitem{vanGrootel2018}
\begin{barticle}
\bauthor{\bsnm{{Van Grootel}}, \binits{V.}},
\bauthor{\bsnm{{Fernandes}}, \binits{C.S.}},
\bauthor{\bsnm{{Gillon}}, \binits{M.}},
\bauthor{\bsnm{{Jehin}}, \binits{E.}},
\bauthor{\bsnm{{Manfroid}}, \binits{J.}},
\bauthor{\bsnm{{Scuflaire}}, \binits{R.}},
\bauthor{\bsnm{{Burgasser}}, \binits{A.J.}},
\bauthor{\bsnm{{Barkaoui}}, \binits{K.}},
\bauthor{\bsnm{{Benkhaldoun}}, \binits{Z.}},
\bauthor{\bsnm{{Burdanov}}, \binits{A.}},
\bauthor{\bsnm{{Delrez}}, \binits{L.}},
\bauthor{\bsnm{{Demory}}, \binits{B.-O.}},
\bauthor{\bsnm{{de Wit}}, \binits{J.}},
\bauthor{\bsnm{{Queloz}}, \binits{D.}},
\bauthor{\bsnm{{Triaud}}, \binits{A.H.M.J.}}:
\batitle{{Stellar Parameters for Trappist-1}}.
\bjtitle{\apj}
\bvolume{853}(\bissue{1}),
\bfpage{30}
(\byear{2018})
{\href{https://arxiv.org/abs/1712.01911}{{arXiv:1712.01911}}}
{[astro-ph.SR]}.
\doiurl{10.3847/1538-4357/aaa023}
\end{barticle}
\endbibitem

\bibitem{Iyer2023}
\begin{barticle}
\bauthor{\bsnm{{Iyer}}, \binits{A.R.}},
\bauthor{\bsnm{{Line}}, \binits{M.R.}},
\bauthor{\bsnm{{Muirhead}}, \binits{P.S.}},
\bauthor{\bsnm{{Fortney}}, \binits{J.J.}},
\bauthor{\bsnm{{Gharib-Nezhad}}, \binits{E.}}:
\batitle{{The SPHINX M-dwarf Spectral Grid. I. Benchmarking New Model
  Atmospheres to Derive Fundamental M-dwarf Properties}}.
\bjtitle{\apj}
\bvolume{944}(\bissue{1}),
\bfpage{41}
(\byear{2023})
{\href{https://arxiv.org/abs/2206.12010}{{arXiv:2206.12010}}}
{[astro-ph.SR]}.
\doiurl{10.3847/1538-4357/acabc2}
\end{barticle}
\endbibitem

\bibitem{Fabrycky2010}
\begin{bchapter}
\bauthor{\bsnm{{Fabrycky}}, \binits{D.C.}}:
\bctitle{{Non-Keplerian Dynamics of Exoplanets}}.
In: \beditor{\bsnm{{Seager}}, \binits{S.}} (ed.)
\bbtitle{Exoplanets},
pp. \bfpage{217}--\blpage{238}
(\byear{2010})
\end{bchapter}
\endbibitem

\bibitem{Winn2010}
\begin{bchapter}
\bauthor{\bsnm{{Winn}}, \binits{J.N.}}:
\bctitle{{Exoplanet Transits and Occultations}}.
In: \beditor{\bsnm{{Seager}}, \binits{S.}} (ed.)
\bbtitle{Exoplanets},
pp. \bfpage{55}--\blpage{77}
(\byear{2010})
\end{bchapter}
\endbibitem

\bibitem{Agol2010}
\begin{barticle}
\bauthor{\bsnm{{Agol}}, \binits{E.}},
\bauthor{\bsnm{{Cowan}}, \binits{N.B.}},
\bauthor{\bsnm{{Knutson}}, \binits{H.A.}},
\bauthor{\bsnm{{Deming}}, \binits{D.}},
\bauthor{\bsnm{{Steffen}}, \binits{J.H.}},
\bauthor{\bsnm{{Henry}}, \binits{G.W.}},
\bauthor{\bsnm{{Charbonneau}}, \binits{D.}}:
\batitle{{The Climate of HD 189733b from Fourteen Transits and Eclipses
  Measured by Spitzer}}.
\bjtitle{\apj}
\bvolume{721}(\bissue{2}),
\bfpage{1861}--\blpage{1877}
(\byear{2010})
{\href{https://arxiv.org/abs/1007.4378}{{arXiv:1007.4378}}}
{[astro-ph.EP]}.
\doiurl{10.1088/0004-637X/721/2/1861}
\end{barticle}
\endbibitem

\bibitem{Lithwick2012}
\begin{barticle}
\bauthor{\bsnm{{Lithwick}}, \binits{Y.}},
\bauthor{\bsnm{{Xie}}, \binits{J.}},
\bauthor{\bsnm{{Wu}}, \binits{Y.}}:
\batitle{{Extracting Planet Mass and Eccentricity from TTV Data}}.
\bjtitle{\apj}
\bvolume{761}(\bissue{2}),
\bfpage{122}
(\byear{2012})
{\href{https://arxiv.org/abs/1207.4192}{{arXiv:1207.4192}}}
{[astro-ph.EP]}.
\doiurl{10.1088/0004-637X/761/2/122}
\end{barticle}
\endbibitem

\bibitem{numpy2020}
\begin{barticle}
\bauthor{\bsnm{{Harris}}, \binits{C.R.}},
\bauthor{\bsnm{{Millman}}, \binits{K.J.}},
\bauthor{\bsnm{{van der Walt}}, \binits{S.J.}},
\bauthor{\bsnm{{Gommers}}, \binits{R.}},
\bauthor{\bsnm{{Virtanen}}, \binits{P.}},
\bauthor{\bsnm{{Cournapeau}}, \binits{D.}},
\bauthor{\bsnm{{Wieser}}, \binits{E.}},
\bauthor{\bsnm{{Taylor}}, \binits{J.}},
\bauthor{\bsnm{{Berg}}, \binits{S.}},
\bauthor{\bsnm{{Smith}}, \binits{N.J.}},
\bauthor{\bsnm{{Kern}}, \binits{R.}},
\bauthor{\bsnm{{Picus}}, \binits{M.}},
\bauthor{\bsnm{{Hoyer}}, \binits{S.}},
\bauthor{\bsnm{{van Kerkwijk}}, \binits{M.H.}},
\bauthor{\bsnm{{Brett}}, \binits{M.}},
\bauthor{\bsnm{{Haldane}}, \binits{A.}},
\bauthor{\bsnm{{del R{\'\i}o}}, \binits{J.F.}},
\bauthor{\bsnm{{Wiebe}}, \binits{M.}},
\bauthor{\bsnm{{Peterson}}, \binits{P.}},
\bauthor{\bsnm{{G{\'e}rard-Marchant}}, \binits{P.}},
\bauthor{\bsnm{{Sheppard}}, \binits{K.}},
\bauthor{\bsnm{{Reddy}}, \binits{T.}},
\bauthor{\bsnm{{Weckesser}}, \binits{W.}},
\bauthor{\bsnm{{Abbasi}}, \binits{H.}},
\bauthor{\bsnm{{Gohlke}}, \binits{C.}},
\bauthor{\bsnm{{Oliphant}}, \binits{T.E.}}:
\batitle{{Array programming with NumPy}}.
\bjtitle{\nat}
\bvolume{585}(\bissue{7825}),
\bfpage{357}--\blpage{362}
(\byear{2020})
{\href{https://arxiv.org/abs/2006.10256}{{arXiv:2006.10256}}}
{[cs.MS]}.
\doiurl{10.1038/s41586-020-2649-2}
\end{barticle}
\endbibitem

\bibitem{matplotlib2007}
\begin{barticle}
\bauthor{\bsnm{{Hunter}}, \binits{J.D.}}:
\batitle{{Matplotlib: A 2D Graphics Environment}}.
\bjtitle{Computing in Science and Engineering}
\bvolume{9}(\bissue{3}),
\bfpage{90}--\blpage{95}
(\byear{2007}).
\doiurl{10.1109/MCSE.2007.55}
\end{barticle}
\endbibitem

\bibitem{astropy2022}
\begin{barticle}
\bauthor{\bsnm{{Astropy Collaboration}}},
\bauthor{\bsnm{{Price-Whelan}}, \binits{A.M.}},
\bauthor{\bsnm{{Lim}}, \binits{P.L.}},
\bauthor{\bsnm{{Earl}}, \binits{N.}},
\bauthor{\bsnm{{Starkman}}, \binits{N.}},
\bauthor{\bsnm{{Bradley}}, \binits{L.}},
\bauthor{\bsnm{{Shupe}}, \binits{D.L.}},
\bauthor{\bsnm{{Patil}}, \binits{A.A.}},
\bauthor{\bsnm{{Corrales}}, \binits{L.}},
\bauthor{\bsnm{{Brasseur}}, \binits{C.E.}},
\bauthor{\bsnm{{N{\"o}the}}, \binits{M.}},
\bauthor{\bsnm{{Donath}}, \binits{A.}},
\bauthor{\bsnm{{Tollerud}}, \binits{E.}},
\bauthor{\bsnm{{Morris}}, \binits{B.M.}},
\bauthor{\bsnm{{Ginsburg}}, \binits{A.}},
\bauthor{\bsnm{{Vaher}}, \binits{E.}},
\bauthor{\bsnm{{Weaver}}, \binits{B.A.}},
\bauthor{\bsnm{{Tocknell}}, \binits{J.}},
\bauthor{\bsnm{{Jamieson}}, \binits{W.}},
\bauthor{\bsnm{{van Kerkwijk}}, \binits{M.H.}},
\bauthor{\bsnm{{Robitaille}}, \binits{T.P.}},
\bauthor{\bsnm{{Merry}}, \binits{B.}},
\bauthor{\bsnm{{Bachetti}}, \binits{M.}},
\bauthor{\bsnm{{G{\"u}nther}}, \binits{H.M.}},
\bauthor{\bsnm{{Aldcroft}}, \binits{T.L.}},
\bauthor{\bsnm{{Alvarado-Montes}}, \binits{J.A.}},
\bauthor{\bsnm{{Archibald}}, \binits{A.M.}},
\bauthor{\bsnm{{B{\'o}di}}, \binits{A.}},
\bauthor{\bsnm{{Bapat}}, \binits{S.}},
\bauthor{\bsnm{{Barentsen}}, \binits{G.}},
\bauthor{\bsnm{{Baz{\'a}n}}, \binits{J.}},
\bauthor{\bsnm{{Biswas}}, \binits{M.}},
\bauthor{\bsnm{{Boquien}}, \binits{M.}},
\bauthor{\bsnm{{Burke}}, \binits{D.J.}},
\bauthor{\bsnm{{Cara}}, \binits{D.}},
\bauthor{\bsnm{{Cara}}, \binits{M.}},
\bauthor{\bsnm{{Conroy}}, \binits{K.E.}},
\bauthor{\bsnm{{Conseil}}, \binits{S.}},
\bauthor{\bsnm{{Craig}}, \binits{M.W.}},
\bauthor{\bsnm{{Cross}}, \binits{R.M.}},
\bauthor{\bsnm{{Cruz}}, \binits{K.L.}},
\bauthor{\bsnm{{D'Eugenio}}, \binits{F.}},
\bauthor{\bsnm{{Dencheva}}, \binits{N.}},
\bauthor{\bsnm{{Devillepoix}}, \binits{H.A.R.}},
\bauthor{\bsnm{{Dietrich}}, \binits{J.P.}},
\bauthor{\bsnm{{Eigenbrot}}, \binits{A.D.}},
\bauthor{\bsnm{{Erben}}, \binits{T.}},
\bauthor{\bsnm{{Ferreira}}, \binits{L.}},
\bauthor{\bsnm{{Foreman-Mackey}}, \binits{D.}},
\bauthor{\bsnm{{Fox}}, \binits{R.}},
\bauthor{\bsnm{{Freij}}, \binits{N.}},
\bauthor{\bsnm{{Garg}}, \binits{S.}},
\bauthor{\bsnm{{Geda}}, \binits{R.}},
\bauthor{\bsnm{{Glattly}}, \binits{L.}},
\bauthor{\bsnm{{Gondhalekar}}, \binits{Y.}},
\bauthor{\bsnm{{Gordon}}, \binits{K.D.}},
\bauthor{\bsnm{{Grant}}, \binits{D.}},
\bauthor{\bsnm{{Greenfield}}, \binits{P.}},
\bauthor{\bsnm{{Groener}}, \binits{A.M.}},
\bauthor{\bsnm{{Guest}}, \binits{S.}},
\bauthor{\bsnm{{Gurovich}}, \binits{S.}},
\bauthor{\bsnm{{Handberg}}, \binits{R.}},
\bauthor{\bsnm{{Hart}}, \binits{A.}},
\bauthor{\bsnm{{Hatfield-Dodds}}, \binits{Z.}},
\bauthor{\bsnm{{Homeier}}, \binits{D.}},
\bauthor{\bsnm{{Hosseinzadeh}}, \binits{G.}},
\bauthor{\bsnm{{Jenness}}, \binits{T.}},
\bauthor{\bsnm{{Jones}}, \binits{C.K.}},
\bauthor{\bsnm{{Joseph}}, \binits{P.}},
\bauthor{\bsnm{{Kalmbach}}, \binits{J.B.}},
\bauthor{\bsnm{{Karamehmetoglu}}, \binits{E.}},
\bauthor{\bsnm{{Ka{\l}uszy{\'n}ski}}, \binits{M.}},
\bauthor{\bsnm{{Kelley}}, \binits{M.S.P.}},
\bauthor{\bsnm{{Kern}}, \binits{N.}},
\bauthor{\bsnm{{Kerzendorf}}, \binits{W.E.}},
\bauthor{\bsnm{{Koch}}, \binits{E.W.}},
\bauthor{\bsnm{{Kulumani}}, \binits{S.}},
\bauthor{\bsnm{{Lee}}, \binits{A.}},
\bauthor{\bsnm{{Ly}}, \binits{C.}},
\bauthor{\bsnm{{Ma}}, \binits{Z.}},
\bauthor{\bsnm{{MacBride}}, \binits{C.}},
\bauthor{\bsnm{{Maljaars}}, \binits{J.M.}},
\bauthor{\bsnm{{Muna}}, \binits{D.}},
\bauthor{\bsnm{{Murphy}}, \binits{N.A.}},
\bauthor{\bsnm{{Norman}}, \binits{H.}},
\bauthor{\bsnm{{O'Steen}}, \binits{R.}},
\bauthor{\bsnm{{Oman}}, \binits{K.A.}},
\bauthor{\bsnm{{Pacifici}}, \binits{C.}},
\bauthor{\bsnm{{Pascual}}, \binits{S.}},
\bauthor{\bsnm{{Pascual-Granado}}, \binits{J.}},
\bauthor{\bsnm{{Patil}}, \binits{R.R.}},
\bauthor{\bsnm{{Perren}}, \binits{G.I.}},
\bauthor{\bsnm{{Pickering}}, \binits{T.E.}},
\bauthor{\bsnm{{Rastogi}}, \binits{T.}},
\bauthor{\bsnm{{Roulston}}, \binits{B.R.}},
\bauthor{\bsnm{{Ryan}}, \binits{D.F.}},
\bauthor{\bsnm{{Rykoff}}, \binits{E.S.}},
\bauthor{\bsnm{{Sabater}}, \binits{J.}},
\bauthor{\bsnm{{Sakurikar}}, \binits{P.}},
\bauthor{\bsnm{{Salgado}}, \binits{J.}},
\bauthor{\bsnm{{Sanghi}}, \binits{A.}},
\bauthor{\bsnm{{Saunders}}, \binits{N.}},
\bauthor{\bsnm{{Savchenko}}, \binits{V.}},
\bauthor{\bsnm{{Schwardt}}, \binits{L.}},
\bauthor{\bsnm{{Seifert-Eckert}}, \binits{M.}},
\bauthor{\bsnm{{Shih}}, \binits{A.Y.}},
\bauthor{\bsnm{{Jain}}, \binits{A.S.}},
\bauthor{\bsnm{{Shukla}}, \binits{G.}},
\bauthor{\bsnm{{Sick}}, \binits{J.}},
\bauthor{\bsnm{{Simpson}}, \binits{C.}},
\bauthor{\bsnm{{Singanamalla}}, \binits{S.}},
\bauthor{\bsnm{{Singer}}, \binits{L.P.}},
\bauthor{\bsnm{{Singhal}}, \binits{J.}},
\bauthor{\bsnm{{Sinha}}, \binits{M.}},
\bauthor{\bsnm{{Sip{\H{o}}cz}}, \binits{B.M.}},
\bauthor{\bsnm{{Spitler}}, \binits{L.R.}},
\bauthor{\bsnm{{Stansby}}, \binits{D.}},
\bauthor{\bsnm{{Streicher}}, \binits{O.}},
\bauthor{\bsnm{{{\v{S}}umak}}, \binits{J.}},
\bauthor{\bsnm{{Swinbank}}, \binits{J.D.}},
\bauthor{\bsnm{{Taranu}}, \binits{D.S.}},
\bauthor{\bsnm{{Tewary}}, \binits{N.}},
\bauthor{\bsnm{{Tremblay}}, \binits{G.R.}},
\bauthor{\bsnm{{de Val-Borro}}, \binits{M.}},
\bauthor{\bsnm{{Van Kooten}}, \binits{S.J.}},
\bauthor{\bsnm{{Vasovi{\'c}}}, \binits{Z.}},
\bauthor{\bsnm{{Verma}}, \binits{S.}},
\bauthor{\bsnm{{de Miranda Cardoso}}, \binits{J.V.}},
\bauthor{\bsnm{{Williams}}, \binits{P.K.G.}},
\bauthor{\bsnm{{Wilson}}, \binits{T.J.}},
\bauthor{\bsnm{{Winkel}}, \binits{B.}},
\bauthor{\bsnm{{Wood-Vasey}}, \binits{W.M.}},
\bauthor{\bsnm{{Xue}}, \binits{R.}},
\bauthor{\bsnm{{Yoachim}}, \binits{P.}},
\bauthor{\bsnm{{Zhang}}, \binits{C.}},
\bauthor{\bsnm{{Zonca}}, \binits{A.}},
\bauthor{\bsnm{{Astropy Project Contributors}}}:
\batitle{{The Astropy Project: Sustaining and Growing a Community-oriented
  Open-source Project and the Latest Major Release (v5.0) of the Core
  Package}}.
\bjtitle{\apj}
\bvolume{935}(\bissue{2}),
\bfpage{167}
(\byear{2022})
{\href{https://arxiv.org/abs/2206.14220}{{arXiv:2206.14220}}}
{[astro-ph.IM]}.
\doiurl{10.3847/1538-4357/ac7c74}
\end{barticle}
\endbibitem

\bibitem{Speagle2020}
\begin{barticle}
\bauthor{\bsnm{{Speagle}}, \binits{J.S.}}:
\batitle{{DYNESTY: a dynamic nested sampling package for estimating Bayesian
  posteriors and evidences}}.
\bjtitle{\mnras}
\bvolume{493}(\bissue{3}),
\bfpage{3132}--\blpage{3158}
(\byear{2020})
{\href{https://arxiv.org/abs/1904.02180}{{arXiv:1904.02180}}}
{[astro-ph.IM]}.
\doiurl{10.1093/mnras/staa278}
\end{barticle}
\endbibitem

\bibitem{Koposov2023}
\begin{botherref}
\oauthor{\bsnm{{Koposov}}, \binits{S.}},
\oauthor{\bsnm{{Speagle}}, \binits{J.}},
\oauthor{\bsnm{{Barbary}}, \binits{K.}},
\oauthor{\bsnm{{Ashton}}, \binits{G.}},
\oauthor{\bsnm{{Bennett}}, \binits{E.}},
\oauthor{\bsnm{{Buchner}}, \binits{J.}},
\oauthor{\bsnm{{Scheffler}}, \binits{C.}},
\oauthor{\bsnm{{Cook}}, \binits{B.}},
\oauthor{\bsnm{{Talbot}}, \binits{C.}},
\oauthor{\bsnm{{Guillochon}}, \binits{J.}},
\oauthor{\bsnm{{Cubillos}}, \binits{P.}},
\oauthor{\bsnm{{Asensio Ramos}}, \binits{A.}},
\oauthor{\bsnm{{Johnson}}, \binits{B.}},
\oauthor{\bsnm{{Lang}}, \binits{D.}},
\oauthor{\bsnm{{Ilya}}},
\oauthor{\bsnm{{Dartiailh}}, \binits{M.}},
\oauthor{\bsnm{{Nitz}}, \binits{A.}},
\oauthor{\bsnm{{McCluskey}}, \binits{A.}},
\oauthor{\bsnm{{Archibald}}, \binits{A.}},
\oauthor{\bsnm{{Deil}}, \binits{C.}},
\oauthor{\bsnm{{Foreman-Mackey}}, \binits{D.}},
\oauthor{\bsnm{{Goldstein}}, \binits{D.}},
\oauthor{\bsnm{{Tollerud}}, \binits{E.}},
\oauthor{\bsnm{{Leja}}, \binits{J.}},
\oauthor{\bsnm{{Kirk}}, \binits{M.}},
\oauthor{\bsnm{{Pitkin}}, \binits{M.}},
\oauthor{\bsnm{{Sheehan}}, \binits{P.}},
\oauthor{\bsnm{{Cargile}}, \binits{P.}},
\oauthor{\bsnm{{Patel}}, \binits{R.}},
\oauthor{\bsnm{{Angus}}, \binits{R.}}:
{joshspeagle/dynesty: v2.1.0}.
Zenodo
(2023).
\doiurl{10.5281/zenodo.7600689}
\end{botherref}
\endbibitem

\end{thebibliography}


\end{document}